\pdfoutput=1
\documentclass[aps,prd,twocolumn,superscriptaddress,preprintnumbers,floatfix,nofootinbib]{revtex4-2}

\usepackage{graphicx}
\usepackage{amsmath}
\usepackage{color}
\usepackage{dcolumn}
\usepackage{tensor}
\usepackage{bm}
\usepackage{makecell}
\usepackage[T1]{fontenc}
\usepackage[utf8]{inputenc}
\usepackage{microtype}
\usepackage{etoolbox}
\usepackage{amssymb}
\usepackage{mathrsfs}
\usepackage{accents}
\usepackage[normalem]{ulem}
\usepackage[table,dvipsnames]{xcolor}
\usepackage[colorlinks,urlcolor=NavyBlue,citecolor=NavyBlue,linkcolor=NavyBlue,pdfusetitle]{hyperref}
\usepackage[all]{hypcap}
\usepackage[inline,shortlabels]{enumitem}
\usepackage{braket}

\usepackage{array}
\usepackage{diagbox}
\usepackage{color}
\usepackage{colortbl}
\usepackage{hhline}
\usepackage{multirow}

\graphicspath{{./}{../python/figures/}}

\newcommand{\scri}{\mathscr{I}}
\newcommand{\NR}{\text{NR}}

\DeclareMathOperator*{\argmin}{arg\,min}
\DeclareMathOperator{\sgn}{sgn}
\renewcommand{\Re}{\text{Re}}
\renewcommand{\Im}{\text{Im}}

\makeatletter

\def\CT@@do@color{%
  \global\let\CT@do@color\relax
  \@tempdima\wd\z@
  \advance\@tempdima\@tempdimb
  \advance\@tempdima\@tempdimc
  \advance\@tempdimb\tabcolsep
  \advance\@tempdimc\tabcolsep
  \advance\@tempdima2\tabcolsep
  \kern-\@tempdimb
  \leaders\vrule
  \hskip\@tempdima\@plus  1fill
  \kern-\@tempdimc
  \hskip-\wd\z@ \@plus -1fill }
\makeatother

\newcommand{\figMirrorDemo}{%
  \begin{figure}
    \centering
    \includegraphics[width=\linewidth]{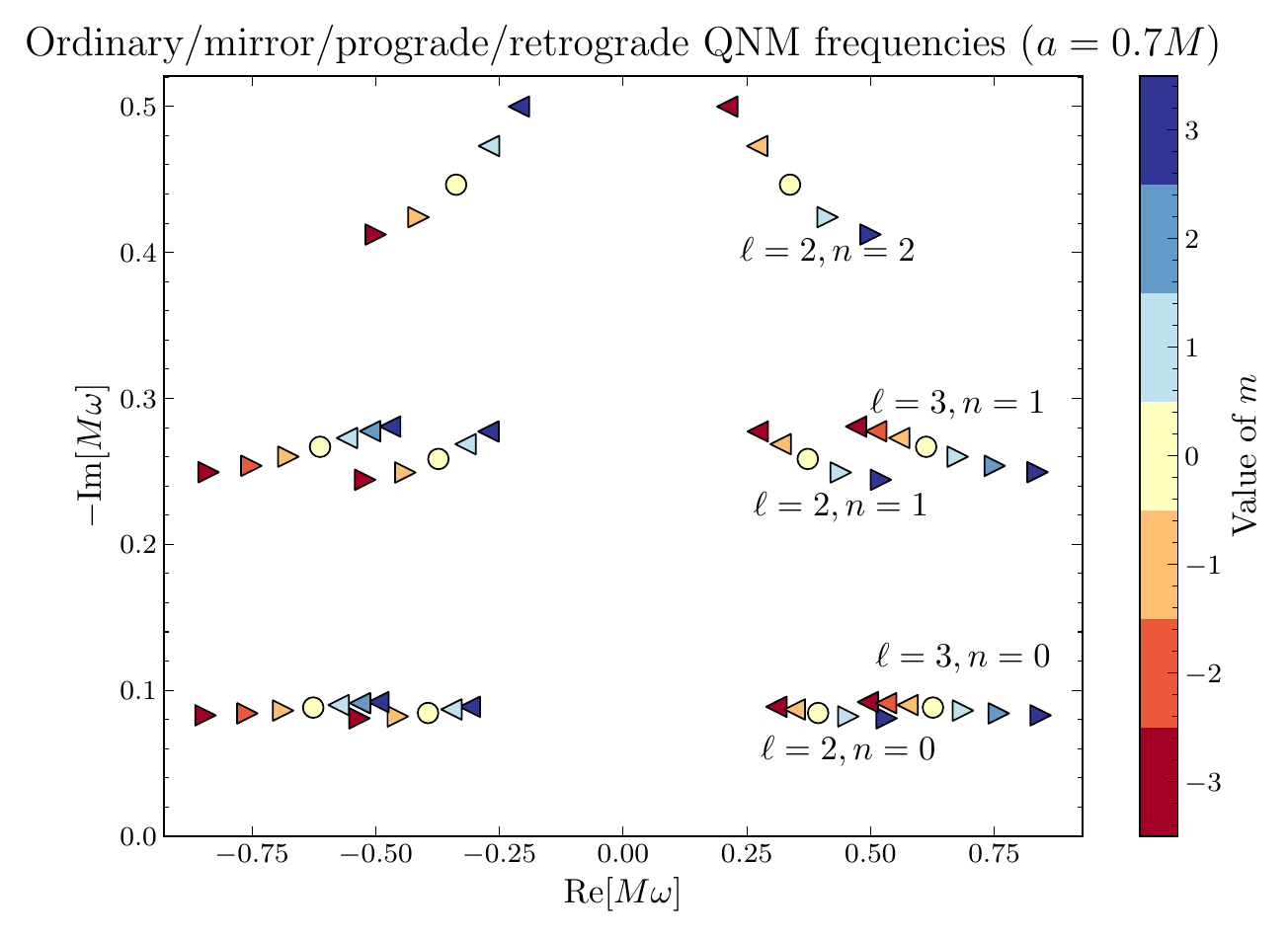}
    \caption{%
      Right-pointing triangles are prograde modes, left-pointing triangles are
      retrograde. Note that prograde and retrograde modes are present
      both in the left half-plane (so-called ``mirror'' modes) and the
      right half-plane (ordinary modes).
    }
    \label{fig:MirrorDemo}
  \end{figure}%
}

\usepackage{orcidlink}

\newcommand{\Cornell}{\affiliation{Cornell Center for Astrophysics and Planetary Science,
		Cornell University, Ithaca, New York 14853, USA}}
\newcommand{\Caltech}{\affiliation{Theoretical Astrophysics 350-17,
		California Institute of Technology, Pasadena, California 91125, USA}}
\newcommand{\OleMiss}{\affiliation{Department of Physics and Astronomy,
		University of Mississippi, University, Mississippi 38677, USA}}
\newcommand{\PennState}{\affiliation{Institute for Gravitation and the Cosmos \& Physics Department, Penn State, University Park, Pennsylvania 16802, USA}}
\newcommand{\MaxPlanck}{\affiliation{Max Planck Institute for Gravitational Physics (Albert Einstein Institute), Am M{\"u}hlenberg 1, Potsdam 14476, Germany}}

\begin{document}

\title{High Precision Ringdown Modeling: Multimode Fits and BMS Frames}

\author{Lorena \surname{Magaña Zertuche}
	\orcidlink{0000-0003-1888-9904}}
\email{lmaganaz@go.olemiss.edu}
\OleMiss
\author{Keefe Mitman
	\orcidlink{0000-0003-0276-3856}}
\email{kmitman@caltech.edu}
\Caltech
\author{Neev Khera
	\orcidlink{0000-0003-3515-2859}}
\email{neevkhera@psu.edu}
\PennState
\author{Leo C. Stein
	\orcidlink{0000-0001-7559-9597}}
\email{lcstein@olemiss.edu}
\OleMiss
\author{\\Michael Boyle
	\orcidlink{0000-0002-5075-5116}}
\Cornell
\author{Nils Deppe
	\orcidlink{0000-0003-4557-4115}}
\Caltech
\author{François Hébert
	\orcidlink{0000-0001-9009-6955}}
\Caltech
\author{Dante A. B. Iozzo
	\orcidlink{0000-0002-7244-1900}}
\Cornell
\author{Lawrence E. Kidder~%
	\orcidlink{0000-0001-5392-7342}}
\Cornell
\author{Jordan Moxon
	\orcidlink{0000-0001-9891-8677}}
\Caltech
\author{Harald P.~Pfeiffer~%
	\orcidlink{0000-0001-9288-519X}}
\MaxPlanck
\author{Mark A. Scheel
	\orcidlink{0000-0001-6656-9134}}
\Caltech
\author{Saul A. Teukolsky
	\orcidlink{0000-0001-9765-4526}}
\Caltech
\Cornell
\author{William Throwe
	\orcidlink{0000-0001-5059-4378}}
\Cornell
\author{Nils Vu
	\orcidlink{0000-0002-5767-3949}}
\MaxPlanck

\hypersetup{pdfauthor={\surname{Magaña Zertuche} et al.}}

\date{\today}

\begin{abstract}
  Quasi-normal mode (QNM) modeling is an invaluable tool for characterizing remnant black holes, studying
  strong gravity, and testing general relativity.
  Only recently have QNM studies begun to focus on multimode fitting
  to numerical relativity strain waveforms.
  As gravitational wave observatories become even more sensitive they
  will be able to resolve higher-order modes.
  Consequently, multimode QNM fits will be critically important, and in turn require a
  more thorough treatment of the asymptotic frame at $\scri^{+}$.
  The first main result of this work is a method for systematically fitting a
  QNM model containing many modes to a numerical waveform produced
  using Cauchy-characteristic extraction (CCE), a waveform extraction technique which is
  known to resolve memory effects.
  We choose the modes to model based on their power contribution to the
  residual between numerical and model waveforms.
  We show that the all-mode strain mismatch improves by a factor of
  $\sim 10^5$ when using multimode fitting as opposed to only fitting the
  $(2,\pm 2,n)$ modes.
  Our most significant result addresses a critical point that has been
  overlooked in the QNM literature: the importance of matching the
  Bondi-van der Burg-Metzner-Sachs (BMS) frame of the numerical waveform to that of the QNM model.
  We show that by mapping the numerical waveforms---which exhibit the memory effect---to a BMS frame known as the
  super rest frame, there is an improvement of $\sim 10^5$ in the
  all-mode strain mismatch compared to using a strain waveform whose BMS
  frame is not fixed.
  Furthermore, we find that by mapping CCE waveforms to the super rest frame, we can obtain
  all-mode mismatches that are, on average, a factor of $\sim 4$ better than using the publicly-available extrapolated waveforms. 
  We illustrate the effectiveness of these modeling enhancements by
  applying them to families of waveforms produced by numerical
  relativity and comparing our results to previous QNM studies.
\end{abstract}

\maketitle

\section{Introduction}

When in a vacuum a black hole can be considered one of the simplest
objects in the universe, since it is fully characterized by its mass
and angular momentum. Despite this simplicity, black holes continue to be challenging to study with a multitude of important and unanswered questions
concerning them~\cite{Barack_2019}. Currently, a rather promising
means of studying black holes is through gravitational wave astronomy,
i.e., using observations of gravitational waves emitted by binary
black hole (BBH), black hole-neutron star (BH-NS), and possibly binary
neutron star (BNS) mergers to study properties of the perturbed
remnant black holes~\cite{LIGOScientific:2016aoc,LIGOScientific:2017vwq,
LIGOScientific:2021djp,LIGOScientific:2021sio}. While the waveform that is emitted during the
merger phase is challenging to model and requires the aid of numerical
simulations~\cite{Pretorius:2005gq,Boyle:2019kee,Jani_2016,Healy_2017,Healy2020},
the radiation emitted by the remnant black hole during its ringdown
phase is expected to oscillate at a certain set of well-understood
frequencies, called quasi-normal mode (QNM) frequencies, until the
remnant black hole settles into a final state of
equilibrium~\cite{Teukolsky:1973}.

Fortunately, these QNM frequencies can be computed using perturbation
theory and are completely determined by the remnant's mass and spin,
thereby allowing for a thorough analysis of the remnant black hole's
properties, provided the QNM model is used properly when fitting to
the observed gravitational waves~\cite{Detweiler:1980gk, Leaver:1985ax, 
Dolan:2009nk, Cook:2014cta, Isi:2021iql}. Often, these QNMs are labeled by the
numbers $\{\ell,m,n\}$, where $(\ell,m)$ are the angular numbers
that correspond to spin-weighted spheroidal harmonics (see~\cite{Press:1973zz}) and $n=0,1,\ldots$
is the overtone number that sorts the QNM frequencies in order of
decreasing damping timescales, with the fundamental $n=0$ mode being
the least-damped mode.

Recently, gravitational wave analysis efforts have shifted their focus from modes 
with $\ell=2$, $m=2$ to multiple $(\ell,m)$ modes through studies that explore 
the effects of overtones, retrograde modes, 
and mode-mixing~\cite{Giesler:2019uxc, Cook:2020otn, London:2014cma,
Baibhav:2018rfk, Berti:2014fga, Dhani:2021vac, Finch:2021iip, Dhani:2020nik, Li:2021wgz}.\footnote{%
By mode-mixing effects, we mean the mixing that occurs when writing
a QNM model in a spherical harmonic basis, rather than its preferred
spheroidal harmonic basis.} Moreover, third-generation, ground-based
detectors such as the Einstein Telescope (ET) and Cosmic Explorer (CE) 
are expected to observe roughly $10^{2} - 10^{4}$ events per year with ringdown
signals that will be strong enough to exhibit various higher-order
mode contributions, which, until now, have not been systematically studied~\cite{Maggiore:2019uih,Baibhav:2019gxm}.
Consequently, being able to include higher-order modes in QNM models is
vitally important for future analyses, since this will be essential for ensuring that the
dynamics of the observed remnant black holes are accurately
captured and measured. In addition, the Laser Interferometer Space Antenna (LISA)
will be even more sensitive to the ringdown phase of compact mergers, thereby
allowing for even more inclusive multi-frequency ringdown studies that
may aid with black hole astronomy and testing
various theories of relativity~\cite{Berti:2005ys, Barausse:2020rsu, Holley-Bockelmann:2020lzp}.

Apart from the clear importance of including overtones, retrograde modes, mode-mixing effects,
and higher-order modes in analyses of gravitational waves that are emitted
during ringdown~\cite{Giesler:2019uxc, Cook:2020otn, London:2014cma,
Baibhav:2018rfk, Berti:2014fga, Dhani:2021vac, Finch:2021iip, Dhani:2020nik, Li:2021wgz}, there is one other crucial component to QNM modeling
that is absent in prior QNM studies: ensuring that the waveforms and the QNM model are
in identical Bondi-van der Burg-Metzner-Sachs (BMS) frames.
Until now, studies that have compared numerical relativity (NR) waveforms to QNMs
have not considered the frame of their waveforms during the ringdown phase. 
While many of them have used NR waveforms whose inspiral phase has been mapped
to the center-of-mass frame~\cite{Boyle:2019kee,Woodford:2019tlo},\footnote{See~\cite{Mitman:2021xkq} for 
an improved way to map to the center-of-mass frame.} this is very different from mapping the
ringdown phase to a certain frame, such as the center-of-mass frame of the remnant BH. Nonetheless, even if one were
to map the remnant to the center-of-mass frame, rather than the inspiraling BHs, this procedure would
still be lacking due to a subtle, but important feature of relativity. Namely, the fact that for asymptotically flat spacetimes the 
symmetry group of future null infinity $\scri^{+}$ is not the Poincaré group, but the
infinite-dimensional BMS group~\cite{Bondi,Sachs}.

Fundamentally, the BMS group is just an extension of the Poincaré
group, in which spacetime translations are replaced by an encompassing collection of transformations known as
\emph{supertranslations}. When working with Bondi coordinates
$(u\equiv t-r,r,\theta,\phi)$, these supertranslations can be
understood rather simply as being direction-dependent time
translations. Namely, a supertranslation transforms the time coordinate via
$u\to u - \alpha(\theta,\phi)$, with $\alpha$  being an arbitrary function. 
Therefore, when fitting QNMs to a waveform, one not only
needs to map to the center-of-mass frame with the remnant BH's spin aligned with the positive $z$-axis, i.e., fixing the Poincaré frame, but they also need to fix
the supertranslation freedom of their waveforms to ensure that
comparisons with QNMs are meaningful. Using an incorrect BMS frame
leads to two effects that are sources of errors in the fits: the
waveform is shifted and settles down to a nonzero value, and
there is a mixing of modes that is distinct from the
spherical-spheroidal mixing mentioned before~\cite{Boyle:2015nqa,Mitman:2021xkq}.

Across this work, we perform QNM fits by including every one of the
important aforementioned components to modeling NR ringdowns
with QNMs, i.e., overtones, retrograde modes, mode-mixing, higher-order modes, and BMS
frame fixing. More specifically, we simultaneously fit various modes
over all angles on the two-sphere while also accounting for the
mode-mixing that occurs because NR waveforms are in a spherical
harmonic basis, while our QNM model is in a spheroidal harmonic
basis. When trying to model such a large number of modes and their
overtones, we must choose which modes to model. We do this
systematically by examining which modes in our model contain the
largest portion of unmodeled power (see Sec.~\ref{sec:GreedyAlgo} for
more detail). Apart from this, we also map our NR waveforms to the BMS
frame that is expected by the QNM model, namely, the super rest frame
(see Sec.~\ref{sec:BMSFramesIntro} for more
detail)~\cite{Mitman:2021xkq,Moreschi:1988pc,Moreschi:1998mw,Dain:2002mj}.\footnote{Note
  that in this work when we refer to mapping a waveform to the super
  rest frame we really mean simultaneously mapping to the
  remnant BH's center-of-mass frame, aligning the remnant BH's spin with the positive $z$-axis, and fixing the supertranslation freedom by mapping to the super rest frame of~\cite{Mitman:2021xkq}.}
We, therefore, create a QNM model by choosing modes based on their 
unmodeled power and, for the first time, fit said model to a NR waveform 
that has been properly mapped to the same BMS frame as is expected 
by the Teukolsky formalism~\cite{Teukolsky:1973}. We find that by carrying
out this procedure, i.e., fitting over the whole two-sphere and
accounting for BMS frames, we can drastically enhance previous
results, such as the GW150914 investigations in Giesler et al.~\cite{Giesler:2019uxc} and
Cook~\cite{Cook:2020otn}, by both reducing mismatches
between NR waveforms and QNM models by a factor of $10^5$ as 
well as improving parameter estimates of the remnant black hole's 
characteristics by more than half an order of magnitude using QNMs.
We tested the effects of multimode modeling and frame fixing with 
14 SXS simulations (see Table~\ref{tab:runs}), which include systems 
of mass ratio 1 and 4, with varying spin configurations, including 
precessions. We also provide an in-depth study on the simulation 
\texttt{SXS:BBH:0305}, a proxy for GW150914.

We present our computations and results as follows. In
Sec.~\ref{sec:Formalism}, we outline the mathematical conventions for
waveform modeling that is used throughout the paper. Furthermore, we also
discuss the reason why fixing the BMS frame is important and present
how we will map our waveforms to the super rest frame. Next, in
Sec.~\ref{sec:GreedyAlgo}, we discuss our greedy algorithm for
choosing modes to include in our QNM model and highlight the
importance of multimode fitting for ringdown modeling. We also show
how multimode fitting affects the mismatch between a NR waveform and
a QNM model. Finally, in Sec.~\ref{sec:BMSFrames} we show the
importance of mapping the remnant black hole to the super rest frame
and the consequences of using the correct BMS frame when fitting to
QNMs.

\section{Formalism}
\label{sec:Formalism}

The remnant black hole formed from a BBH merger is well-described as a
supertranslated Kerr metric with (potentially large) perturbations, which decay with time.
As shown by Teukolsky in 1973~\cite{Teukolsky:1973}, applying
perturbation theory to the Kerr spacetime, one acquires a decoupled
``master equation'' that describes linear curvature perturbations of
the black hole.  The Teukolsky partial differential equation also
separates into temporal, radial, and angular ordinary differential
equations. The oscillatory yet exponentially decaying modal solutions 
are the QNMs, and the spacetime after merger is modeled as a linear 
superposition of many of these QNMs. Imposing appropriate conditions 
of decay (at null infinity, $\scri$) and regularity
(at the horizon and poles) quantizes the allowed complex QNM
frequencies~\cite{Leaver:1985ax}.  One finds the frequencies,
separation constants, and angular wavefunctions simultaneously, either
via Leaver's original approach of infinite continued
fractions~\cite{Leaver:1985ax}, or a more recent spectral eigenvalue
approach of Cook and Zalutskiy~\cite{Cook:2014cta, Cook:2020otn}.  The
spectral approach finds the angular mode shapes---the spin-weighted
spheroidal harmonics---as a decomposition in spin-weighted spherical
harmonics. This decomposition has been employed before 
in~\cite{London:2014cma, Berti:2014fga, Lim:2019xrb, Dhani:2020nik, Cook:2020otn, Li:2021wgz}. We obtain 
the QNM frequencies and spherical-spheroidal decomposition coefficients 
from the open-source Python package {\tt qnm}, which uses a Leaver 
solver for the radial sector, with the spectral eigenvalue approach for the 
angular sector~\cite{Leaver:1985ax, Stein:2019mop}. For more details on 
the implementation see~\cite{Stein:2019mop} and references therein.

In this study, we are exclusively interested in working with complex
waveforms, like the strain $h$,\footnote{We explicitly define the
  strain as described in Appendix C of~\cite{Boyle:2019kee}.} which
are decomposed into spin-weight $-2$ spherical harmonic bases and live
on future null infinity $\scri^{+}$. We write these waveforms as,
e.g.,
\begin{align}
\label{eq:hlm}
h(u, \theta, \phi)=\sum\limits_{\ell\geq 2,|m|\leq\ell}h_{\ell m}(u) \ {}_{-2}Y_{\ell m}(\theta,\phi).
\end{align}
The spin-weighted spherical harmonics of fixed spin-weight $s$ form a
complete and orthonormal basis on the
two-sphere~\cite{Goldberg:1966uu, Newman:1966ub},
\begin{align}
\label{eq:orthog}
\int_{S^{2}} ({}_{s}Y_{\ell m})^{*} \ {}_{s}Y_{\ell' m'} \, d\Omega=\delta_{\ell \ell'} \delta_{m m'},
\end{align}
where ${}^{*}$ denotes complex conjugation and the differential
$d\Omega = d\cos\theta\,d\phi$ is the standard volume element on the
two-sphere. 

\subsection{QNM formalism}

For a given black hole spin $|a|<M$, and choice of angular numbers
$(\ell, m)$, there are an infinite number of QNM frequencies
satisfying the boundary conditions.  These are all in the lower
half-plane, $\Im[\omega]<0$, as required by stability.  The symmetry
of the equations means that if $\omega_{\ell m}$ is a QNM frequency,
then so is $-\omega_{\ell , -m}^{*}$; this is a ``mirror'' symmetry
between the left and right half-planes.  This leads to the
nomenclature (see also Table I of~\cite{Li:2021wgz})
\begin{itemize}
\item ``ordinary'' mode: $\Re[\omega] > 0$,
\item ``mirror'' mode: $\Re[\omega] < 0$.
\end{itemize}
Because of this symmetry, much of the QNM literature has focused on
just the ordinary modes, since the mirror modes can be recovered via
the transformation $\{ m\to -m, \omega \to -\omega^{*} \}$.  Within
each family, there are still an infinite number of overtones.
Thus to uniquely identify each solution, we label
the mode
$
  \omega^{p}_{\ell m n}
$
with
$\{ \ell, m, n, p \}$, where
$n=0, 1,\ldots$ is the overtone number, ordered by the magnitude of
$\Im[\omega]$, and $p = \sgn(\Re[\omega])$ is either $\pm 1$. 
The least-damped $n=0$ mode is often called 
the fundamental mode or zeroth tone while the $n>0$ modes are referred 
to as overtones~\cite{Isi:2021iql}.

Moreover, there are prograde and retrograde modes in both the right and
left (mirror) half-planes~\cite{London:2018nxs,Lim:2019xrb,Isi:2021iql,Li:2021wgz}. 
A QNM is labeled prograde if its
wavefronts circulate around the BH in the same sense as its rotation.
Because a QNM solution goes as $\propto \exp(-i\omega t + i m \phi)$, we
see that surfaces of constant phase circulate in the positive $\phi$ direction when
\begin{itemize}
\item ``prograde'' mode: $\sgn(m)=+\sgn(\Re[\omega])$,
\end{itemize}
and in the negative $\phi$ direction when
\begin{itemize}
\item ``retrograde'' mode: $\sgn(m)=-\sgn(\Re[\omega])$.
\end{itemize}
Modes with $m=0$ cannot be labeled as either prograde or retrograde.
This is demonstrated in Fig.~\ref{fig:MirrorDemo}.

\figMirrorDemo

When a perturber is corotating with the spin of the black hole, it
dominantly excites the prograde modes.  For most binary coalescences,
the remnant spin ends up with a positive projection onto the direction
of the orbital angular momentum at plunge; thus the prograde modes are
expected to be most important. In this study, we found that the power of $m\neq 0$
retrograde modes was a very small fraction of the total power, but we nonetheless
include them to provide a more complete picture of QNM modeling and attain 
marginally higher accuracies. Note though that for $m=0$ modes,
neither of the pair of mirror modes is dominant, so both must be included in the fits.

In addition to the frequency, there is also an associated angular mode
distribution for a given QNM, which is given by a spin-weighted
spheroidal harmonic function ${}_{s} S_{\ell m}(\theta,\phi; c)$ that
solves the separated angular equation~\cite{Teukolsky:1973}.
Here $\theta$ and $\phi$ represent the polar and azimuthal angles, and
crucially this is a coordinate system where the black hole is at
rest and its spin vector is along the $\theta=0$ direction.

The complex oblateness parameter $c=a \omega_{\ell m n}$ is determined
by both the QNM frequency and the BH's spin parameter $a=|J|/M$, with
$0 \le a <M$ for a horizon to exist.  When $c=0$, a spheroidal
harmonic reduces to a spherical harmonic.
For a fixed value of $c$ which is purely real or imaginary,
we obtain a complete and orthonormal basis of
oblate or prolate spheroidal harmonics. However, we have complex values of 
$c$, and a different $c$ associated to each QNM, so they may no longer 
form a complete basis.

Therefore, we choose to work in the complete basis of spin-weighted
spherical harmonics.
Each spheroidal harmonic can be decomposed as a series of
spherical harmonics with the same $m$ but different
$\ell$ as
\begin{align}
\label{eq:spheroidal-spherical-decomp}
{}_{s} S_{\ell' m}(\theta,\phi; c)=\sum\limits_{\ell}C_{\ell\ell' m}(c) \ {}_{s}Y_{\ell m}(\theta,\phi),
\end{align}
where the $C_{\ell \ell' m}(c)$ functions are called the spherical-spheroidal
mixing coefficients~\cite{Berti:2014fga}.
Here we follow the conventions of~\cite{Cook:2014cta}, as implemented
in~\cite{Stein:2019mop}.  These conventions are that $\sum_{\ell}
|C_{\ell \ell' m}|^{2} =1$, and that $C_{\ell \ell' m}$ is purely real
when $\ell=\ell'$.
In the case where $c=0$, we then have $C_{\ell\ell'm}(0)=\delta_{\ell\ell'}$, i.e.,
${}_{s}S_{\ell' m}(\theta,\phi; 0)={}_{s}Y_{\ell'
  m}(\theta,\phi)$.

With these spheroidal harmonics in mind, we can now write the general
ansatz for the strain of a ringing black hole at future null infinity,
$h^{\text{Q}}(u,\theta,\phi)$.  This ansatz is simply a linear
combination of QNMs
\begin{align}
\label{eq:h-QNM-spheroidal}
h^{Q}(u,\theta,\phi)&=\sum_{\ell',m,n,p} \mathcal{A}^{p}_{\ell' m n} e^{-i\omega^{p}_{\ell' m n}\left(u-u_{0}\right)}\nonumber\\
&\phantom{\sum_{\ell',m,n,p}[}{}_{-2} S_{\ell' m}(\theta,\phi;a\omega^{p}_{\ell'mn}),
\end{align}
where the $\mathcal{A}^{p}_{\ell' mn}$ are complex amplitudes for
each QNM and $u_{0}$ is a freely-specified start time of the QNM model. Although
these amplitudes transform in a simple way under translations of the retarded time coordinate
$u$ and rotations about the $z$-axis, they are not rotated by the Wigner-$D$
matrix under more general rotations.

Now we insert the spherical-spheroidal decomposition from
Eq.~\eqref{eq:spheroidal-spherical-decomp} into the spheroidal ansatz
of Eq.~\eqref{eq:h-QNM-spheroidal}, rewriting it as
\begin{align}
\label{eq:full-analytical}
h^{Q}(u,\theta,\phi)&=\sum_{\ell',m,n,p}\Bigg[ \mathcal{A}^{p}_{\ell' m n} e^{-i\omega^{p}_{\ell' m n}\left(u-u_{0}\right)}\nonumber\\
&\phantom{=.\sum_{\ell' m n}[}\sum_{\ell} C_{\ell\ell'm}(a\omega^{p}_{\ell'mn}){}_{-2} Y_{\ell m}(\theta,\phi)\Bigg].
\end{align}
Since this is now in the spin-weighted spherical harmonic
basis, it is ideal for modeling numerical relativity results. 
Writing the QNM ansatz $h^{\text{Q}}(u,\theta,\phi)$
in spin-weighted spherical harmonics as in
Eq.~\eqref{eq:hlm} and (using completeness) matching up the coefficients of
${}_{-2} Y_{\ell m}(\theta,\phi)$, we readily find that the spherical
mode-decomposed analytical QNM model is
\begin{align}
  \label{eq:analytical}
  h^{Q}_{\ell m}(u) = \sum_{\ell', n, p}
  \mathcal{A}^{p}_{\ell' m n} e^{-i\omega^{p}_{\ell' m n}\left(u-u_{0}\right)}
  C_{\ell\ell'm}(a\omega^{p}_{\ell'mn})
  \,.
\end{align}
These $h^{Q}_{\ell m}$'s do indeed rotate under the 
Wigner-$D$ matrix since they are expressed in 
the spin-weighted spherical harmonic basis~\cite{Boyle:2016tjj}. 
Such a rotation is explicitly shown by Eq.~(12) of~\cite{Cook:2020otn} 
with a couple of subtle differences between that paper and this one.
In this work, we use the conventions of~\cite{Cook:2014cta}, where 
$C$ is the spherical-spheroidal mixing coefficient.  Additionally, we
do not express $\omega^-$ or $C(a\omega^-)$ in terms of their positive frequency
counterparts, via
\begin{align}
  C_{\ell \ell' m} \left(a\omega_{\ell m n}^-\right) &=
  C_{\ell \ell' m} \left(-a
    \left(
      \omega^{+}_{\ell , -m, n}
    \right)^{*}\right) \,\\
  C_{\ell \ell' m} \left(a\omega_{\ell m n}^-\right)
  &= (-1)^{\ell + \ell'} C^*_{\ell , \ell' , -m} \left(a\omega^{+}_{\ell , -m, n}\right)
  \,.
\end{align}
Using this identity, we can restate our Eq.~\eqref{eq:analytical} to
look like Cook's Eq.~(12)~\cite{Cook:2020otn}.
Regardless of the way one writes down this mode-decomposed
analytical model, it allows us to consider a 
ringing black hole with its spin axis oriented in any direction 
by rotating $h^{Q}_{\ell m}$.

\subsection{Importance of BMS frames}
\label{sec:BMSFramesIntro}

One important takeaway from the functional form of the QNM model
$h_{\ell m}^{Q}(u)$ in Eq.~\eqref{eq:analytical} is the fact that
\begin{align}
\lim\limits_{u\rightarrow+\infty}h_{\ell m}^{Q}(u)=0.
\end{align}
That is, $h_{\ell m}^{Q}(u)$ tends to zero at late times, approaching $i^{+}$.
Consequently, whenever we fit this model to a waveform,
our waveform should also decay to zero as the retarded time approaches
$+\infty$ in order to obtain reasonable results.
 
What some readers may not be familiar with is that gravitational waves
need not be, and often are not, zero as $u\to +\infty$, due to an
effect which is commonly called \emph{gravitational
  memory}~\cite{Zeldovich_1974, Thorne_1987, Christodoulou_1991,
  Thorne_1992}. Fundamentally, gravitational memory can be understood
as a persistent physical change to spacetime that is induced by the
passage of transient radiation. While there are various types of
gravitational memory effects with varying properties
(see~\cite{Mitman:2020pbt,Grant:2021hga} for more thorough explanations), the one that
will strongly impact our ability to model the ringdown of a black hole
with QNMs is the \emph{displacement memory} effect. This is because,
unlike the other gravitational memories, the displacement memory
uniquely corresponds to an overall net change in the strain between
the two points $i^{0}$ and $i^{+}$; that is, the difference
$\Delta = h(u \to +\infty) - h(u \to -\infty)$. Consequently, the
strain need not return to zero, provided that the memory is nonzero
and the strain's value at $i^{0}$ does not cancel the memory's
value. Fortunately, it turns out that this important problem
regarding gravitational memory is only present if one does not account for
the frame that a perturbed black hole should be in for proper QNM
modeling.

As has been understood since the 1960s, the symmetry group
of asymptotic infinity is not the usual Poincaré group, but a group
with a richer structure called the Bondi-van der Burg-Metzner-Sachs (BMS)
group~\cite{Bondi,Sachs}. The BMS group is a semidirect product
of the usual Lorentz group with an infinite-dimensional group of
transformations called supertranslations, which are angle-dependent
time advances/delays that contain the familiar spacetime translations 
as a subgroup. Fundamentally, supertranslations act on
the Bondi coordinates $(u,r,\theta,\phi)$ as
\begin{align}
u'=u-\alpha(\theta,\phi).
\end{align}
and the strain as
\begin{align}
\label{eq:strainsupertranslation}
h'(u',\theta,\phi)&=h(u',\theta,\phi)-\bar{\eth}^{2}\alpha(\theta,\phi)\nonumber\\
&=\sum\limits_{k=0}^{\infty}\frac{1}{k!}\left(-\alpha(\theta,\phi)\frac{\partial}{\partial u}\right)^{k}h(u,\theta,\phi)-\bar{\eth}^{2}\alpha(\theta,\phi).
\end{align}
where $\bar{\eth}$ is the conjugate of the Geroch-Held-Penrose differential spin-weight operator~\cite{GHP1973}.
Above
\begin{align}
\alpha(\theta,\phi)\equiv\sum\limits_{\ell\geq0,|m|\leq\ell}\alpha_{\ell m}Y_{\ell m}(\theta,\phi)
\end{align}
with
\begin{align}
\alpha_{\ell m}=(-1)^{m}\bar{\alpha}_{\ell,-m}
\end{align}
is a real function which characterizes the supertranslation. The $\ell=0$ component of $\alpha(\theta,\phi)$
is a time translation, the $\ell=1$ components are space translations,
and the $\ell\geq2$ are proper supertranslations.
From Eq.~\eqref{eq:strainsupertranslation}, one can easily realize that under the action of a supertranslation the strain experiences two types of changes.
First, the strain is changed by the angle-dependent constant $\bar{\eth}^{2}\alpha(\theta,\phi)$. Apart from this, however,
because the retarded time changes as $u'=u-\alpha(\theta,\phi)$, we also expand about $u$ to express the transformed strain directly in terms of the strain in the original frame. We see from Eq.~\eqref{eq:strainsupertranslation} that this involves multiplying the time derivatives of the strain with powers of $\alpha(\theta,\phi)$. Consequently, the strain will experience mode-mixing in addition to changing by an angle-dependent constant. Furthermore, if one imagines taking a time-derivative of Eq.~\eqref{eq:strainsupertranslation} then it can be seen that the
news will also experience mode-mixing due to the supertranslation's effect on the retarded time.

Therefore, because of these extra symmetries, whenever we
examine a system that is radiating gravitational waves
it is insufficient to specify just a Poincaré frame, e.g., the remnant BH's center-of-mass frame; we instead
need to specify the entire BMS frame, i.e., how the system's supertranslation freedom is being fixed in addition to the usual Poincaré transformations.

In~\cite{Mitman:2021xkq} this task of specifying a system's BMS frame
was performed for the first time by mapping numerical waveforms from
BBH systems to the post-Newtonian (PN) BMS frame, i.e., the frame that
PN waveforms are in. When fitting the ringdown phase of waveforms to
Eq.~\eqref{eq:analytical}, mapping waveforms to the PN BMS frame is
not the appropriate BMS frame choice, because this frame corresponds to
the strain going to zero at early times (when approaching $i^{0}$),
rather than at late times (when approaching $i^{+}$).
Instead, we should be mapping our waveforms to what is
called the \emph{nice section}~\cite{Dain:2002mj} or the \emph{super
  rest frame}~\cite{Mitman:2021xkq}
at $i^{+}$. This is because when Teukolsky
found the linear equations that describe the dynamical gravitational
perturbations of a rotating black hole~\cite{Teukolsky:1973}, i.e.,
the equations that give rise to QNMs,
he implicitly worked in the BMS frame adapted to the stationary
background metric~\cite{Flanagan:2015pxa}, i.e., the super rest
frame. However, black holes in nature or the remnant black holes
produced in numerical simulations are supertranslated relative to this
preferred frame.  As a result, we need to map
these black holes to the frame that Teukolsky worked in.

As outlined in~\cite{Mitman:2021xkq}, the way to map a system to the
super rest frame is to use the Moreschi supermomentum, which is an
extension of the usual Bondi four-momentum,
\begin{align}
\Psi^{\text{M}}(u,\theta,\phi)=\sum\limits_{\ell\geq0,|m|\leq\ell}\Psi_{\ell m}^{\text{M}}(u)Y_{\ell m}(\theta,\phi),
\end{align}
where
\begin{align}
\label{eq:supermomentum}
\Psi_{\ell m}^{\text{M}}(u)=-\frac{1}{\sqrt{4\pi}}\int_{S^{2}}Y_{\ell m}\left[\Psi_{2}+\sigma\dot{\bar{\sigma}}+\eth^{2}\sigma\right]\,d\Omega,
\end{align}
$\Psi_{2}$ is one of the Weyl scalars, and $\sigma$ is the
shear.\footnote{%
  Note that here and in Eq.~\eqref{eq:supermomentum} we are
  specifically working with the Moreschi-Boyle
  convention~\cite{Mitman:2021xkq, Boyle:2015nqa, iozzo2020improving,
    Moreschi1986}, i.e., in comparison to the numerical formulation of
  the strain and the Weyl scalars we simply have
  $h^{\text{\tt NR}}=2\bar{\sigma}$ and
  $\Psi_{i}^{\text{\tt NR}}=\frac{1}{2}(-\sqrt{2})^{i}\Psi_{i}$.}
Ideally, to map to the super rest frame we would want to minimize the
Moreschi supermomentum as $u \to +\infty$.  But, since our simulations
do not go all the way to $i^{+}$, we can instead minimize the
Moreschi supermomentum during a late portion of the ringdown phase.
Specifically, we construct the BMS frame of our waveforms via the following:
\begin{itemize}
	\item take the boost velocity and space translation, i.e., the $\ell=1$ components of the supertranslation, to be
	the transformations that minimize the remnant's center-of-mass charge over the late time
	window $u\in [u_{\text{peak}}+150M, u_{\text{peak}}+350 M]$~\cite{Mitman:2021xkq};
	\item take the $2\leq\ell\leq4$ modes of the supertranslation to be the transformations that minimize the
	$L^{2}$ norm of the $2\leq\ell\leq4$ modes of $\Psi^{\text{M}}$ over the late time
	window $u\in [u_{\text{peak}}+150M, u_{\text{peak}}+350 M]$ ~\cite{Mitman:2021xkq};
	\item fix the system's rotation freedom by aligning the remnant BH's spin with the positive $z$-axis.
\end{itemize}
These calculations for fixing the BMS frame require the system's strain as well as the four Weyl scalars $\Psi_{1-4}$~\cite{Mitman:2021xkq}. Note that while only the strain, $\Psi_{1}$, and $\Psi_{2}$ are needed to compute the BMS charges, $\Psi_{3}$ and $\Psi_{4}$ are also needed to transform $\Psi_{1}$ and $\Psi_{2}$ during the frame fixing procedure. We obtain the $\Psi_{3}$ and $\Psi_{4}$ Weyl scalars independently from the strain through our Cauchy-characteristic extraction. Above $u_{\text{peak}}$ is the time at which the $L^2$ norm of the strain achieves its maximum value. By performing this frame-fixing procedure, we transform to a waveform with its BMS frame fixed so
that it can be modeled by Eq.~\eqref{eq:analytical}. Note that the window $u\in [u_{\text{peak}}+150M, u_{\text{peak}}+350 M]$ is chosen as such because it is roughly the $200M$ before the earliest end time of our simulations. We find that our results are fairly independent of this time window, provided that it starts beyond $u\approx(u_{\text{peak}}+100M)$. A quantitative description of how the results are affected by the choice of time window is beyond the scope of this paper. Nevertheless, an analysis of when to map to the super rest frame will be presented in future work.

\subsection{QNM fitting procedure}
\label{sec:fittingprocedure}
Given a numerical waveform $h_{\ell m}^{\NR}(u)$ and the functional
form of $h_{\ell m}^{Q}(u)$, we can consider the
problem of fitting for the complex QNM amplitudes $\mathcal{A}^{p}_{\ell' m n}$.
To do this, we first need an inner product on the space of spin-weight
$s$ waveforms on $\scri^{+}$. For waveforms $a$ and $b$, the natural
inner product is defined as
\begin{align}
\label{eq:inner-product}
\langle a, b \rangle \equiv \int_{u_0}^{u_f} du \int_{S^{2}} d\Omega\,a^{*}(u,\theta,\phi) b(u,\theta,\phi),
\end{align}
Where $[u_0, u_f]$ is the interval of time where we would like 
to fit the waveform with a QNM model.
Both waveforms can be decomposed into $a_{\ell m}$ and $b_{\ell' m'}$
as in Eq.~\eqref{eq:hlm}. By applying the orthogonality relationship
of Eq.~\eqref{eq:orthog} to collapse the double sum to a single sum,
the inner product on $\scri^{+}$ then becomes
\begin{align}
\label{eq:decomp-inner-product}
\langle a,b\rangle=\int_{u_0}^{u_f} du \sum_{\ell,m} a^{*}_{\ell m}(u) b_{\ell m}(u)=\sum_{\ell,m}\langle a_{lm}, b_{lm} \rangle_{u},
\end{align}
where
\begin{align}
  \label{eq:L2-usual}
  \langle f,g \rangle_{u} \equiv \int_{u_0}^{u_f} f^{*}(u) g(u) du
\end{align}
is the usual $L^{2}$
inner product for complex functions on the real line. However, because
our study only considers modes with $\ell\leq 4$, the inner product
that we use in the rest of paper is in fact given by
\begin{align}
\label{eq:decomp-inner-product_L4}
\langle a,b\rangle=\sum_{\ell\leq4, m}\langle a_{lm}, b_{lm} \rangle_{u},
\end{align}
which we henceforth call the all-mode inner product, keeping in mind
that here `all' means all the modes included in the NR waveform.

From this inner product we construct the mismatch $\mathcal{M}$, a figure of merit commonly used
in the literature, as follows:
\begin{align}
\label{eq:mismatch}
\mathcal{M}(a,b)\equiv1 - \mathcal{O}(a,b),
\end{align}
where $\mathcal{O}(a,b)$ is the overlap,
\begin{align}
\label{eq:overlap}
\mathcal{O}(a,b)\equiv\text{Re}\left[\frac{\langle a, b \rangle}{\sqrt{\langle a, a \rangle\langle b, b \rangle}}\right].
\end{align}

Consequently, for a NR waveform $h^{\NR}(u,\theta,\phi)$
expressed by its spin-weighted spherical harmonic coefficients
$h_{\ell m}^{\NR}$, we can quantify the effectiveness of a fit $h^Q(\vec{\lambda})$
by calculating the all-mode mismatches 
$\mathcal{M}(h^{\NR}, h^{Q}(\vec{\lambda}))$. Here $\vec{\lambda}$ is the set of
free parameters of the fit. When we focus on a single mode $(\ell, m)$, however, we instead
use the single mode mismatch $\mathcal{M}(h_{\ell m}^{\NR}, h_{\ell m}^{Q}(\vec{\lambda}))$.

Nonetheless, we do not find the optimal parameters $\vec{\lambda}_{\text{opt}}$
by directly minimizing this figure of merit. Instead we first calculate the residual
\begin{equation}
     R \equiv h^{\NR} - h^{Q},
\end{equation}
and then compute the squared norm of the residual, $\langle R, R \rangle$, 
as the figure of merit that we want to minimize. One can show
that because the norm of $h^{Q}(\vec{\lambda})$ can be independently varied,
 minimizing the norm of the residual also minimizes
the mismatch. However the problem of minimizing the former is manifestly linear 
in nature for the QNM amplitudes, and is not degenerate in the norm.
Therefore we find the optimal parameters $\vec{\lambda}_{\text{opt}}$ by
\begin{align}
  \vec{\lambda}_{\text{opt}} = \argmin_{\vec{\lambda}} \langle R, R \rangle \quad \text{or} \quad 
  \vec{\lambda}_{\text{opt}} = \argmin_{\vec{\lambda}} \langle \dot{R}, \dot{R} \rangle
  \,,
\end{align}
where we use the second choice if we want to work in the domain of the news $\mathcal{N} = \dot{h}$.  
Although one could also consider working in the $\Psi_4$ domain, 
our analyses focus on the strain, 
since it is the physical quantity that the gravitational-wave 
detectors measure and on the news, since the 
power is naturally defined by it.

Now, $\vec{\lambda}$ can take on one of two forms:
$\vec{\lambda}=\{\mathcal{A}^{p}_{\ell'mn}\}$, or
$\vec{\lambda}=\{\{\mathcal{A}^{p}_{\ell'mn}\}, M, a\}$, where $\mathcal{A}^{p}_{\ell'mn}$ are the
QNM amplitudes from Eq.~\eqref{eq:analytical}, and $M$ and $a$ are the
mass and spin of the remnant black hole.  In the former we use the
remnant black hole's mass and spin obtained from the simulation (see below), and solve for
$\vec{\lambda}=\{\mathcal{A}^{p}_{\ell'mn}\}$ using NumPy's linear least square method~\cite{Harris:2020xlr}. On the other hand when
$\vec{\lambda}=\{\{\mathcal{A}^{p}_{\ell'mn}\}, M, a\}$, the remnant properties are
deduced by fitting the waveform. Here we perform a
least-squares minimization between the NR waveform and
the QNM model using SciPy's Nelder-Mead algorithm~\cite{Virtanen:2019joe, NelderMead} to
find the remnant BH's mass and spin and simultaneously use
the linear least square method to determine the amplitudes. We also note
that since we rotate the remnant BH's 
spin direction to be aligned with the positive $z$-axis, there is no mixing 
of the $m$ modes. Therefore, we can fit the QNM amplitudes for 
each value of $m$ independently.

Finally, it should be noted that when we are only solving for the
amplitudes, i.e., $\vec{\lambda}=\{\mathcal{A}^{p}_{\ell'mn}\}$, we obtain
the mass and the spin of the remnant from $\scri^+$ rather than
the apparent horizon. That is, following the work of~\cite{Iozzo:2021vnq,
  Mitman:2021xkq}, we use Poincaré charges to obtain the remnant's
mass and spin via Eqs.~(11) and (15) of~\cite{Iozzo:2021vnq}. The mass and spin are taken to be the values of the charges at the last
available time step.

\subsection{Numerical waveforms}

\begin{table*}
  \label{tab:runs}
  \centering
  \renewcommand{\arraystretch}{1.2}
  \begin{tabular}{@{}l@{\hspace*{7mm}}c@{\hspace*{7mm}}c@{\hspace*{7mm}}c@{}c@{}c@{}c@{\hspace*{7mm}}c@{}c@{}c@{}c@{}}
    \Xhline{3\arrayrulewidth}
    Name & CCE radius & $q$ & $\chi_{A}$:\, & $(\hat{x},\,$ & $\hat{y},\,$ & $\hat{z})$ & $\chi_{B}$:\, & $(\hat{x},\,$ & $\hat{y},\,$ & $\hat{z})$\\
    \hline
    \texttt{q1\_nospin} & 292 & $1.0$ & & $(0,\,$ & $0,\,$ &  $0)$ & & $(0,\,$ & $0,\,$ & $0)$ \\
    \texttt{q1\_aligned\_chi0\_2} & 261 & $1.0$ & & $(0,\,$ & $0,\,$ & $0.2)$ & & $(0,\,$ & $0,\,$ & $0.2)$ \\
    \texttt{q1\_aligned\_chi0\_4} & 250 & $1.0$ & & $(0,\,$ & $0,\,$ & $0.4)$ & & $(0,\,$ & $0,\,$ & $0.4)$ \\
    \texttt{q1\_aligned\_chi0\_6} & 236 & $1.0$ & & $(0,\,$ & $0,\,$ & $0.6)$ & & $(0,\,$ & $0,\,$ & $0.6)$ \\
    \texttt{q1\_antialigned\_chi0\_2} & 274 & $1.0$ & & $(0,\,$ & $0,\,$ & $0.2)$ & & $(0,\,$ & $0,\,$ & $-0.2)$ \\
    \texttt{q1\_antialigned\_chi0\_4} & 273 & $1.0$ & & $(0,\,$ & $0,\,$ & $0.4)$ & & $(0,\,$ & $0,\,$ & $-0.4)$ \\
    \texttt{q1\_antialigned\_chi0\_6} & 270 & $1.0$ & & $(0,\,$ & $0,\,$ & $0.6)$ & & $(0,\,$ & $0,\,$ & $-0.6)$ \\
    \texttt{q1\_precessing} & 305 & $1.0$ & & $(0.487,\,$ & $0.125,\,$ & $-0.327)$ & & $(-0.190,\,$ & $0.051,\,$ & $-0.227)$ \\
    \texttt{q1\_superkick} & 270 & $1.0$ & & $(0.6,\,$ & $0,\,$ & $0)$ & & $(-0.6,\,$ & $0,\,$ & $0)$ \\
    \texttt{q4\_nospin} & 235 & $4.0$ & & $(0,\,$ & $0,\,$ & $0)$ & & $(0,\,$ & $0,\,$ & $0)$ \\
    \texttt{q4\_aligned\_chi0\_4} & 222 & $4.0$ & & $(0,\,$ & $0,\,$ & $0.4)$ & & $(0,\,$ & $0,\,$ & $0.4)$ \\
    \texttt{q4\_antialigned\_chi0\_4} & 223 & $4.0$ & & $(0,\,$ & $0,\,$ & $0.4)$ & & $(0,\,$ & $0,\,$ & $-0.4)$ \\
    \texttt{q4\_precessing} & 237 & $4.0$ & & $(0.487,\,$ & $0.125,\,$ & $-0.327)$ & & $(-0.190,\,$ & $0.051,\,$ & $-0.227)$ \\
    \texttt{SXS:BBH:0305} (GW150914) & 267 & $1.221$ && $(0,\,$ & $0,\,$ & $0.330)$ & & $(0,\,$ & $0,\,$ & $-0.440)$ \\
    \Xhline{3\arrayrulewidth}
  \end{tabular}
  \caption{Parameters of the BBH mergers used in our results. The mass ratio is $q=M_A/M_B$, and the initial dimensionless spins of the two black holes are $\chi_A$ and $\chi_B$. These simulations have been made publicly available at~\cite{ExtCCECatalog,SXSCatalog}.}
\end{table*}

For the following results, we numerically evolved a set of 14 binary
black hole mergers with many mass ratios and spin configurations using
the Spectral Einstein Code (SpEC)~\cite{SpECCode}. We list the
important parameters of these various BBH systems in
Table~\ref{tab:runs}. Each simulation contains roughly 19 orbits prior
to merger and is evolved until the waves from ringdown leave the
computational domain. Unlike the evolutions in the SXS catalog, the
full set of Weyl scalars and the strain have been extracted from these
runs and the waveforms have been computed using the extrapolation
technique described in~\cite{Iozzo:2020jcu} and the
Cauchy-characteristic extraction (CCE) procedure that is outlined
in~\cite{Moxon:2020gha,Moxon:2021gbv}. Extrapolation is performed with the
python module \texttt{scri}~\cite{scri_url, Boyle:2013nka,
  Boyle:2015nqa, Boyle:2014ioa} and CCE is run with SpECTRE's CCE
module~\cite{Moxon:2020gha,Moxon:2021gbv,CodeSpECTRE}.

For the CCE extractions, the four world tubes that are available have
radii that are equally spaced between $2\lambdabar_{0}$ and
$21\lambdabar_{0}$, where $\lambdabar_0\equiv 1/\omega_0$ is the initial
reduced gravitational wavelength as determined by the orbital
frequency of the binary from the initial data. Based on the recent
work of~\cite{Mitman:2020bjf}, however, we choose to use only the
waveforms that correspond to the world tube with the second-smallest
radius, since these waveforms have been shown to minimally violate the
BMS balance laws. For clarity, we provide the world tube radius used
for each system in Table~\ref{tab:runs}. All of these 14 BBH systems'
waveforms have been made publicly available
at~\cite{ExtCCECatalog,SXSCatalog}.

As mentioned above, the asymptotic strain waveforms are computed using
two methods: extrapolation and CCE. The first method utilizes
Regge-Wheeler-Zerilli (RWZ) extraction to compute the strain on a
series of concentric spheres of constant coordinate radius and then
proceeds to extrapolate these values to future null infinity
$\scri^{+}$ using $1/r$ approximations~\cite{Sarbach:2001qq,
  Regge:1957td, Zerilli:1970se, Boyle:2019kee, Iozzo:2020jcu,
  Boyle:2009vi}.  This is the strain that can be found in the public SXS
catalog. The other and more faithful extraction method, which is known
as CCE, computes the strain by using the world tube data provided by a
Cauchy evolution as the inner boundary data for a nonlinear evolution
of the Einstein field equations on null
hypersurfaces extending all the way to
$\scri^{+}$~\cite{Moxon:2020gha,Moxon:2021gbv}. CCE requires freely
specifying the strain on the initial null hypersurface labeled
$u=0$. Like~\cite{Mitman:2020pbt, Mitman:2020bjf, Mitman:2021xkq}, we
choose this field to match the value and the first radial derivative
of $h$ from the Cauchy data on the world tube using the ansatz
\begin{align}
h(u=0,r,\theta^{A})=\frac{A(\theta^{A})}{r}+\frac{B(\theta^{A})}{r^{3}},
\end{align}
where the two coefficients $A(\theta^{A})$ and $B(\theta^{A})$ are
fixed by the Cauchy data on the world tube.

Lastly, when performing our analyses, we predominantly use the code
\texttt{scri}~\cite{scri_url, Boyle:2013nka, Boyle:2014ioa,
  Boyle:2015nqa} to compute Poincaré charges and transform our
asymptotic waveform quantities to the super rest frame
using the procedures outlined in Sec.~II and Appendix A
of~\cite{Mitman:2021xkq}.  Our waveforms only include the
$\ell\leq4$ modes since these are the modes included in the BMS frame
fixing procedure. We also only model our waveforms up to $u_f=u_{\text{peak}}+90\,M$ as in~\cite{Giesler:2019uxc}.

\section{On which modes to include}
\label{sec:GreedyAlgo}

The importance of using multiple waveform modes to capture the physics
of a remnant black hole---considering both multiple angular $(\ell,m)$
modes as well multiple overtones---has been studied
extensively~\cite{Giesler:2019uxc, London:2014cma, Cook:2020otn, Forteza:2021wfq, Li:2021wgz}.  When
constructing a QNM model it is crucial that we are able to choose as
many modes as necessary to accurately model our system, without
overfitting or introducing degeneracy.  Because
manually choosing an arbitrary number of modes without knowing which
modes are important to include is objectionable, we have written a
greedy algorithm that provides us with an efficiently low number of modes
needed to model the ringdown waveform to a requested
precision. Consequently, we can reduce the number of modes that are
needed to capture the most physics and also identify the most
physically-relevant modes.

\subsection{Greedy algorithm}
\label{sec:greedy-algorithm}

\begin{figure}[t]
	\centering
	\includegraphics[width=\linewidth]{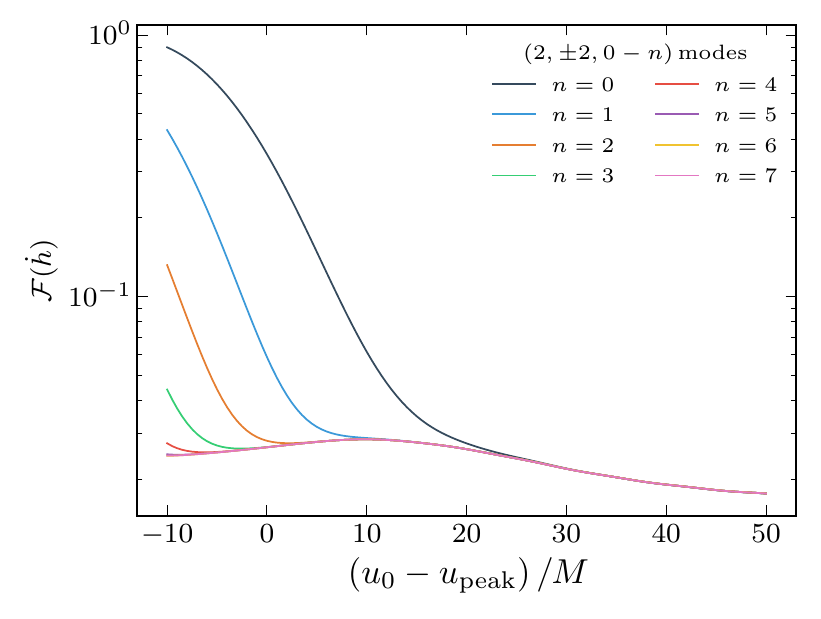}
	\caption{%
		Fraction of unmodeled power that is obtained when comparing a QNM model built from the $(2,\pm2,0-n)$ mode(s) to a CCE 
		strain waveform as a function of the QNM model start time $u_{0}$. We compute the fraction of unmodeled power in the waveform 
		using the news waveforms,
		i.e., by using Eq.~\eqref{eq:powerfractionnews}. This includes the power that is unmodeled
		because of neglecting higher modes in the QNM model. \\
		BBH merger: \texttt{SXS:BBH:0305}.}
	\label{fig:UnmodeledPower_from_L2_M2}
\end{figure}

\begin{figure}[t]
	\centering
	\includegraphics[width=\linewidth]{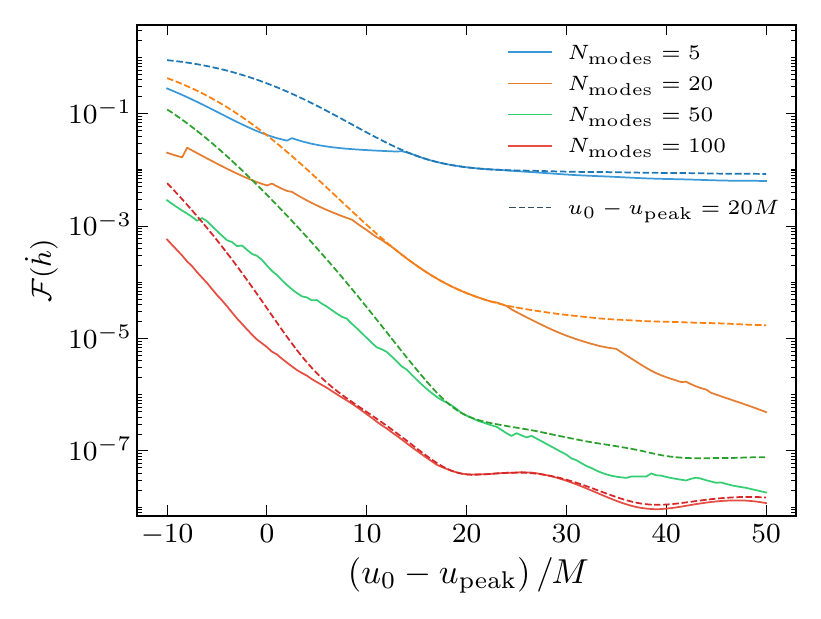}
	\caption{%
		Fraction of unmodeled power (solid) that is obtained when comparing QNM models
		built by our greedy algorithm with various number of modes to a CCE strain waveform.
		Again, the power, which is plotted and used in our
		algorithm to pick modes to model, is computed using the news waveforms, i.e., by using
		Eq.~\eqref{eq:powerfractionnews}. The dashed curves are QNM fits using the fixed set of 
		modes determined by the greedy algorithm at the time $u_{0} -
		u_{\text{peak}}=20 M$. On the other hand, the solid curves have their set of modes 
		determined for each $u_{0}$ independently, which causes these curves to not be smooth. \\
		BBH merger: \texttt{SXS:BBH:0305}.}
	\label{fig:UnmodeledPower_vs_NModes}
\end{figure}

The greedy algorithm that we implement is iterative, adding the
prograde and retrograde modes at each iteration.  The data at iteration $i$ is a
collection of $i$ mode labels $(\ell', m, n, p)$ and the parameter
vector of length $i$,
\begin{align}
  \vec{\lambda}^{(i)} = \{ \mathcal{A}^{p}_{\ell' m n} \}
  \,,
\end{align}
corresponding to those modes.

The greedy algorithm can be summarized as follows.
\begin{enumerate}[(I)]
\item Begin with an empty list of modes and amplitudes.
\item \label{item:greedy_form_resid}
  At each iteration $i$, form the residual
  \begin{align}
    \label{eq:Rdef}
    R^{(i)} \equiv h^{\NR} - h^{Q,i}
    \,,
  \end{align}
  between the NR waveform and $h^{Q,i}$, which is built from the $i$
  amplitudes $\{\mathcal{A}^{p}_{\ell' m n} \}$.  If instead working in
  the news domain, we form the residual as the difference of the news waveforms
  \begin{align}
  \label{eq:Rnewsdef}
  \dot{R}^{(i)} \equiv \dot{h}^{\NR} - \dot{h}^{Q,i}
  \,.
  \end{align}
\item Compute the power in each mode of the residual,
  \begin{align}
    P_{\ell m}^{(i)}(R)\equiv \langle R^{(i)}_{\ell m}, R^{(i)}_{\ell m} \rangle_{u}
    \,,
  \end{align}
  using the usual $L^{2}$ metric in Eq.~\eqref{eq:L2-usual}. Analogously,
  when working in the news domain we use $P_{\ell m}^{(i)}(\dot{R})$ as the 
  power instead.  Notice that, $\sum_{\ell,m} P_{\ell
    m}^{(i)}(\dot{R})$ is proportional to the physical gravitational-wave
  luminosity of the residual waveform.
\item Rank the $(\ell,m)$ modes in the residuals by their powers
  $P_{\ell m}^{(i)}$, and identify the mode $(\bar{\ell},\bar{m})$ with the
  largest residual power.
  \begin{enumerate}[(1)]
  \item \label{item:greedy_which_mode}
    If no QNMs with the $(\ell=\bar{\ell},m=\bar{m})$ mode are present in
    the parameter list, add the prograde and retrograde modes
    $(\bar{\ell}, \bar{m}, 0, \pm1)$.
  \item If some QNMs with $(\ell=\bar{\ell},m=\bar{m})$ are already in
    the parameter list, identify the smallest $\bar{n}$ not present in
    the parameter list. If $\bar{n} \leq n_{\max}$ for some max overtone number
    $n_{\max}$, add this next mode with the prograde and retrograde modes
    $(\bar{\ell}, \bar{m}, \bar{n}, \pm1)$. For this paper we chose a max overtone number of $n_{\max}=7$.
  \item If all modes $(\bar{\ell},\bar{m})$ with $0\le n \le n_{\max}$
    are already in the parameter list, set $(\bar{\ell},\bar{m})$ as
    the mode of the residual that is the next loudest in the list of
    $P_{\ell m}^{(i)}$.  Return to step~\ref{item:greedy_which_mode}
    to find which mode to include.
  \end{enumerate}
 Although here at each step we are adding 2 QNMs $(\bar{\ell}, \bar{m}, \bar{n}, \pm1)$,
 we group these modes together and count them as adding one mode. 
\item After identifying the next mode(s) to include, re-solve
  the linear least squares problem to determine the optimal values of
  $\{ \mathcal{A}^{p}_{\ell' m n} \}$.
\item Compute the fraction of residual power to target waveform power in the
strain domain
\begin{align}
 \label{eq:powerfraction}
\mathcal{F}(h)\equiv\frac{\langle R^{(i)}, R^{(i)}\rangle}{\langle h^{\text{NR}},h^{\text{NR}}\rangle}
\end{align}
or the news domain
\begin{align}
\label{eq:powerfractionnews}
\mathcal{F}(\dot{h})\equiv\frac{\langle \dot{R}^{(i)}, \dot{R}^{(i)}\rangle}{\langle \dot{h}^{\text{NR}},\dot{h}^{\text{NR}}\rangle}
\,,
\end{align}
where $R$ and $\dot{R}$ are defined in Eq.~\eqref{eq:Rdef} and Eq.~\eqref{eq:Rnewsdef}.
Both of these choices are well motivated, but we primarily use Eq.~\eqref{eq:powerfractionnews},
since this corresponds to the physical gravitational-wave luminosity of the residual waveform.
\item Terminate if either $\mathcal{F} < \mathcal{F}_{\text{target}}$
  for some target residual power fraction, or if the number of modes $i = N_{\max}$: a
  maximum number of modes to include.
\item Return to step~\ref{item:greedy_form_resid} and repeat.
\end{enumerate}

A study using multimode fitting to investigate the performance of three 
different fitting methods across three different sets of 
modes---$\{(2,2)\}, \{(2,2),(3,2)\},\{(2,2),(3,2),(4,2)\}$---was recently
carried out in~\cite{Cook:2020otn}.  Results show that when fitting for 
more than just the dominant $(2,2)$ mode, all fitting methods converge. 
However, these methods are tested using only a limited set of modes, 
whereas our greedy algorithm can use all modes
(see~\cite{Cook:2020otn} for more details).

Earlier studies of multimode fitting have been carried out in one of 
two ways: some models have been executed by manually choosing 
a set of modes~\cite{Giesler:2019uxc, Cook:2020otn, Finch:2021iip} 
while others use greedy algorithms to pick which modes to 
model~\cite{London:2014cma, Li:2021wgz}. However, instead of 
focusing on a single $(\ell,m)$ mode as in~\cite{London:2014cma} 
or greedily picking only the angular numbers $(\ell, m)$ as 
in~\cite{Li:2021wgz}, we implement a physically well motivated, 
multimode greedy algorithm where each QNM with labels $(\ell, m, n)$ 
is picked greedily. At the moment, checks to compare greedy algorithms 
with one another have not been performed. This would be interesting to 
examine in the future.

\subsection{Importance of multiple modes in modeling}
In the remaining parts of this section, we use the simulation 
\texttt{SXS:BBH:0305}, which corresponds to GW150914, (see Table~\ref{tab:runs}) to study the
importance of multimode fitting. We begin by applying our QNM
modeling procedure to the $(2,\pm 2)$ modes with up to 7 overtones. In Fig.~\ref{fig:UnmodeledPower_from_L2_M2}
we show the fraction of unmodeled power as a function of
$u_0-u_{\text{peak}}$
using $n$ number of overtones in the QNM model. For computing the
unmodeled power, we use the CCE waveform, in the super rest frame, and
measure the fraction of unmodeled power in the news domain with
Eq.~\eqref{eq:powerfractionnews}. From this plot, one can easily observe the
importance of including overtones in the model. By using just the
$n=0$ mode, one can only model about $65\%$ of the power starting at 
$u_{0}=u_{\text{peak}}$. With all 7 of the overtones included, the modeled power
improves to roughly $97\%$ of the total power. Note, however, that these numbers
will vary depending on the time that one chooses to model the QNMs.

\begin{figure}[t]
	\centering
	\includegraphics[width=\linewidth]{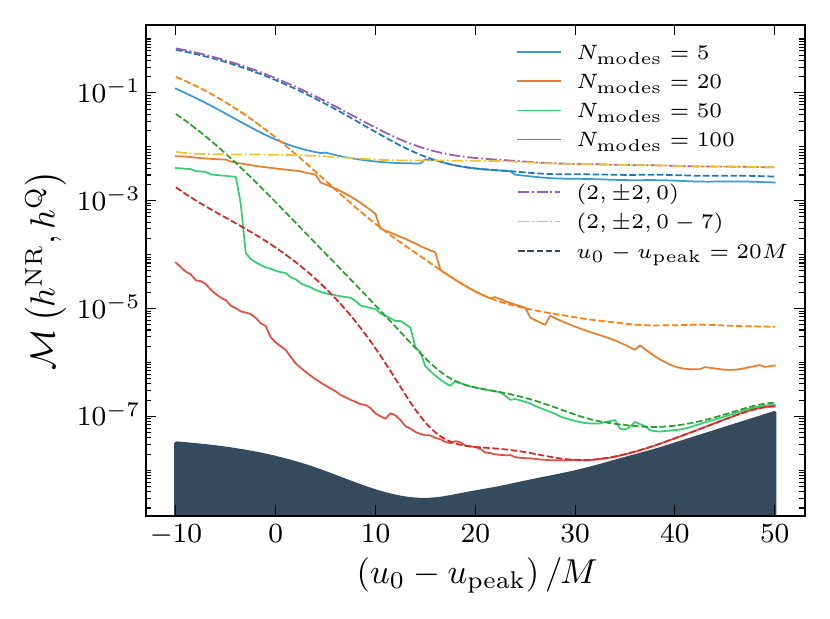}
	\caption{%
		All-mode mismatches between the CCE waveform and a QNM model
		fitting $N$ number of modes. The solid curves correspond to QNM
		models with a varying number of modes that are modeled. The
		dashed curves are QNM amplitude fits using the modes from
		$u_{0} - u_{\text{peak}}=20 M$. The dash-dotted curves represent
		the all-mode mismatches from just the $(2,\pm2)$ modes---one
		with the $n=0$ tone and another with the $n=0-7$ tones. 
		Finally, the top of the black region illustrates the mismatch
		between the highest and the next-highest resolution waveforms to
		provide a reference for the numerical error that is present in our strain waveform.\\
		BBH merger: \texttt{SXS:BBH:0305}.}
	\label{fig:Mismatches_vs_NModes}
\end{figure}

Accounting for overtones, however, is only one of the important
components for correctly modeling a waveform with QNMs. Although the
$(2,\pm2)$ modes are the most important to use due to their dominance,
including higher-order modes is crucial to more
accurately describe the ringdown phase. This is especially true for systems that
may not exhibit symmetries, e.g., having mass ratio one. Overall,
higher-order modes contain less power and overtones have shorter damping times relative to
the $(2,\pm2)$ modes and $n=0$ modes. Therefore, their importance within the QNM model
lessens as we reach later stages of ringdown. Nevertheless, they
exhibit a considerable amount of power at early ringdown times and are
essential in capturing the remaining power stored in a gravitational
wave. To highlight this, in Fig.~\ref{fig:UnmodeledPower_vs_NModes} we
compute the fraction of unmodeled power for models that include
$N\in\{5, 20, 50, 100\}$ modes as a function of the QNM model's start time $u_{0}$.
We remind the reader that $N$ counts the number of pairs of prograde and retrograde $(\ell,m,n)$
modes included by the greedy algorithm, e.g., the set $\{(2,2,0,\pm1), (2,2,1,\pm1), (3,2,0,\pm1)\}$
corresponds to $N=3$. In this plot, the solid curves correspond to running
the greedy algorithm independently for each $u_{0}$, while the dashed curves just use the
fixed set of modes that are obtained by the greedy algorithm at $u_{0}-u_{\text{peak}}=20M$.
The jaggedness of the solid curves illustrates the fact that the 
greedy algorithm's choice of modes for the QNM model is not 
a smooth function of the model start time $u_{0}$. Moreover, we have 
performed a minimal test of the greedy algorithm to ensure that the 
results do not depend too sensitively on the initial mode content. We do this by giving the 
greedy algorithm an initial set of modes to fit before it adds the modes it
has ranked. Slight changes in results only arise when modeling a small 
number of modes, e.g., $N=5$ at times before $u_0=u_{\text{peak}}$. 
This is not surprising since overtones play an important role at early 
times. Consequently, replacing an overtone with a higher harmonic at such times 
would slightly worsen our model by increasing the fraction of unmodeled 
power. For a higher number of modes $N$ and later times, however, no 
detectable change occurs.

Using the solid curves, we find that at $u_{0}=u_{\text{peak}}$ the power captured in the model is nearly 96\% with 5
modes, which is a rather comparable result to using the $(2,\pm2,0-7)$ modes. With 20 modes over 99\%
of the power is captured. With 100 modes, we are modeling 99.999\% of
the power. Again, for this plot we are using the CCE waveform for
\texttt{SXS:BBH:0305} and are performing computations of the power in the news
domain by using Eq.~\eqref{eq:powerfractionnews}. The order in which all 168 modes of this waveform are included is shown in Table.~\ref{tab:modes}.

\begin{table*}
  \label{tab:modes}
  \centering
  \resizebox{1.92\columnwidth}{!}{%
  \begin{tabular}{|c|*{25}{c|}}
    \Xhline{3\arrayrulewidth}
    \backslashbox{$\phantom{m}$}{$\ell$} & \multicolumn{8}{c|}{2} & \multicolumn{8}{c|}{3} & \multicolumn{8}{c|}{4} \\\hline
    \backslashbox{$m$}{$n$} & 0 & 1 & 2 & 3 & 4 & 5 & 6 & 7 & 0 & 1 & 2 & 3 & 4 & 5 & 6 & 7 & 0 & 1 & 2 & 3 & 4 & 5 & 6 & 7 \\\hline
     -4 & \cellcolor[HTML]{C0C0C0} & \cellcolor[HTML]{C0C0C0} & \cellcolor[HTML]{C0C0C0} & \cellcolor[HTML]{C0C0C0} & \cellcolor[HTML]{C0C0C0} & \cellcolor[HTML]{C0C0C0} & \cellcolor[HTML]{C0C0C0} & \cellcolor[HTML]{C0C0C0} & \cellcolor[HTML]{C0C0C0} & \cellcolor[HTML]{C0C0C0} & \cellcolor[HTML]{C0C0C0} & \cellcolor[HTML]{C0C0C0} & \cellcolor[HTML]{C0C0C0} & \cellcolor[HTML]{C0C0C0} & \cellcolor[HTML]{C0C0C0} & \cellcolor[HTML]{C0C0C0} & \cellcolor[rgb]{0.5667, 0.7833, 0.7369} 12 & \cellcolor[rgb]{0.5667, 0.7833, 0.7292} 14 & \cellcolor[rgb]{0.5667, 0.7833, 0.7137} 18 & \cellcolor[rgb]{0.5667, 0.7833, 0.6634} 31 & \cellcolor[rgb]{0.5667, 0.7833, 0.6363} 38 & \cellcolor[rgb]{0.5667, 0.7833, 0.5899} 50 & \cellcolor[rgb]{0.617, 0.7833, 0.5667} 69 & \cellcolor[rgb]{0.6982, 0.7833, 0.5667} 90 \\\hline -3 & \cellcolor[HTML]{C0C0C0} & \cellcolor[HTML]{C0C0C0} & \cellcolor[HTML]{C0C0C0} & \cellcolor[HTML]{C0C0C0} & \cellcolor[HTML]{C0C0C0} & \cellcolor[HTML]{C0C0C0} & \cellcolor[HTML]{C0C0C0} & \cellcolor[HTML]{C0C0C0} & \cellcolor[rgb]{0.5667, 0.7833, 0.7446} 10 & \cellcolor[rgb]{0.5667, 0.7833, 0.7214} 16 & \cellcolor[rgb]{0.5667, 0.7833, 0.6557} 33 & \cellcolor[rgb]{0.5667, 0.7833, 0.617} 43 & \cellcolor[rgb]{0.6711, 0.7833, 0.5667} 83 & \cellcolor[rgb]{0.7833, 0.7833, 0.5667} 112 & \cellcolor[rgb]{0.7833, 0.6247, 0.5667} 153 & \cellcolor[rgb]{0.7833, 0.5667, 0.5667} 168 & \cellcolor[rgb]{0.5667, 0.7833, 0.6131} 44 & \cellcolor[rgb]{0.5667, 0.7833, 0.5976} 48 & \cellcolor[rgb]{0.5667, 0.7833, 0.5667} 56 & \cellcolor[rgb]{0.6247, 0.7833, 0.5667} 71 & \cellcolor[rgb]{0.733, 0.7833, 0.5667} 99 & \cellcolor[rgb]{0.7833, 0.7679, 0.5667} 116 & \cellcolor[rgb]{0.7833, 0.7137, 0.5667} 130 & \cellcolor[rgb]{0.7833, 0.6286, 0.5667} 152 \\\hline -2 & \cellcolor[rgb]{0.5667, 0.7833, 0.7756} 2 & \cellcolor[rgb]{0.5667, 0.7833, 0.7679} 4 & \cellcolor[rgb]{0.5667, 0.7833, 0.7601} 6 & \cellcolor[rgb]{0.5667, 0.7833, 0.6866} 25 & \cellcolor[rgb]{0.5667, 0.7833, 0.5705} 55 & \cellcolor[rgb]{0.7524, 0.7833, 0.5667} 104 & \cellcolor[rgb]{0.7833, 0.6943, 0.5667} 135 & \cellcolor[rgb]{0.7833, 0.6131, 0.5667} 156 & \cellcolor[rgb]{0.5667, 0.7833, 0.6827} 26 & \cellcolor[rgb]{0.5667, 0.7833, 0.675} 28 & \cellcolor[rgb]{0.5667, 0.7833, 0.6518} 34 & \cellcolor[rgb]{0.5667, 0.7833, 0.6015} 47 & \cellcolor[rgb]{0.6595, 0.7833, 0.5667} 80 & \cellcolor[rgb]{0.7833, 0.7756, 0.5667} 114 & \cellcolor[rgb]{0.7833, 0.644, 0.5667} 148 & \cellcolor[rgb]{0.7833, 0.5821, 0.5667} 164 & \cellcolor[rgb]{0.6402, 0.7833, 0.5667} 75 & \cellcolor[rgb]{0.6479, 0.7833, 0.5667} 77 & \cellcolor[rgb]{0.675, 0.7833, 0.5667} 84 & \cellcolor[rgb]{0.7253, 0.7833, 0.5667} 97 & \cellcolor[rgb]{0.764, 0.7833, 0.5667} 107 & \cellcolor[rgb]{0.7833, 0.7601, 0.5667} 118 & \cellcolor[rgb]{0.7833, 0.6634, 0.5667} 143 & \cellcolor[rgb]{0.7833, 0.5976, 0.5667} 160 \\\hline -1 & \cellcolor[rgb]{0.5667, 0.7833, 0.7098} 19 & \cellcolor[rgb]{0.5667, 0.7833, 0.6982} 22 & \cellcolor[rgb]{0.5667, 0.7833, 0.6324} 39 & \cellcolor[rgb]{0.5899, 0.7833, 0.5667} 62 & \cellcolor[rgb]{0.7408, 0.7833, 0.5667} 101 & \cellcolor[rgb]{0.7833, 0.7253, 0.5667} 127 & \cellcolor[rgb]{0.7833, 0.6557, 0.5667} 145 & \cellcolor[rgb]{0.7833, 0.5938, 0.5667} 161 & \cellcolor[rgb]{0.5938, 0.7833, 0.5667} 63 & \cellcolor[rgb]{0.6054, 0.7833, 0.5667} 66 & \cellcolor[rgb]{0.6324, 0.7833, 0.5667} 73 & \cellcolor[rgb]{0.7137, 0.7833, 0.5667} 94 & \cellcolor[rgb]{0.7717, 0.7833, 0.5667} 109 & \cellcolor[rgb]{0.7833, 0.7485, 0.5667} 121 & \cellcolor[rgb]{0.7833, 0.706, 0.5667} 132 & \cellcolor[rgb]{0.7833, 0.6789, 0.5667} 139 & \cellcolor[rgb]{0.7833, 0.7408, 0.5667} 123 & \cellcolor[rgb]{0.7833, 0.733, 0.5667} 125 & \cellcolor[rgb]{0.7833, 0.7021, 0.5667} 133 & \cellcolor[rgb]{0.7833, 0.6866, 0.5667} 137 & \cellcolor[rgb]{0.7833, 0.6711, 0.5667} 141 & \cellcolor[rgb]{0.7833, 0.6402, 0.5667} 149 & \cellcolor[rgb]{0.7833, 0.6054, 0.5667} 158 & \cellcolor[rgb]{0.7833, 0.5744, 0.5667} 166 \\\hline 0 & \cellcolor[rgb]{0.5667, 0.7833, 0.7563} 7 & \cellcolor[rgb]{0.5667, 0.7833, 0.7524} 8 & \cellcolor[rgb]{0.5667, 0.7833, 0.7021} 21 & \cellcolor[rgb]{0.5667, 0.7833, 0.644} 36 & \cellcolor[rgb]{0.5667, 0.7833, 0.6247} 41 & \cellcolor[rgb]{0.5783, 0.7833, 0.5667} 59 & \cellcolor[rgb]{0.6557, 0.7833, 0.5667} 79 & \cellcolor[rgb]{0.7176, 0.7833, 0.5667} 95 & \cellcolor[rgb]{0.5667, 0.7833, 0.5821} 52 & \cellcolor[rgb]{0.5667, 0.7833, 0.5783} 53 & \cellcolor[rgb]{0.5744, 0.7833, 0.5667} 58 & \cellcolor[rgb]{0.5821, 0.7833, 0.5667} 60 & \cellcolor[rgb]{0.6015, 0.7833, 0.5667} 65 & \cellcolor[rgb]{0.6518, 0.7833, 0.5667} 78 & \cellcolor[rgb]{0.7021, 0.7833, 0.5667} 91 & \cellcolor[rgb]{0.7485, 0.7833, 0.5667} 103 & \cellcolor[rgb]{0.6827, 0.7833, 0.5667} 86 & \cellcolor[rgb]{0.6866, 0.7833, 0.5667} 87 & \cellcolor[rgb]{0.6905, 0.7833, 0.5667} 88 & \cellcolor[rgb]{0.706, 0.7833, 0.5667} 92 & \cellcolor[rgb]{0.7369, 0.7833, 0.5667} 100 & \cellcolor[rgb]{0.7601, 0.7833, 0.5667} 106 & \cellcolor[rgb]{0.7795, 0.7833, 0.5667} 111 & \cellcolor[rgb]{0.7833, 0.7524, 0.5667} 120 \\\hline 1 & \cellcolor[rgb]{0.5667, 0.7833, 0.706} 20 & \cellcolor[rgb]{0.5667, 0.7833, 0.6943} 23 & \cellcolor[rgb]{0.5667, 0.7833, 0.6286} 40 & \cellcolor[rgb]{0.586, 0.7833, 0.5667} 61 & \cellcolor[rgb]{0.7446, 0.7833, 0.5667} 102 & \cellcolor[rgb]{0.7833, 0.7214, 0.5667} 128 & \cellcolor[rgb]{0.7833, 0.6518, 0.5667} 146 & \cellcolor[rgb]{0.7833, 0.5899, 0.5667} 162 & \cellcolor[rgb]{0.5976, 0.7833, 0.5667} 64 & \cellcolor[rgb]{0.6092, 0.7833, 0.5667} 67 & \cellcolor[rgb]{0.6286, 0.7833, 0.5667} 72 & \cellcolor[rgb]{0.7098, 0.7833, 0.5667} 93 & \cellcolor[rgb]{0.7756, 0.7833, 0.5667} 110 & \cellcolor[rgb]{0.7833, 0.7446, 0.5667} 122 & \cellcolor[rgb]{0.7833, 0.7098, 0.5667} 131 & \cellcolor[rgb]{0.7833, 0.675, 0.5667} 140 & \cellcolor[rgb]{0.7833, 0.7369, 0.5667} 124 & \cellcolor[rgb]{0.7833, 0.7292, 0.5667} 126 & \cellcolor[rgb]{0.7833, 0.6982, 0.5667} 134 & \cellcolor[rgb]{0.7833, 0.6827, 0.5667} 138 & \cellcolor[rgb]{0.7833, 0.6673, 0.5667} 142 & \cellcolor[rgb]{0.7833, 0.6363, 0.5667} 150 & \cellcolor[rgb]{0.7833, 0.6092, 0.5667} 157 & \cellcolor[rgb]{0.7833, 0.5783, 0.5667} 165 \\\hline 2 & \cellcolor[rgb]{0.5667, 0.7833, 0.7795} 1 & \cellcolor[rgb]{0.5667, 0.7833, 0.7717} 3 & \cellcolor[rgb]{0.5667, 0.7833, 0.764} 5 & \cellcolor[rgb]{0.5667, 0.7833, 0.6905} 24 & \cellcolor[rgb]{0.5667, 0.7833, 0.5744} 54 & \cellcolor[rgb]{0.7563, 0.7833, 0.5667} 105 & \cellcolor[rgb]{0.7833, 0.6905, 0.5667} 136 & \cellcolor[rgb]{0.7833, 0.617, 0.5667} 155 & \cellcolor[rgb]{0.5667, 0.7833, 0.6789} 27 & \cellcolor[rgb]{0.5667, 0.7833, 0.6711} 29 & \cellcolor[rgb]{0.5667, 0.7833, 0.6479} 35 & \cellcolor[rgb]{0.5667, 0.7833, 0.6054} 46 & \cellcolor[rgb]{0.6634, 0.7833, 0.5667} 81 & \cellcolor[rgb]{0.7833, 0.7717, 0.5667} 115 & \cellcolor[rgb]{0.7833, 0.6479, 0.5667} 147 & \cellcolor[rgb]{0.7833, 0.586, 0.5667} 163 & \cellcolor[rgb]{0.6363, 0.7833, 0.5667} 74 & \cellcolor[rgb]{0.644, 0.7833, 0.5667} 76 & \cellcolor[rgb]{0.6789, 0.7833, 0.5667} 85 & \cellcolor[rgb]{0.7214, 0.7833, 0.5667} 96 & \cellcolor[rgb]{0.7679, 0.7833, 0.5667} 108 & \cellcolor[rgb]{0.7833, 0.7562, 0.5667} 119 & \cellcolor[rgb]{0.7833, 0.6595, 0.5667} 144 & \cellcolor[rgb]{0.7833, 0.6015, 0.5667} 159 \\\hline 3 & \cellcolor[HTML]{C0C0C0} & \cellcolor[HTML]{C0C0C0} & \cellcolor[HTML]{C0C0C0} & \cellcolor[HTML]{C0C0C0} & \cellcolor[HTML]{C0C0C0} & \cellcolor[HTML]{C0C0C0} & \cellcolor[HTML]{C0C0C0} & \cellcolor[HTML]{C0C0C0} & \cellcolor[rgb]{0.5667, 0.7833, 0.7485} 9 & \cellcolor[rgb]{0.5667, 0.7833, 0.7253} 15 & \cellcolor[rgb]{0.5667, 0.7833, 0.6595} 32 & \cellcolor[rgb]{0.5667, 0.7833, 0.6208} 42 & \cellcolor[rgb]{0.6673, 0.7833, 0.5667} 82 & \cellcolor[rgb]{0.7833, 0.7795, 0.5667} 113 & \cellcolor[rgb]{0.7833, 0.6208, 0.5667} 154 & \cellcolor[rgb]{0.7833, 0.5705, 0.5667} 167 & \cellcolor[rgb]{0.5667, 0.7833, 0.6092} 45 & \cellcolor[rgb]{0.5667, 0.7833, 0.5938} 49 & \cellcolor[rgb]{0.5705, 0.7833, 0.5667} 57 & \cellcolor[rgb]{0.6208, 0.7833, 0.5667} 70 & \cellcolor[rgb]{0.7292, 0.7833, 0.5667} 98 & \cellcolor[rgb]{0.7833, 0.764, 0.5667} 117 & \cellcolor[rgb]{0.7833, 0.7176, 0.5667} 129 & \cellcolor[rgb]{0.7833, 0.6324, 0.5667} 151 \\\hline 4 & \cellcolor[HTML]{C0C0C0} & \cellcolor[HTML]{C0C0C0} & \cellcolor[HTML]{C0C0C0} & \cellcolor[HTML]{C0C0C0} & \cellcolor[HTML]{C0C0C0} & \cellcolor[HTML]{C0C0C0} & \cellcolor[HTML]{C0C0C0} & \cellcolor[HTML]{C0C0C0} & \cellcolor[HTML]{C0C0C0} & \cellcolor[HTML]{C0C0C0} & \cellcolor[HTML]{C0C0C0} & \cellcolor[HTML]{C0C0C0} & \cellcolor[HTML]{C0C0C0} & \cellcolor[HTML]{C0C0C0} & \cellcolor[HTML]{C0C0C0} & \cellcolor[HTML]{C0C0C0} & \cellcolor[rgb]{0.5667, 0.7833, 0.7408} 11 & \cellcolor[rgb]{0.5667, 0.7833, 0.733} 13 & \cellcolor[rgb]{0.5667, 0.7833, 0.7176} 17 & \cellcolor[rgb]{0.5667, 0.7833, 0.6673} 30 & \cellcolor[rgb]{0.5667, 0.7833, 0.6402} 37 & \cellcolor[rgb]{0.5667, 0.7833, 0.586} 51 & \cellcolor[rgb]{0.6131, 0.7833, 0.5667} 68 & \cellcolor[rgb]{0.6943, 0.7833, 0.5667} 89 \\
    \Xhline{3\arrayrulewidth}
  \end{tabular}%
  }
  \centering
  \raisebox{-.498\height}{\includegraphics[scale=.33]{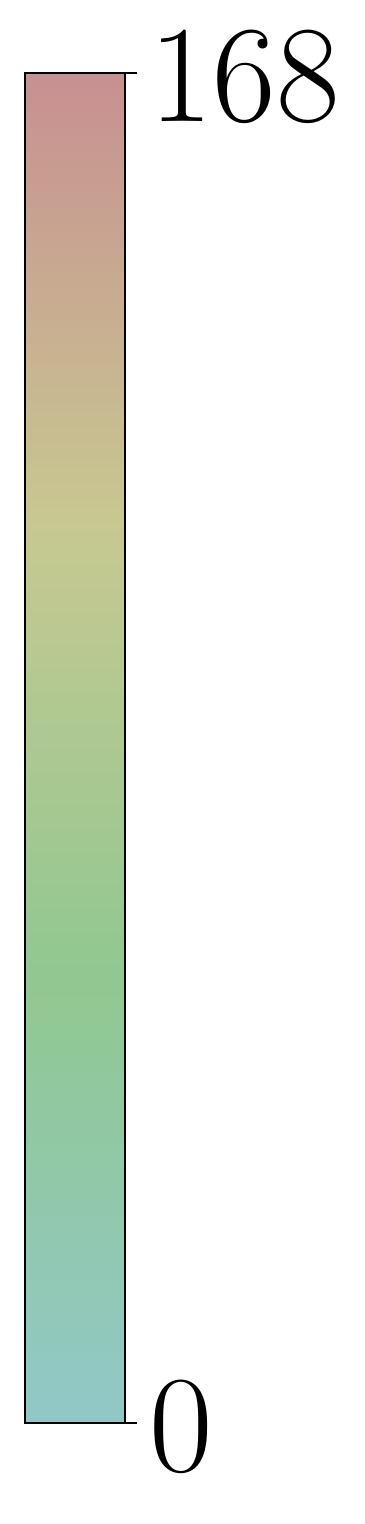}}
  \caption{%
    Order in which 168 $(\ell,m,n)$ modes are added to the QNM model
    by the greedy algorithm for \texttt{SXS:BBH:0305} with the QNM model's start time $u_{0}$ taken to be $u_{\text{peak}}$, i.e., the peak of the $L^{2}$ norm of the strain.}
\end{table*}

Apart from the fraction of unmodeled power, we also calculate the 
mismatches between the CCE strain and the model using varying number 
of modes, as shown by the solid curves in
Fig.~\ref{fig:Mismatches_vs_NModes}. As a reference, we also provide
two dash-dotted curves showing the all-mode mismatches from just the 
$(2,\pm2)$ modes---one with the $n=0$ tone and another with the $n=0-7$ 
tones. Moreover, we show the mismatch between the highest and
the next-highest resolution waveforms via the top of the black region to
illustrate that every mismatch curve is above our numerical error, and thus, 
there is no concern for overfitting to numerical noise.

\begin{figure*}
	\centering
	\includegraphics[width=.65\linewidth]{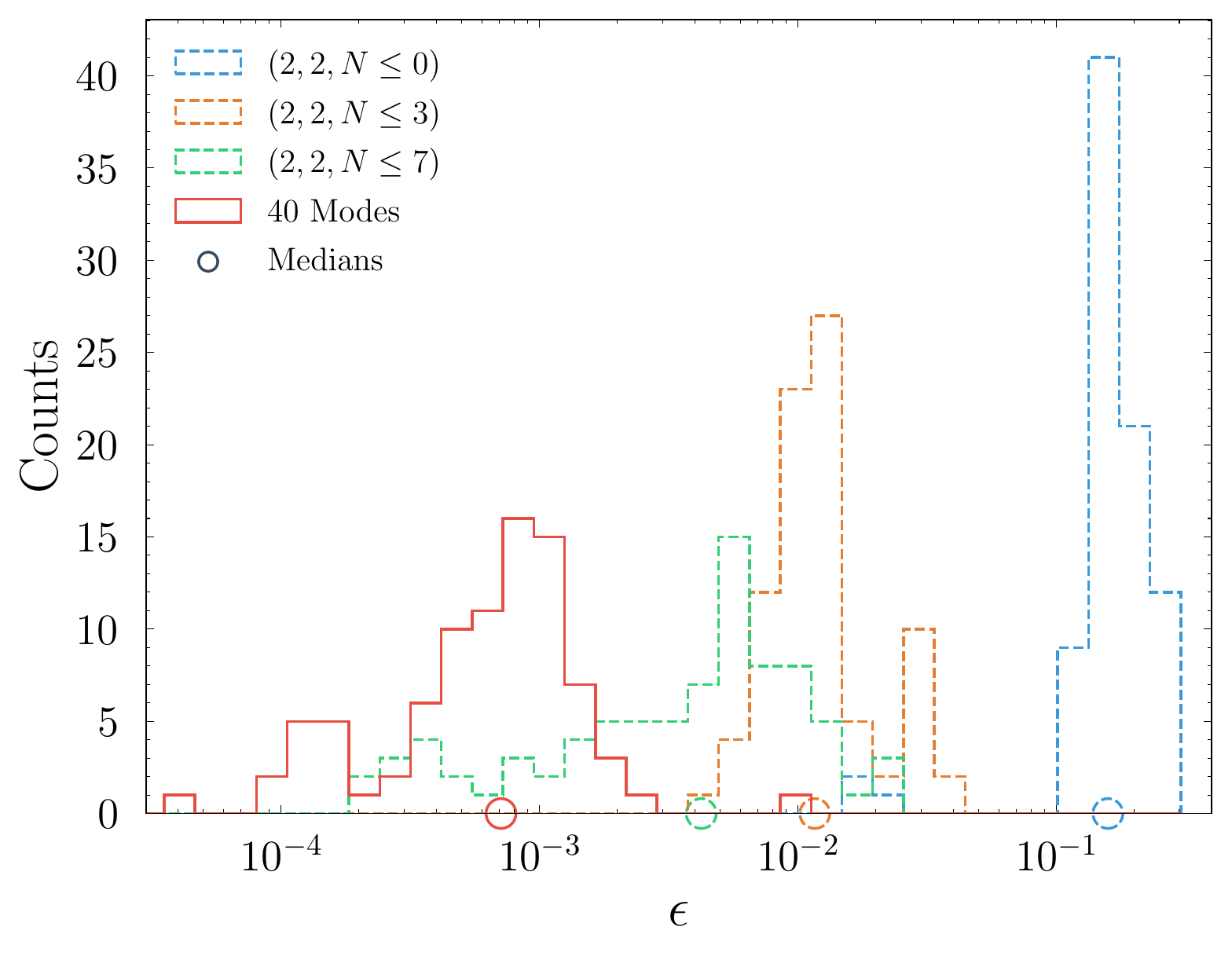}
	\caption{%
		Various distributions of the epsilons (see Eq.~\eqref{eq:epsilon})
		that have been obtained by minimizing the residual between the
		strain waveforms from 86 SXS simulations used in the \texttt{NRHyb3dq8}
		surrogate and a QNM model that has either been built with the $(2,2,0)$ mode (blue), the $(2,2,0-3)$ modes (orange), the $(2,2,0-7)$ modes 
		(green), or the 40 modes that are chosen by the
		multimode algorithm (red) (see Sec.~\ref{sec:greedy-algorithm}). On the
		$\epsilon$-axis, we also provide the median values of epsilon for each distribution.
		The starting time for the distributions created using the $(2,2)$ mode and various overtones is taken to be $u_{0}-u_{\text{peak}}=0M$ while for the distribution created using 40 modes as chosen by the greedy algorithm it is $u_0-u_\text{peak}=20M$.}
	\label{fig:Epsilon_Analysis}
\end{figure*}

Our most important finding regarding multimode fitting, however, is that by using multimode
fitting rather than just the $(2,2)$ mode with its $n=0$ tone and the first 7 overtone
modes we can significantly improve our ability to extract the
remnant's mass and spin using a QNM model. To confirm that our QNM model is able to
faithfully represent the full numerical simulation, rather than just
the waveform, we perform a minimization of the 
residual between the QNM model and the NR strain waveform, with the
remnant's mass and spin as free parameters. As a measure of the error
in the mass and spin found by our NR/QNM mismatch minimization
procedure, we use
\begin{align}
\label{eq:epsilon}
\epsilon=\sqrt{(\delta M/M)^{2}+(\delta\chi)^{2}},
\end{align}
where the terms $\delta M$ and $\delta\chi$ are the differences
between the minimization results and the remnant values obtained by
computing the Poincaré charges that correspond to the strain and
Weyl scalars produced by the simulation (see
Sec.~\ref{sec:fittingprocedure} and Eqs.~(11) and (15)
of~\cite{Iozzo:2021vnq}). Our results from this procedure are shown in
Fig.~\ref{fig:Epsilon_Analysis}.

For this analysis, we used 86 SXS simulations that were incorporated
into the \texttt{NRHyb3dq8} surrogate~\cite{Varma:2018mmi}. In this figure, there are
four histograms that show the epsilon values obtained from either
using the $(2,2,0)$ mode, the $(2,2,0-3)$ modes, the $(2,2,0-7)$ modes, or from
using the first 40 modes that are chosen by our
multimode algorithm, which is summarized in
Sec.~\ref{sec:greedy-algorithm}. For each of these histograms, we also plot
the median of the epsilons on the $\epsilon$-axis. However, as can be seen by using
just 40 modes, we can improve the median epsilon estimate across these
simulations by a more than half an order of magnitude: specifically, the
median epsilon that is obtained from the $(2,2)$ mode with up to 7 overtones
is $4.23\times10^{-3}$ while the median epsilon obtained from using 40
modes is $7.10\times10^{-4}$. Furthermore, we find that the median
epsilon only improves as one includes more than 40 modes, but
eventually reaches a minimum value of $2.68\times10^{-4}$ when using
every mode available in the waveform. We attribute this inability to
push the epsilon precision any lower to the fact that the higher-order
modes of the waveform are more influenced by numerical error and also have more
nonlinear contributions, which are not captured by the linear QNM
model (see, e.g., Fig.~2 of~\cite{Mitman:2020pbt}). Finally, we should also note that when using modes other than just
the $(2,2)$ mode, the QNM model needs to start at a time later than the usual
$u_{0} - u_{\text{peak}}=0$. This is because other modes, such as $m=0$ modes, exhibit nonlinearities related to memory effects that cannot be represented by the QNM model. Consequently, for the histogram created with 40 modes that we
show in Fig.~\ref{fig:Epsilon_Analysis}, we start our fits at
$u_{0} - u_{\text{peak}}=20 M$.

\section{Consequences of working in the super rest frame}
\label{sec:BMSFrames}

\begin{figure*}[t]
	\centering
	\includegraphics[width=\linewidth]{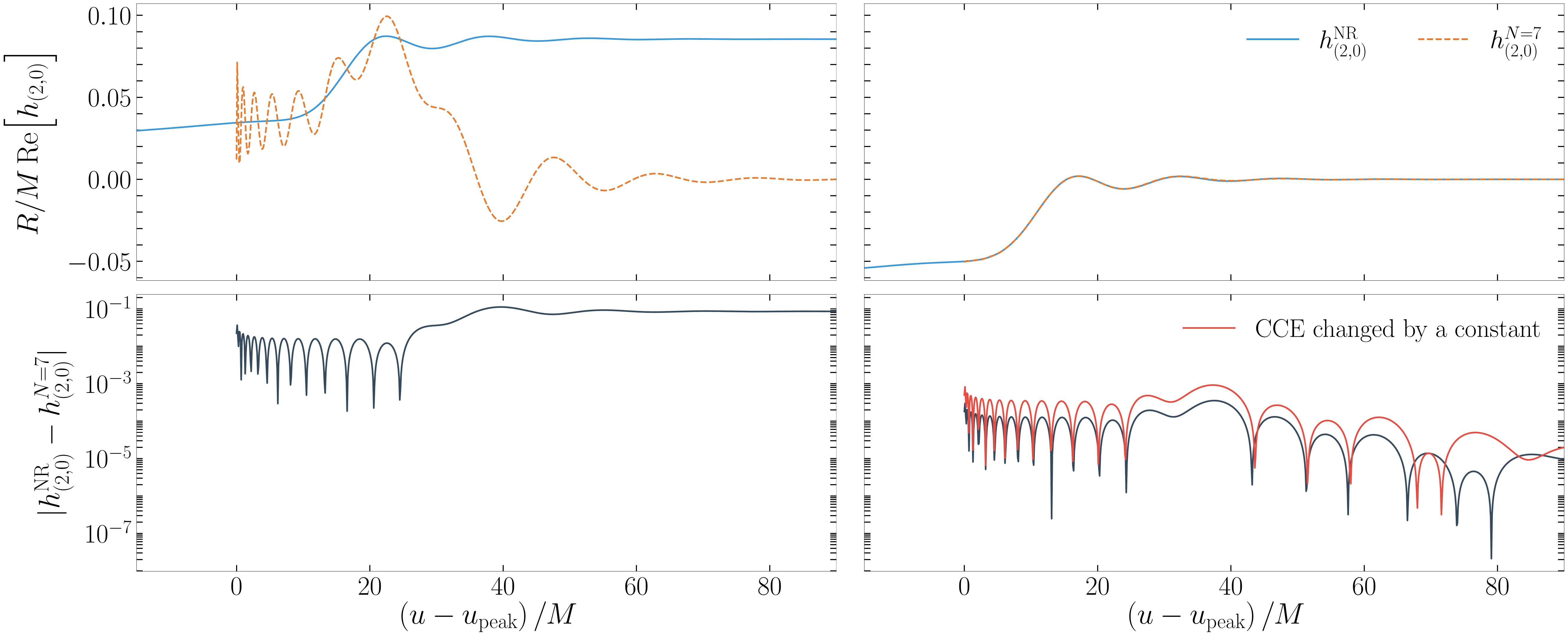}
	\caption{%
		Comparison between the real component of the $(2,0)$ mode of a
		CCE waveform and the QNM model built from the $(2,0)$ mode with $n=0$ and 7 
        overtones. The upper panels show both waveforms, while the lower
		panels show the residual between the two. In the plots on the
		left, we are using a NR waveform in the center-of-mass frame of the remnant BH, while in
		the plots on the right we have mapped the NR waveform to the super
		rest frame using the method outlined in Sec. II and Appendix A
		of~\cite{Mitman:2021xkq}. In the bottom right plot, we also show a residual curve
		in red, whose NR waveform has been mapped to the center-of-mass frame of the remnant BH and 
		changed by a constant so that it obtains a final value of zero.
		We include this curve to illustrate that by performing a supertranslation,
		rather than changing the strain by a constant, one can obtain much more accurate QNM fits
		due to the mode-mixing that is induced by supertranslations~\cite{Boyle:2015nqa}.\\
		BBH merger: \texttt{SXS:BBH:0305}.}
	\label{fig:L2_M0}
\end{figure*}

At this point, we now wish to illustrate the importance of using
waveforms that are in the super rest frame and clarify some points of
disagreement that have been present in recent works regarding
QNMs. As a reminder, in this work by super rest frame we mean the frame in which the boost velocity and space translation are fixed by minimizing the center-of-mass charge, the $2\leq\ell\leq4$ supertranslations are fixed by minimizing the $L^{2}$ norm of the $2\leq\ell\leq4$ modes of the Moreschi supermomentum (see Sec.~\ref{sec:BMSFramesIntro}), and the rotation is fixed by aligning the remnant BH's spin with the positive $z$-axis. First, we simply show the most prominent effect that working in
the incorrect BMS frame has on QNMs. In Fig.~\ref{fig:L2_M0}, we show
two plots. Both are comparisons between the real component of the
strain $(2,0)$ mode of the CCE waveform that corresponds to the
GW150914 event and a QNM model for the same mode with 7
overtones. However, the plots on the left use the waveform in the remnant BH's center-of-mass frame
whereas the plots on the right use the waveform once it
has been mapped to the super rest frame. As can be seen, the primary
difference between these two curves is that the curve on the left
approaches some nonzero value as $u\rightarrow+\infty$ while the curve
on the right instead approaches zero. Consequently, the QNM model in
the left plot completely fails, while the QNM model in the right plot
is what we would expect to see based on QNM fits to the
$(2,2)$ mode, e.g., Fig.~2 of~\cite{Giesler:2019uxc}. Again, the reason for this stark contrast in results is
due to the supertranslation freedom that is present in our asymptotic
waveforms. If one does not map their system to the super rest frame,
i.e., if one does not make their system resemble a Kerr black hole in its canonical BMS frame---rather
than a supertranslated Kerr black hole---then the QNM model
fails to represent the waveform.

\begin{figure*}[t]
	\centering
	\includegraphics[width=\linewidth]{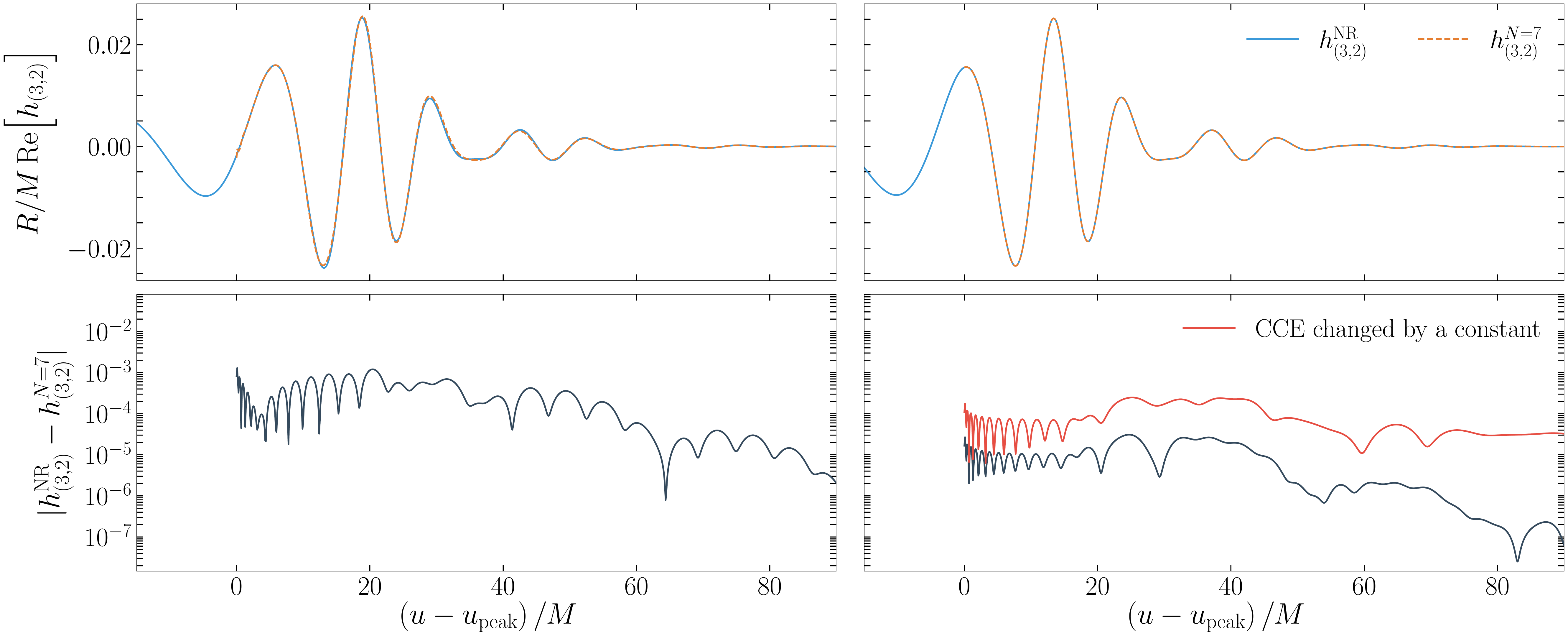}
	\caption{%
		The same as Fig.~\ref{fig:L2_M0}, but for the (3,2) mode. Note that when building this QNM model we have included not only the $(3,2,0-7)$ modes, but also the $(2,2,0-7)$ modes because these modes are needed to accurately represent the $(3,2)$ mode due to the spherical-spheroidal mixing that occurs when changing the basis of the QNM model.}
	\label{fig:L3_M2}
\end{figure*}

\begin{figure*}[t]
	\centering
	\includegraphics[width=\linewidth]{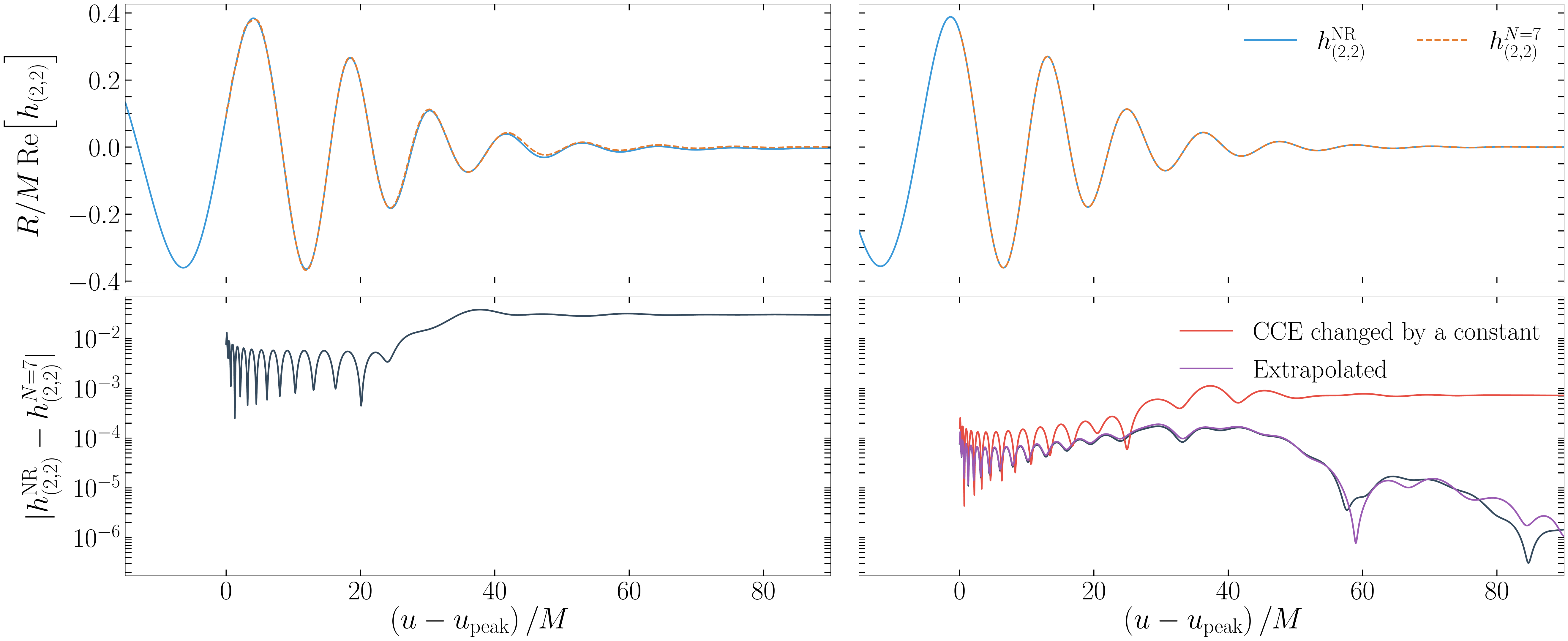}
	\caption{%
		The same as Fig.~\ref{fig:L2_M0}, but for the (2,2) mode. We also include the purple curve, which illustrates the previous result obtained by~\cite{Dhani:2020nik,Dhani:2021vac} when using an extrapolated waveform which has been changed by a constant so that its final value is zero.}
	\label{fig:L2_M2}
\end{figure*}

Apart from this, in the bottom right plot of Fig.~\ref{fig:L2_M0} we also provide the
red curve to highlight the importance of performing supertranslations, i.e.,
transforming the coordinates as well as the waveform objects,
rather than just changing the strain by a constant. This curve also shows the residual between a
NR waveform and its corresponding best-fit QNM model, but the NR waveform has been
changed by a constant so that its final value is zero, as was performed in Giesler et al.~\cite{Giesler:2019uxc}. As can be seen, while the error in this QNM fit
is comparable to that of the NR waveform whose BMS frame has been properly fixed, it is still off by
nearly an order of magnitude. Therefore, even though changing the strain by a constant is simpler than
performing a BMS transformation, applying a supertranslation produces a
much better QNM fit because it also fixes unwanted mode-mixing that occurs due to the supertranslation also changing the retarded time~\cite{Boyle:2015nqa}.

\begin{figure}[t]
	\centering
	\includegraphics[width=\linewidth]{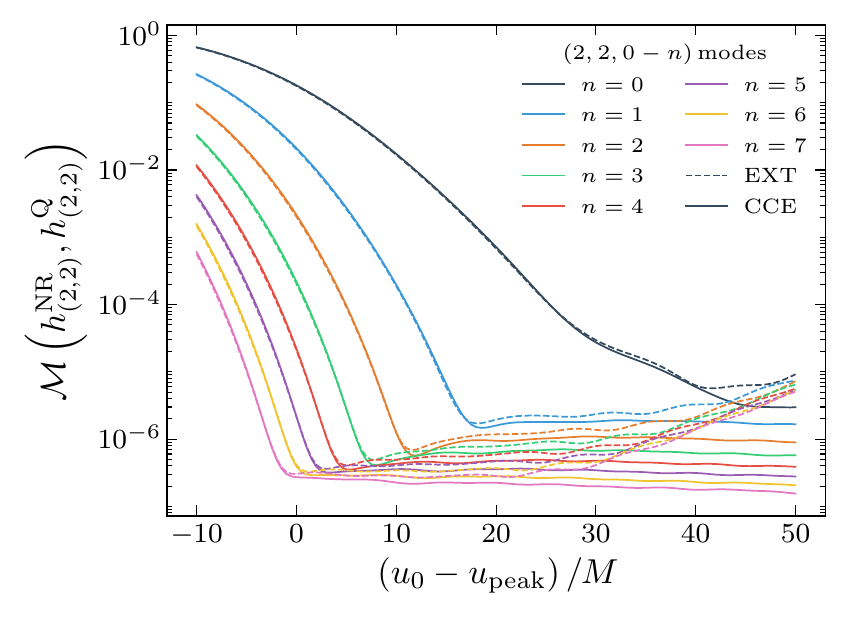}
	\caption{%
		Comparing the mismatch curves for the $(2,2)$ mode and its
		overtones obtained from the extrapolated waveform used
		in~\cite{Giesler:2019uxc} and~\cite{Cook:2020otn} (dashed curves)
		as well as the corresponding CCE waveform, after it was mapped to
		the super rest frame (solid curves). This plot also serves the
		purpose of clarifying why Fig.~1 of~\cite{Giesler:2019uxc} and
		Fig.~2 of~\cite{Cook:2020otn} are
		different. In~\cite{Giesler:2019uxc}, they performed an 
		ad hoc subtraction of their waveform, while in~\cite{Cook:2020otn} no such change to the
		extrapolated waveform was performed. Note that for this plot we only include the prograde modes in our QNM model
		to remain consistent with the results of~\cite{Giesler:2019uxc} and~\cite{Cook:2020otn}.\\
		BBH merger: \texttt{SXS:BBH:0305}.}
	\label{fig:Mismatch_from_L2_M2}
\end{figure}

While this effect is most prominent in the strain $(2,0)$
mode, it is also present in other modes, such as the $(3,2)$ as shown by Fig.~\ref{fig:L3_M2},
and even the more-commonly used strain $(2,2)$
mode, as shown by Fig.~\ref{fig:L2_M2}. Note that in Fig.~\ref{fig:L2_M2} we also provide the purple curve in the bottom right plot, which shows the previous result obtained by~\cite{Dhani:2020nik,Dhani:2021vac} when using an extrapolated waveform that has been changed by a constant so that its final value is zero. By comparing the black and purple curves in Fig.~\ref{fig:L2_M2}, one can see that previous studies that have used the $(2,2)$ mode from the extrapolated waveforms are only slightly impacted by mapping to the super rest frame.
However, we will see later that for higher modes this is not true.

In Fig.~\ref{fig:Mismatch_from_L2_M2}, we show the mismatch in
the $(2,2)$ mode between a numerical waveform and a QNM model, with
varying numbers of overtones, as a function of the QNM model's start
time $u_{0}$. Ultimately, this plot is a recreation of Fig.~1
in~\cite{Giesler:2019uxc} or Fig.~2 in~\cite{Cook:2020otn}, but with
the intent of clarifying why the figures from those two papers are in
clear contrast with one another, despite using the same SXS
waveform. In~\cite{Giesler:2019uxc}, their plot more closely resembles
our solid curves, which have been created using a CCE waveform that
has been mapped to the super rest frame. In~\cite{Cook:2020otn}, their
plot is identical to our dashed curves, which have been created using
the publicly available extrapolated waveform that can be found in the
SXS Catalog~\cite{Boyle:2019kee, SXSCatalog}. This is the waveform used in~\cite{Giesler:2019uxc,Cook:2020otn}, without one important change.
What is different about the data used in~\cite{Giesler:2019uxc} is that
they performed an ad hoc subtraction of their waveform to send it to $0$ as $u\rightarrow+\infty$. In~\cite{Cook:2020otn} and in the dashed curves of
Fig.~\ref{fig:Mismatch_from_L2_M2}, this subtraction was not
performed, hence the worsening of the mismatch that can be seen as
$u_{0}$ increases. The reason why our solid curves more closely
resemble the curves seen in~\cite{Giesler:2019uxc} is because we have
mapped our waveform to the super rest frame using supertranslations,
rather than changing the waveform by a constant. While
the two actions have similar effects, supertranslations also affect the
coordinates, which is not true of changing the waveform by a
constant. This is illustrated by the red curve in the bottom right plot
of Figs.~\ref{fig:L2_M0},~\ref{fig:L3_M2}, and~\ref{fig:L2_M2}. Therefore, Fig.~\ref{fig:Mismatch_from_L2_M2} clearly
illustrates the importance of mapping to the super rest frame, even
for modes such as the $(2,2)$ mode where such effects were thought to
be negligible.

\begin{figure*}
	\centering
	\includegraphics[width=0.75\linewidth]{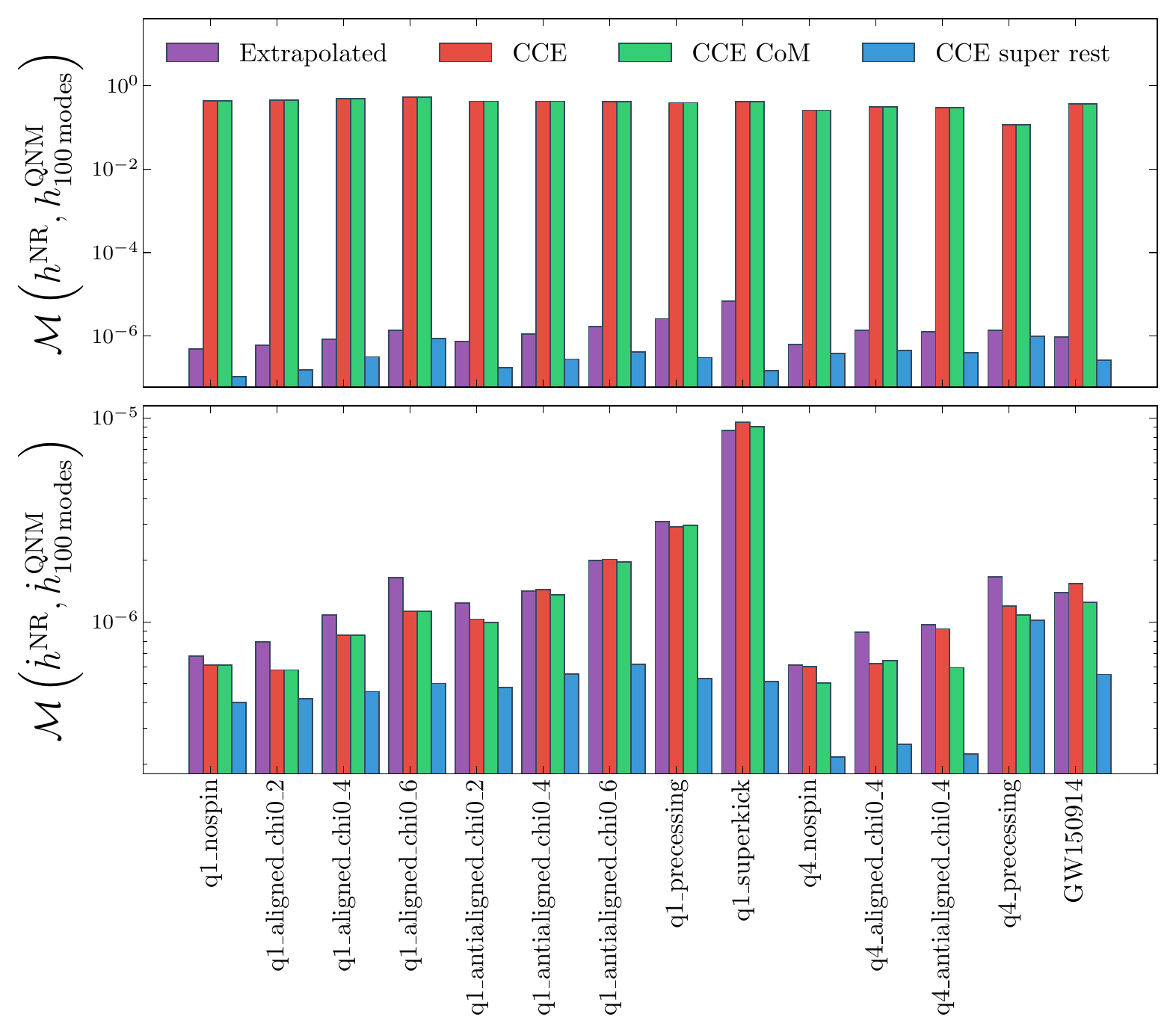}
	\caption{%
		Examining the mismatch between a NR waveform and a QNM model that
		is built from 100 modes as a function of the BMS frame that the
		numerical waveform is mapped to. The QNM model start time $u_{0}$ is taken to be the time at which the $L^{2}$ norm of the news takes on its maximum value. We show four bars that
		correspond to the extrapolated waveform (EXT) and the CCE waveform in three different BMS frames: the arbitrary BMS frame
		that the output of CCE is in, the remnant BH's center-of-mass frame, and the
		super rest frame. In the top plot, we show the mismatch between
		the strain waveforms, while in the bottom plot we show the
		mismatch between the news waveforms. The parameters of the 14
		binary black holes mergers that appear on the horizontal axis can be
		found in Table~\ref{tab:runs}.}
	\label{fig:BMS_Frame_Comparison}
\end{figure*}

Based on the results that are shown in Fig.~\ref{fig:L2_M0}, one's immediate response to this
issue of BMS frames might be to simply fit the QNM model to the news
instead of the strain, seeing as the displacement memory effect is not
present in the news. To counter this proposal, however, we provide
Fig.~\ref{fig:BMS_Frame_Comparison}, which shows the mismatch between
numerical waveforms and QNM models built from 100 modes for a wide range of
systems whose parameters can be found in Table~\ref{tab:runs}. In the
top panel, we are performing our QNM fits in the strain domain, while
in the bottom panel we are performing our QNM fits in the news
domain. For each panel, we also show four types of mismatch comparisons: when the numerical
waveforms are extrapolated waveforms (EXT) and when the CCE waveforms (i) are not mapped to a certain BMS frame, (ii) have been mapped to
just the center-of-mass (CoM) frame, or (iii) have been mapped to the the super rest frame
using the procedures outlined in~\cite{Mitman:2021xkq}. As can be seen
in the top panel, mapping to the super rest frame, on average,
improves the mismatches by 5 orders of magnitude. This, however,
should not come as a surprise seeing as this result is predominantly
due to supertranslating away the offset in the strain induced by the
gravitational memory effect, e.g., what is shown in Fig.~\ref{fig:L2_M0}. What
might be surprising is what is shown in the bottom
panel: namely that mapping to the super rest frame, on average,
also improves the mismatches in the news domain by a factor of 4. Due to this, we now realize
that mapping to the super rest frame is even important in the news or $\Psi_{4}$ domains where there is no memory effect.
This phenomenon is due to the mode mixing that occurs due to the 
change in coordinates of the system by the supertranslations.

\begin{figure*}
	\centering
	\includegraphics[width=0.75\linewidth]{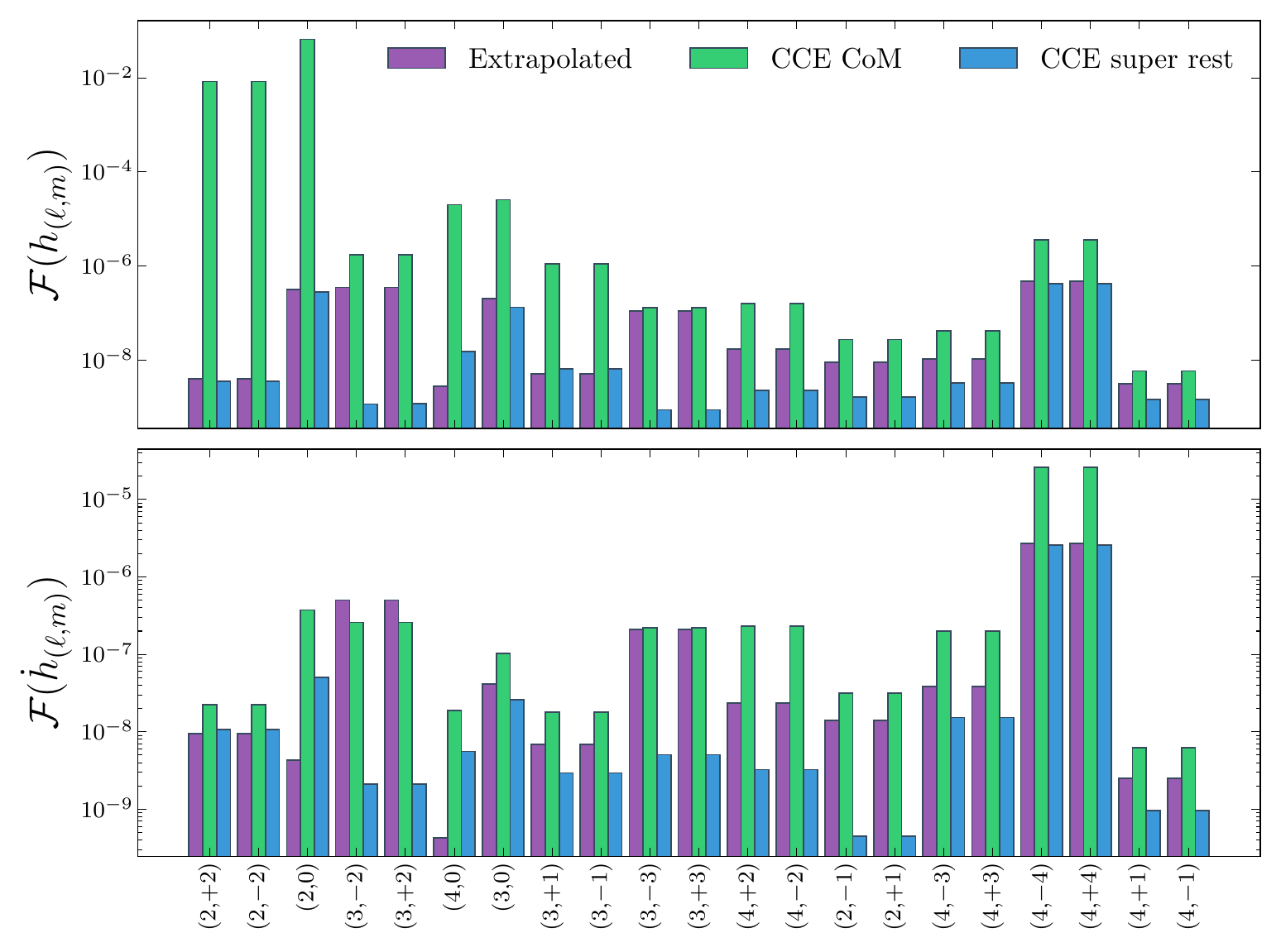}
	\caption{%
		Examining the fraction of unmodeled power between a NR waveform and a QNM model that
n		is built from 168 modes as a function of mode for extrapolated (EXT) and CCE waveforms. The QNM model start time $u_{0}$ is taken to be $u_{\text{peak}}$.
		We show three bars that
		correspond to the extrapolated waveform, the CCE waveform in the remnant BH's center-of-mass frame, and the CCE waveform in
		super rest frame. In the top plot, we show the fraction of unmodeled power between
		the strain waveforms, while in the bottom plot we show the
		fraction of unmodeled power between the news waveforms. The modes are organized in terms
		of the largest relative difference in the fraction of unmodeled power in the strain domain between the
		center-of-mass and super rest frame CCE waveforms.\\
		BBH merger: \texttt{SXS:BBH:0305}.
		}
	\label{fig:UnmodeledPower_vs_Mode}
\end{figure*}

Last, to help illustrate what brings about these changes in the mismatch as a function of frame, we present Fig.~\ref{fig:UnmodeledPower_vs_Mode}. In Fig.~\ref{fig:UnmodeledPower_vs_Mode} we show how the
fraction of unmodeled power varies as a function of mode for strain and news waveforms
in the center-of-mass or super rest frame for the simulation \texttt{SXS:BBH:0305}. More specifically,
for each waveform we build a QNM model using every available mode and then we compute
the fraction of unmodeled power between the numerical waveform and the QNM model using Eq.~\eqref{eq:powerfraction} (top plot) or Eq.~\eqref{eq:powerfractionnews} (bottom plot) with the residual, i.e., Eq.~\eqref{eq:Rdef} or~Eq.~\eqref{eq:Rnewsdef}, only involving the corresponding mode of the waveform and the QNM model. We organize the modes in terms of the largest relative difference in the fraction of unmodeled power in the strain domain between the center-of-mass and the super rest frame waveforms. As can be seen, in the strain domain the modes that are most strongly impacted by the super rest frame are the $(2,\pm2)$ modes, the $(3,\pm2)$ modes, and the $m=0$ modes. This occurs for many reasons. For the $m=0$ modes, this is most naturally understood by realizing that these modes often exhibit more memory effects than others and thus require the supertranslations to reduce the offset normally found in the ringdown phase of these modes. Put differently, these modes are strongly influenced by the $\bar{\eth}^{2}\alpha(\theta,\phi)$ factor in Eq.~\eqref{eq:strainsupertranslation}. For the $(3,\pm2)$ modes, the reason why these modes are impacted is because of the mode-mixing that occurs due to Taylor expanding the strain in the supertranslated coordinate system about the original coordinate system. For the supertranslations that we apply to map to the super rest frame, the most dominant mode is the $(2,0)$ mode. Consequently, since the dominant modes of the news are the $(2,\pm2)$ modes, the mode of the new strain that will be most influenced by the supertranslation's mode mixing is the mode corresponding to the product of the $Y_{(2,0)}$ and $\phantom{}_{-2}Y_{(2,\pm2)}$ functions, which happens to be the $(3,\pm2)$ mode. This can be seen directly by making use of the spin-weighted spherical harmonic triple integral identity:
\begin{widetext}
\begin{equation}
\label{eq:triple}
\int_{S^{2}}\phantom{}_{s_{1}}Y_{\ell_{1}m_{1}}\,\phantom{}_{s_{2}}Y_{\ell_{2}m_{2}}\,\phantom{}_{s_{3}}Y_{\ell_{3}m_{3}}=\sqrt{\frac{(2\ell_{1}+1)(2\ell_{2}+1)(2\ell_{3}+1)}{4\pi}}\begin{pmatrix}\ell_{1}&\ell_{2}&\ell_{3}\\m_{1}&m_{2}&m_{3}\end{pmatrix}\begin{pmatrix}\ell_{1}&\ell_{2}&\ell_{3}\\-s_{1}&-s_{2}&-s_{3}\end{pmatrix}
\end{equation}
\end{widetext}
for $s_{1}+s_{2}+s_{3}=0$, and then computing the corresponding Wigner $3-j$ symbols to see which modes are excited~\cite{Campbell:1970ww}. Last, for the $(2,\pm2)$ modes, this is because these two modes of the strain experience an unexpected initial offset due to transient effects arising in the CCE evolution~\cite{Mitman:2020pbt,Moxon:2020gha,Moxon:2021gbv}. Meanwhile, in the news domain, by closer inspection one finds that the modes most strongly influenced by mapping to the super rest frame are the $(3,\pm2)$, $(4,\pm2)$, and $(2,\pm1)$ modes. Like the case of the $(3,\pm2)$ modes in the strain domain, this is because these modes also experience considerable changes due to supertranslation mode-mixing effects, as can be verified with Eq.~\eqref{eq:triple}. The other important thing to note regarding Fig.~\ref{fig:UnmodeledPower_vs_Mode}, as well as Fig.~\ref{fig:BMS_Frame_Comparison}, is that by mapping CCE waveforms to the super rest frame, we can always produce a better mismatch between the numerical waveform and the QNM model than if we were using an extrapolated waveform.

\section{Conclusion}

Across this study, we have developed a QNM model which simultaneously
fits multiple modes over all angles and times using NR waveforms that
have been mapped to the super rest frame. First, we showed that for 
\texttt{SXS:BBH:0305} the amount of power captured in the QNM model 
when fitting to only the dominant strain $(2,\pm2)$ modes is below 65\% 
of the whole numerical waveform's power over all modes. Moreover, we also found that 
including the $(2,\pm2)$ overtone modes does not dramatically increase 
the power modeled, except at times very close to $u_{0}=u_{\text{peak}}$, in which case a 
50\% improvement over the fundamental mode can be seen by including 
7 overtones. To increase the amount of modeled power, one must instead 
rely on higher-order modes.

Choosing which higher-order modes to include in the QNM model is a
nontrivial task. Therefore, we developed a greedy algorithm that
picks which modes to include based on the fraction of unmodeled power
in them. We find that by including just 5 modes, we can model
96\% of a waveform's power and by including just 20 modes, we can push
that number up to 99\%. We also find an all-angles mismatch 
improvement by a factor of $10^5$ when using multimode fitting as 
compared to using the $(2,\pm 2,n)$ modes. Furthermore, we also showed the practical
importance of this higher-order mode power modeling improvement:
obtaining more accurate estimates of the remnant's mass and spin. With
40 modes we found that we can, on average, obtain mass and spin
estimates that yield an epsilon value (see Eq.~\eqref{eq:epsilon})
that is more than half an order of magnitude better than what can be obtained
by using the $(2,2)$ mode with 7 overtones. While we found that we can
further improve estimates by including even more modes, the minimum
median epsilon that we computed was only 62\% less than that obtained
by using 40 modes. We attribute this to the fact that by including
higher-order modes, there are more nonlinearities that the QNM model
has to try and fit (see, e.g., Fig.~2 of~\cite{Mitman:2020pbt}).\footnote{In 
Fig.~2 of~\cite{Mitman:2020pbt} the blue and green curves in the middle plot 
represent nonlinearities, which would not be captured by the QNM model.}

Greedy algorithms can suffer instabilities due to degeneracy in the underlying 
model~\cite{Guyon2003AnIT}. However, we know that the overlaps 
between different QNM modes with distinct angular indices $(\ell,m)$ are 
small for the spins that we are considering because the angular part of 
the QNM $(l,m,n,p)$ is dominantly in the $(l,m)$ spherical harmonic. 
Therefore, any degeneracy we might expect is likely only due to the overtone 
number~\cite{Berti:2014fga}. Since the overtones are chosen sequentially, 
and not greedily, our algorithm should not suffer from such degeneracy 
problems. Nevertheless, this is an important check to keep in mind and 
would be interesting to examine in the future.

Lastly, we also illustrated the importance of using
waveforms that have been mapped to the same BMS frame as that of the
QNM model. As shown in Figs.~\ref{fig:L2_M0},~\ref{fig:L3_M2},~\ref{fig:L2_M2},~\ref{fig:Mismatch_from_L2_M2},~\ref{fig:BMS_Frame_Comparison}, and~\ref{fig:UnmodeledPower_vs_Mode} if one does not map their
waveforms to the super rest frame then being in the wrong BMS frame
makes it problematic to model the ringdown part of a waveform
with QNMs. This is because when the Teukolsky equation is separated, the 
coordinate system used corresponds to the super rest frame at $\scri^+$.
Thus the QNM ansatz in Eq.~\eqref{eq:h-QNM-spheroidal} is only valid in this
frame. On the other hand, numerical simulations have a history of radiated
gravitational waves that cause their frame to typically deviate significantly
from this frame.  
Consequently, the waveforms emitted by these ringing
black holes need to be mapped to the super rest frame, if they are to be correctly
modeled by QNMs. Furthermore, this importance
of BMS frames extends beyond accounting for the memory effect in the
strain. In Fig.~\ref{fig:BMS_Frame_Comparison}, we showed that while
mapping to the super rest frame is most important for modeling the
strain, it also plays a nontrivial role in modeling the news because
supertranslations also change the Bondi coordinates and can thus reduce 
supertranslation-induced mode-mixing in the news. Overall, we found that previous studies that focus on modeling the $(2,2)$ mode from extrapolated waveforms are only slightly impacted by mapping to the super rest frame. However, with the inclusion of more modes or memory effects, fixing the BMS frame before fitting QNMs is crucial.

As is illustrated by the fact that future ground-based detectors like
the Einstein Telescope and Cosmic Explorer are expected to observe
$10^{2} - 10^{4}$ events per year with strong ringdown signals,
including higher-order modes and BMS frame fixing will undoubtedly be
important for correctly modeling such ringdown signals with QNMs. These modeling enhancements
should therefore also help with measuring
properties of the remnant black holes as well as testing Einstein's
theory of relativity~\cite{Maggiore:2019uih,Baibhav:2019gxm, Isi:2019aib}. While BMS frame fixing may
not prove to be directly useful for LIGO/Virgo observations,\footnote{Because the detector
measures the waveform at a single point on the sky only, the supertranslation will only
shift the waveform.} if the ringdown phase of NR waveforms is to be used to study 
remnant BHs and model their amplitudes then fixing the BMS frame will certainly be 
important, as illustrated in this work.
Furthermore, while we have presented a template for
improving QNM models by comparing QNMs against numerical relativity
waveforms, it would be very interesting to see our work applied to the
observations already collected by LIGO and Virgo.

\section*{Acknowledgments}
We thank Max Isi for fruitful discussions, Matt Giesler for sharing his work, which helped us clarify
the differences between our results and those
of~\cite{Giesler:2019uxc} and~\cite{Cook:2020otn}, and Arnab Dhani
for detailed discussions on the methods used in his work~\cite{Dhani:2020nik}. We also thank Greg Cook, Arnab Dhani, Matt Giesler, Max Isi, and Xiang Li for reviewing an earlier version of this manuscript. Calculations were
performed with the Wheeler cluster at the California Institute of
Technology (Caltech), which is supported by the Sherman Fairchild
Foundation and by Caltech.
The work of L.M.Z. was partially supported by the MSSGC Graduate
Research Fellowship, awarded through the NASA Cooperative Agreement
80NSSC20M0101. Part of this research was performed while L.M.Z. was 
visiting the Institute for Pure and Applied Mathematics (IPAM), which is 
supported by the National Science Foundation (Grant No. DMS-1925919)
The work of K.M was partially supported by NSF Grants No.~PHY-2011961,
No.~PHY-2011968, and No.~OAC-1931266.
The work of N.K. was partially supported by NSF Grant No.~PHY-1806356, Grant No.~UN2017-92945 from the Urania Stott Fund of the Pittsburgh Foundation, the Eberly research funds of Penn State at Penn State and the Mebus fellowship.
The work of L.C.S. was partially supported by NSF CAREER Award
PHY--2047382.
All plots were made using the python package \texttt{matplotlib}~\cite{Hunter:2007ouj}.

\def\bibsection{\section*{References}}

\bibliography{qnm_fits}

\begin{thebibliography}{80}%
\makeatletter
\providecommand \@ifxundefined [1]{%
 \@ifx{#1\undefined}
}%
\providecommand \@ifnum [1]{%
 \ifnum #1\expandafter \@firstoftwo
 \else \expandafter \@secondoftwo
 \fi
}%
\providecommand \@ifx [1]{%
 \ifx #1\expandafter \@firstoftwo
 \else \expandafter \@secondoftwo
 \fi
}%
\providecommand \natexlab [1]{#1}%
\providecommand \enquote  [1]{``#1''}%
\providecommand \bibnamefont  [1]{#1}%
\providecommand \bibfnamefont [1]{#1}%
\providecommand \citenamefont [1]{#1}%
\providecommand \href@noop [0]{\@secondoftwo}%
\providecommand \href [0]{\begingroup \@sanitize@url \@href}%
\providecommand \@href[1]{\@@startlink{#1}\@@href}%
\providecommand \@@href[1]{\endgroup#1\@@endlink}%
\providecommand \@sanitize@url [0]{\catcode `\\12\catcode `\$12\catcode
  `\&12\catcode `\#12\catcode `\^12\catcode `\_12\catcode `\%12\relax}%
\providecommand \@@startlink[1]{}%
\providecommand \@@endlink[0]{}%
\providecommand \url  [0]{\begingroup\@sanitize@url \@url }%
\providecommand \@url [1]{\endgroup\@href {#1}{\urlprefix }}%
\providecommand \urlprefix  [0]{URL }%
\providecommand \Eprint [0]{\href }%
\providecommand \doibase [0]{https://doi.org/}%
\providecommand \selectlanguage [0]{\@gobble}%
\providecommand \bibinfo  [0]{\@secondoftwo}%
\providecommand \bibfield  [0]{\@secondoftwo}%
\providecommand \translation [1]{[#1]}%
\providecommand \BibitemOpen [0]{}%
\providecommand \bibitemStop [0]{}%
\providecommand \bibitemNoStop [0]{.\EOS\space}%
\providecommand \EOS [0]{\spacefactor3000\relax}%
\providecommand \BibitemShut  [1]{\csname bibitem#1\endcsname}%
\let\auto@bib@innerbib\@empty
\bibitem [{\citenamefont {Barack}\ \emph {et~al.}(2019)\citenamefont {Barack},
  \citenamefont {Cardoso}, \citenamefont {Nissanke}, \citenamefont {Sotiriou},
  \citenamefont {Askar}, \citenamefont {Belczynski}, \citenamefont {Bertone},
  \citenamefont {Bon}, \citenamefont {Blas}, \citenamefont {Brito},\ and\
  \citenamefont {et~al.}}]{Barack_2019}%
  \BibitemOpen
  \bibfield  {author} {\bibinfo {author} {\bibfnamefont {L.}~\bibnamefont
  {Barack}}, \bibinfo {author} {\bibfnamefont {V.}~\bibnamefont {Cardoso}},
  \bibinfo {author} {\bibfnamefont {S.}~\bibnamefont {Nissanke}}, \bibinfo
  {author} {\bibfnamefont {T.~P.}\ \bibnamefont {Sotiriou}}, \bibinfo {author}
  {\bibfnamefont {A.}~\bibnamefont {Askar}}, \bibinfo {author} {\bibfnamefont
  {C.}~\bibnamefont {Belczynski}}, \bibinfo {author} {\bibfnamefont
  {G.}~\bibnamefont {Bertone}}, \bibinfo {author} {\bibfnamefont
  {E.}~\bibnamefont {Bon}}, \bibinfo {author} {\bibfnamefont {D.}~\bibnamefont
  {Blas}}, \bibinfo {author} {\bibfnamefont {R.}~\bibnamefont {Brito}},\ and\
  \bibinfo {author} {\bibnamefont {et~al.}},\ }\bibfield  {title} {\bibinfo
  {title} {Black holes, gravitational waves and fundamental physics: a
  roadmap},\ }\href {https://doi.org/10.1088/1361-6382/ab0587} {\bibfield
  {journal} {\bibinfo  {journal} {Classical and Quantum Gravity}\ }\textbf
  {\bibinfo {volume} {36}},\ \bibinfo {pages} {143001} (\bibinfo {year}
  {2019})}\BibitemShut {NoStop}%
\bibitem [{\citenamefont {Abbott}\ \emph {et~al.}(2016)\citenamefont {Abbott}
  \emph {et~al.}}]{LIGOScientific:2016aoc}%
  \BibitemOpen
  \bibfield  {author} {\bibinfo {author} {\bibfnamefont {B.~P.}\ \bibnamefont
  {Abbott}} \emph {et~al.} (\bibinfo {collaboration} {LIGO Scientific,
  Virgo}),\ }\bibfield  {title} {\bibinfo {title} {{Observation of
  Gravitational Waves from a Binary Black Hole Merger}},\ }\href
  {https://doi.org/10.1103/PhysRevLett.116.061102} {\bibfield  {journal}
  {\bibinfo  {journal} {Phys. Rev. Lett.}\ }\textbf {\bibinfo {volume} {116}},\
  \bibinfo {pages} {061102} (\bibinfo {year} {2016})},\ \Eprint
  {https://arxiv.org/abs/1602.03837} {arXiv:1602.03837 [gr-qc]} \BibitemShut
  {NoStop}%
\bibitem [{\citenamefont {Abbott}\ \emph {et~al.}(2017)\citenamefont {Abbott}
  \emph {et~al.}}]{LIGOScientific:2017vwq}%
  \BibitemOpen
  \bibfield  {author} {\bibinfo {author} {\bibfnamefont {B.~P.}\ \bibnamefont
  {Abbott}} \emph {et~al.} (\bibinfo {collaboration} {LIGO Scientific,
  Virgo}),\ }\bibfield  {title} {\bibinfo {title} {{GW170817: Observation of
  Gravitational Waves from a Binary Neutron Star Inspiral}},\ }\href
  {https://doi.org/10.1103/PhysRevLett.119.161101} {\bibfield  {journal}
  {\bibinfo  {journal} {Phys. Rev. Lett.}\ }\textbf {\bibinfo {volume} {119}},\
  \bibinfo {pages} {161101} (\bibinfo {year} {2017})},\ \Eprint
  {https://arxiv.org/abs/1710.05832} {arXiv:1710.05832 [gr-qc]} \BibitemShut
  {NoStop}%
\bibitem [{\citenamefont {Abbott}\ \emph
  {et~al.}(2021{\natexlab{a}})\citenamefont {Abbott} \emph
  {et~al.}}]{LIGOScientific:2021djp}%
  \BibitemOpen
  \bibfield  {author} {\bibinfo {author} {\bibfnamefont {R.}~\bibnamefont
  {Abbott}} \emph {et~al.} (\bibinfo {collaboration} {LIGO Scientific, VIRGO,
  KAGRA}),\ }\bibfield  {title} {\bibinfo {title} {{GWTC-3: Compact Binary
  Coalescences Observed by LIGO and Virgo During the Second Part of the Third
  Observing Run}},\ }\Eprint {https://arxiv.org/abs/2111.03606}
  {arXiv:2111.03606 [gr-qc]} \BibitemShut {NoStop}%
\bibitem [{\citenamefont {Abbott}\ \emph
  {et~al.}(2021{\natexlab{b}})\citenamefont {Abbott} \emph
  {et~al.}}]{LIGOScientific:2021sio}%
  \BibitemOpen
  \bibfield  {author} {\bibinfo {author} {\bibfnamefont {R.}~\bibnamefont
  {Abbott}} \emph {et~al.} (\bibinfo {collaboration} {LIGO Scientific, VIRGO,
  KAGRA}),\ }\bibfield  {title} {\bibinfo {title} {{Tests of General Relativity
  with GWTC-3}},\ }\Eprint {https://arxiv.org/abs/2112.06861} {arXiv:2112.06861
  [gr-qc]} \BibitemShut {NoStop}%
\bibitem [{\citenamefont {Pretorius}(2005)}]{Pretorius:2005gq}%
  \BibitemOpen
  \bibfield  {author} {\bibinfo {author} {\bibfnamefont {F.}~\bibnamefont
  {Pretorius}},\ }\bibfield  {title} {\bibinfo {title} {{Evolution of binary
  black hole spacetimes}},\ }\href
  {https://doi.org/10.1103/PhysRevLett.95.121101} {\bibfield  {journal}
  {\bibinfo  {journal} {Phys. Rev. Lett.}\ }\textbf {\bibinfo {volume} {95}},\
  \bibinfo {pages} {121101} (\bibinfo {year} {2005})},\ \Eprint
  {https://arxiv.org/abs/gr-qc/0507014} {arXiv:gr-qc/0507014} \BibitemShut
  {NoStop}%
\bibitem [{\citenamefont {Boyle}\ \emph {et~al.}(2019)\citenamefont {Boyle}
  \emph {et~al.}}]{Boyle:2019kee}%
  \BibitemOpen
  \bibfield  {author} {\bibinfo {author} {\bibfnamefont {M.}~\bibnamefont
  {Boyle}} \emph {et~al.},\ }\bibfield  {title} {\bibinfo {title} {{The SXS
  Collaboration catalog of binary black hole simulations}},\ }\href
  {https://doi.org/10.1088/1361-6382/ab34e2} {\bibfield  {journal} {\bibinfo
  {journal} {Class. Quant. Grav.}\ }\textbf {\bibinfo {volume} {36}},\ \bibinfo
  {pages} {195006} (\bibinfo {year} {2019})},\ \Eprint
  {https://arxiv.org/abs/1904.04831} {arXiv:1904.04831 [gr-qc]} \BibitemShut
  {NoStop}%
\bibitem [{\citenamefont {Jani}\ \emph {et~al.}(2016)\citenamefont {Jani},
  \citenamefont {Healy}, \citenamefont {Clark}, \citenamefont {London},
  \citenamefont {Laguna},\ and\ \citenamefont {Shoemaker}}]{Jani_2016}%
  \BibitemOpen
  \bibfield  {author} {\bibinfo {author} {\bibfnamefont {K.}~\bibnamefont
  {Jani}}, \bibinfo {author} {\bibfnamefont {J.}~\bibnamefont {Healy}},
  \bibinfo {author} {\bibfnamefont {J.~A.}\ \bibnamefont {Clark}}, \bibinfo
  {author} {\bibfnamefont {L.}~\bibnamefont {London}}, \bibinfo {author}
  {\bibfnamefont {P.}~\bibnamefont {Laguna}},\ and\ \bibinfo {author}
  {\bibfnamefont {D.}~\bibnamefont {Shoemaker}},\ }\bibfield  {title} {\bibinfo
  {title} {Georgia tech catalog of gravitational waveforms},\ }\href
  {https://doi.org/10.1088/0264-9381/33/20/204001} {\bibfield  {journal}
  {\bibinfo  {journal} {Classical and Quantum Gravity}\ }\textbf {\bibinfo
  {volume} {33}},\ \bibinfo {pages} {204001} (\bibinfo {year}
  {2016})}\BibitemShut {NoStop}%
\bibitem [{\citenamefont {Healy}\ \emph {et~al.}(2017)\citenamefont {Healy},
  \citenamefont {Lousto}, \citenamefont {Zlochower},\ and\ \citenamefont
  {Campanelli}}]{Healy_2017}%
  \BibitemOpen
  \bibfield  {author} {\bibinfo {author} {\bibfnamefont {J.}~\bibnamefont
  {Healy}}, \bibinfo {author} {\bibfnamefont {C.~O.}\ \bibnamefont {Lousto}},
  \bibinfo {author} {\bibfnamefont {Y.}~\bibnamefont {Zlochower}},\ and\
  \bibinfo {author} {\bibfnamefont {M.}~\bibnamefont {Campanelli}},\ }\bibfield
   {title} {\bibinfo {title} {The rit binary black hole simulations catalog},\
  }\href {https://doi.org/10.1088/1361-6382/aa91b1} {\bibfield  {journal}
  {\bibinfo  {journal} {Classical and Quantum Gravity}\ }\textbf {\bibinfo
  {volume} {34}},\ \bibinfo {pages} {224001} (\bibinfo {year}
  {2017})}\BibitemShut {NoStop}%
\bibitem [{\citenamefont {Healy}\ and\ \citenamefont
  {Lousto}(2020)}]{Healy2020}%
  \BibitemOpen
  \bibfield  {author} {\bibinfo {author} {\bibfnamefont {J.}~\bibnamefont
  {Healy}}\ and\ \bibinfo {author} {\bibfnamefont {C.~O.}\ \bibnamefont
  {Lousto}},\ }\bibfield  {title} {\bibinfo {title} {Third rit binary black
  hole simulations catalog},\ }\href
  {http://dx.doi.org/10.1103/PhysRevD.102.104018} {\bibfield  {journal}
  {\bibinfo  {journal} {Phys. Rev. D}\ }\textbf {\bibinfo {volume} {102}},\
  \bibinfo {pages} {104018} (\bibinfo {year} {2020})},\ \Eprint
  {https://arxiv.org/abs/2007.07910} {arXiv:2007.07910 [gr-qc]} \BibitemShut
  {NoStop}%
\bibitem [{\citenamefont {{Teukolsky}}(1973)}]{Teukolsky:1973}%
  \BibitemOpen
  \bibfield  {author} {\bibinfo {author} {\bibfnamefont {S.~A.}\ \bibnamefont
  {{Teukolsky}}},\ }\bibfield  {title} {\bibinfo {title} {{Perturbations of a
  Rotating Black Hole. I. Fundamental Equations for Gravitational,
  Electromagnetic, and Neutrino-Field Perturbations}},\ }\href
  {https://doi.org/10.1086/152444} {\bibfield  {journal} {\bibinfo  {journal}
  {\apj}\ }\textbf {\bibinfo {volume} {185}},\ \bibinfo {pages} {635} (\bibinfo
  {year} {1973})}\BibitemShut {NoStop}%
\bibitem [{\citenamefont {Detweiler}(1980)}]{Detweiler:1980gk}%
  \BibitemOpen
  \bibfield  {author} {\bibinfo {author} {\bibfnamefont {S.~L.}\ \bibnamefont
  {Detweiler}},\ }\bibfield  {title} {\bibinfo {title} {{Black Holes and
  Gravitational Waves. III. The Resonant Frequencies of Rotating Holes}},\
  }\href {https://doi.org/10.1086/158109} {\bibfield  {journal} {\bibinfo
  {journal} {Astrophys. J.}\ }\textbf {\bibinfo {volume} {239}},\ \bibinfo
  {pages} {292} (\bibinfo {year} {1980})}\BibitemShut {NoStop}%
\bibitem [{\citenamefont {Leaver}(1985)}]{Leaver:1985ax}%
  \BibitemOpen
  \bibfield  {author} {\bibinfo {author} {\bibfnamefont {E.~W.}\ \bibnamefont
  {Leaver}},\ }\bibfield  {title} {\bibinfo {title} {{An Analytic
  representation for the quasi normal modes of Kerr black holes}},\ }\href
  {https://doi.org/10.1098/rspa.1985.0119} {\bibfield  {journal} {\bibinfo
  {journal} {Proc. Roy. Soc. Lond. A}\ }\textbf {\bibinfo {volume} {402}},\
  \bibinfo {pages} {285} (\bibinfo {year} {1985})}\BibitemShut {NoStop}%
\bibitem [{\citenamefont {Dolan}\ and\ \citenamefont
  {Ottewill}(2009)}]{Dolan:2009nk}%
  \BibitemOpen
  \bibfield  {author} {\bibinfo {author} {\bibfnamefont {S.~R.}\ \bibnamefont
  {Dolan}}\ and\ \bibinfo {author} {\bibfnamefont {A.~C.}\ \bibnamefont
  {Ottewill}},\ }\bibfield  {title} {\bibinfo {title} {{On an Expansion Method
  for Black Hole Quasinormal Modes and Regge Poles}},\ }\href
  {https://doi.org/10.1088/0264-9381/26/22/225003} {\bibfield  {journal}
  {\bibinfo  {journal} {Class. Quant. Grav.}\ }\textbf {\bibinfo {volume}
  {26}},\ \bibinfo {pages} {225003} (\bibinfo {year} {2009})},\ \Eprint
  {https://arxiv.org/abs/0908.0329} {arXiv:0908.0329 [gr-qc]} \BibitemShut
  {NoStop}%
\bibitem [{\citenamefont {Cook}\ and\ \citenamefont
  {Zalutskiy}(2014)}]{Cook:2014cta}%
  \BibitemOpen
  \bibfield  {author} {\bibinfo {author} {\bibfnamefont {G.~B.}\ \bibnamefont
  {Cook}}\ and\ \bibinfo {author} {\bibfnamefont {M.}~\bibnamefont
  {Zalutskiy}},\ }\bibfield  {title} {\bibinfo {title} {{Gravitational
  perturbations of the Kerr geometry: High-accuracy study}},\ }\href
  {https://doi.org/10.1103/PhysRevD.90.124021} {\bibfield  {journal} {\bibinfo
  {journal} {Phys. Rev. D}\ }\textbf {\bibinfo {volume} {90}},\ \bibinfo
  {pages} {124021} (\bibinfo {year} {2014})},\ \Eprint
  {https://arxiv.org/abs/1410.7698} {arXiv:1410.7698 [gr-qc]} \BibitemShut
  {NoStop}%
\bibitem [{\citenamefont {Isi}\ and\ \citenamefont {Farr}(2021)}]{Isi:2021iql}%
  \BibitemOpen
  \bibfield  {author} {\bibinfo {author} {\bibfnamefont {M.}~\bibnamefont
  {Isi}}\ and\ \bibinfo {author} {\bibfnamefont {W.~M.}\ \bibnamefont {Farr}},\
  }\bibfield  {title} {\bibinfo {title} {{Analyzing black-hole ringdowns}},\
  }\Eprint {https://arxiv.org/abs/2107.05609} {arXiv:2107.05609 [gr-qc]}
  \BibitemShut {NoStop}%
\bibitem [{\citenamefont {Press}\ and\ \citenamefont
  {Teukolsky}(1973)}]{Press:1973zz}%
  \BibitemOpen
  \bibfield  {author} {\bibinfo {author} {\bibfnamefont {W.~H.}\ \bibnamefont
  {Press}}\ and\ \bibinfo {author} {\bibfnamefont {S.~A.}\ \bibnamefont
  {Teukolsky}},\ }\bibfield  {title} {\bibinfo {title} {{Perturbations of a
  Rotating Black Hole. II. Dynamical Stability of the Kerr Metric}},\ }\href
  {https://doi.org/10.1086/152445} {\bibfield  {journal} {\bibinfo  {journal}
  {Astrophys. J.}\ }\textbf {\bibinfo {volume} {185}},\ \bibinfo {pages} {649}
  (\bibinfo {year} {1973})}\BibitemShut {NoStop}%
\bibitem [{\citenamefont {Giesler}\ \emph {et~al.}(2019)\citenamefont
  {Giesler}, \citenamefont {Isi}, \citenamefont {Scheel},\ and\ \citenamefont
  {Teukolsky}}]{Giesler:2019uxc}%
  \BibitemOpen
  \bibfield  {author} {\bibinfo {author} {\bibfnamefont {M.}~\bibnamefont
  {Giesler}}, \bibinfo {author} {\bibfnamefont {M.}~\bibnamefont {Isi}},
  \bibinfo {author} {\bibfnamefont {M.~A.}\ \bibnamefont {Scheel}},\ and\
  \bibinfo {author} {\bibfnamefont {S.}~\bibnamefont {Teukolsky}},\ }\bibfield
  {title} {\bibinfo {title} {{Black Hole Ringdown: The Importance of
  Overtones}},\ }\href {https://doi.org/10.1103/PhysRevX.9.041060} {\bibfield
  {journal} {\bibinfo  {journal} {Phys. Rev. X}\ }\textbf {\bibinfo {volume}
  {9}},\ \bibinfo {pages} {041060} (\bibinfo {year} {2019})},\ \Eprint
  {https://arxiv.org/abs/1903.08284} {arXiv:1903.08284 [gr-qc]} \BibitemShut
  {NoStop}%
\bibitem [{\citenamefont {Cook}(2020)}]{Cook:2020otn}%
  \BibitemOpen
  \bibfield  {author} {\bibinfo {author} {\bibfnamefont {G.~B.}\ \bibnamefont
  {Cook}},\ }\bibfield  {title} {\bibinfo {title} {{Aspects of multimode Kerr
  ringdown fitting}},\ }\href {https://doi.org/10.1103/PhysRevD.102.024027}
  {\bibfield  {journal} {\bibinfo  {journal} {Phys. Rev. D}\ }\textbf {\bibinfo
  {volume} {102}},\ \bibinfo {pages} {024027} (\bibinfo {year} {2020})},\
  \Eprint {https://arxiv.org/abs/2004.08347} {arXiv:2004.08347 [gr-qc]}
  \BibitemShut {NoStop}%
\bibitem [{\citenamefont {London}\ \emph {et~al.}(2014)\citenamefont {London},
  \citenamefont {Shoemaker},\ and\ \citenamefont {Healy}}]{London:2014cma}%
  \BibitemOpen
  \bibfield  {author} {\bibinfo {author} {\bibfnamefont {L.}~\bibnamefont
  {London}}, \bibinfo {author} {\bibfnamefont {D.}~\bibnamefont {Shoemaker}},\
  and\ \bibinfo {author} {\bibfnamefont {J.}~\bibnamefont {Healy}},\ }\bibfield
   {title} {\bibinfo {title} {{Modeling ringdown: Beyond the fundamental
  quasinormal modes}},\ }\href {https://doi.org/10.1103/PhysRevD.90.124032}
  {\bibfield  {journal} {\bibinfo  {journal} {Phys. Rev. D}\ }\textbf {\bibinfo
  {volume} {90}},\ \bibinfo {pages} {124032} (\bibinfo {year} {2014})},\
  \bibinfo {note} {[Erratum: Phys.Rev.D 94, 069902 (2016)]},\ \Eprint
  {https://arxiv.org/abs/1404.3197} {arXiv:1404.3197 [gr-qc]} \BibitemShut
  {NoStop}%
\bibitem [{\citenamefont {Baibhav}\ and\ \citenamefont
  {Berti}(2019)}]{Baibhav:2018rfk}%
  \BibitemOpen
  \bibfield  {author} {\bibinfo {author} {\bibfnamefont {V.}~\bibnamefont
  {Baibhav}}\ and\ \bibinfo {author} {\bibfnamefont {E.}~\bibnamefont
  {Berti}},\ }\bibfield  {title} {\bibinfo {title} {{Multimode black hole
  spectroscopy}},\ }\href {https://doi.org/10.1103/PhysRevD.99.024005}
  {\bibfield  {journal} {\bibinfo  {journal} {Phys. Rev. D}\ }\textbf {\bibinfo
  {volume} {99}},\ \bibinfo {pages} {024005} (\bibinfo {year} {2019})},\
  \Eprint {https://arxiv.org/abs/1809.03500} {arXiv:1809.03500 [gr-qc]}
  \BibitemShut {NoStop}%
\bibitem [{\citenamefont {Berti}\ and\ \citenamefont
  {Klein}(2014)}]{Berti:2014fga}%
  \BibitemOpen
  \bibfield  {author} {\bibinfo {author} {\bibfnamefont {E.}~\bibnamefont
  {Berti}}\ and\ \bibinfo {author} {\bibfnamefont {A.}~\bibnamefont {Klein}},\
  }\bibfield  {title} {\bibinfo {title} {{Mixing of spherical and spheroidal
  modes in perturbed Kerr black holes}},\ }\href
  {https://doi.org/10.1103/PhysRevD.90.064012} {\bibfield  {journal} {\bibinfo
  {journal} {Phys. Rev. D}\ }\textbf {\bibinfo {volume} {90}},\ \bibinfo
  {pages} {064012} (\bibinfo {year} {2014})},\ \Eprint
  {https://arxiv.org/abs/1408.1860} {arXiv:1408.1860 [gr-qc]} \BibitemShut
  {NoStop}%
\bibitem [{\citenamefont {Dhani}\ and\ \citenamefont
  {Sathyaprakash}(2021)}]{Dhani:2021vac}%
  \BibitemOpen
  \bibfield  {author} {\bibinfo {author} {\bibfnamefont {A.}~\bibnamefont
  {Dhani}}\ and\ \bibinfo {author} {\bibfnamefont {B.~S.}\ \bibnamefont
  {Sathyaprakash}},\ }\bibfield  {title} {\bibinfo {title} {{Overtones, mirror
  modes, and mode-mixing in binary black hole mergers}},\ }\Eprint
  {https://arxiv.org/abs/2107.14195} {arXiv:2107.14195 [gr-qc]} \BibitemShut
  {NoStop}%
\bibitem [{\citenamefont {Finch}\ and\ \citenamefont
  {Moore}(2021)}]{Finch:2021iip}%
  \BibitemOpen
  \bibfield  {author} {\bibinfo {author} {\bibfnamefont {E.}~\bibnamefont
  {Finch}}\ and\ \bibinfo {author} {\bibfnamefont {C.~J.}\ \bibnamefont
  {Moore}},\ }\bibfield  {title} {\bibinfo {title} {{Modeling the ringdown from
  precessing black hole binaries}},\ }\href
  {https://doi.org/10.1103/PhysRevD.103.084048} {\bibfield  {journal} {\bibinfo
   {journal} {Phys. Rev. D}\ }\textbf {\bibinfo {volume} {103}},\ \bibinfo
  {pages} {084048} (\bibinfo {year} {2021})},\ \Eprint
  {https://arxiv.org/abs/2102.07794} {arXiv:2102.07794 [gr-qc]} \BibitemShut
  {NoStop}%
\bibitem [{\citenamefont {Dhani}(2021)}]{Dhani:2020nik}%
  \BibitemOpen
  \bibfield  {author} {\bibinfo {author} {\bibfnamefont {A.}~\bibnamefont
  {Dhani}},\ }\bibfield  {title} {\bibinfo {title} {{Importance of mirror modes
  in binary black hole ringdown waveform}},\ }\href
  {https://doi.org/10.1103/PhysRevD.103.104048} {\bibfield  {journal} {\bibinfo
   {journal} {Phys. Rev. D}\ }\textbf {\bibinfo {volume} {103}},\ \bibinfo
  {pages} {104048} (\bibinfo {year} {2021})},\ \Eprint
  {https://arxiv.org/abs/2010.08602} {arXiv:2010.08602 [gr-qc]} \BibitemShut
  {NoStop}%
\bibitem [{\citenamefont {Li}\ \emph {et~al.}(2022)\citenamefont {Li},
  \citenamefont {Sun}, \citenamefont {Lo}, \citenamefont {Payne},\ and\
  \citenamefont {Chen}}]{Li:2021wgz}%
  \BibitemOpen
  \bibfield  {author} {\bibinfo {author} {\bibfnamefont {X.}~\bibnamefont
  {Li}}, \bibinfo {author} {\bibfnamefont {L.}~\bibnamefont {Sun}}, \bibinfo
  {author} {\bibfnamefont {R.~K.~L.}\ \bibnamefont {Lo}}, \bibinfo {author}
  {\bibfnamefont {E.}~\bibnamefont {Payne}},\ and\ \bibinfo {author}
  {\bibfnamefont {Y.}~\bibnamefont {Chen}},\ }\bibfield  {title} {\bibinfo
  {title} {{Angular emission patterns of remnant black holes}},\ }\href
  {https://doi.org/10.1103/PhysRevD.105.024016} {\bibfield  {journal} {\bibinfo
   {journal} {Phys. Rev. D}\ }\textbf {\bibinfo {volume} {105}},\ \bibinfo
  {pages} {024016} (\bibinfo {year} {2022})},\ \Eprint
  {https://arxiv.org/abs/2110.03116} {arXiv:2110.03116 [gr-qc]} \BibitemShut
  {NoStop}%
\bibitem [{\citenamefont {Maggiore}\ \emph {et~al.}(2020)\citenamefont
  {Maggiore} \emph {et~al.}}]{Maggiore:2019uih}%
  \BibitemOpen
  \bibfield  {author} {\bibinfo {author} {\bibfnamefont {M.}~\bibnamefont
  {Maggiore}} \emph {et~al.},\ }\bibfield  {title} {\bibinfo {title} {{Science
  Case for the Einstein Telescope}},\ }\href
  {https://doi.org/10.1088/1475-7516/2020/03/050} {\bibfield  {journal}
  {\bibinfo  {journal} {JCAP}\ }\textbf {\bibinfo {volume} {03}},\ \bibinfo
  {pages} {050}},\ \Eprint {https://arxiv.org/abs/1912.02622} {arXiv:1912.02622
  [astro-ph.CO]} \BibitemShut {NoStop}%
\bibitem [{\citenamefont {Baibhav}\ \emph {et~al.}(2019)\citenamefont
  {Baibhav}, \citenamefont {Berti}, \citenamefont {Gerosa}, \citenamefont
  {Mapelli}, \citenamefont {Giacobbo}, \citenamefont {Bouffanais},\ and\
  \citenamefont {Di~Carlo}}]{Baibhav:2019gxm}%
  \BibitemOpen
  \bibfield  {author} {\bibinfo {author} {\bibfnamefont {V.}~\bibnamefont
  {Baibhav}}, \bibinfo {author} {\bibfnamefont {E.}~\bibnamefont {Berti}},
  \bibinfo {author} {\bibfnamefont {D.}~\bibnamefont {Gerosa}}, \bibinfo
  {author} {\bibfnamefont {M.}~\bibnamefont {Mapelli}}, \bibinfo {author}
  {\bibfnamefont {N.}~\bibnamefont {Giacobbo}}, \bibinfo {author}
  {\bibfnamefont {Y.}~\bibnamefont {Bouffanais}},\ and\ \bibinfo {author}
  {\bibfnamefont {U.~N.}\ \bibnamefont {Di~Carlo}},\ }\bibfield  {title}
  {\bibinfo {title} {{Gravitational-wave detection rates for compact binaries
  formed in isolation: LIGO/Virgo O3 and beyond}},\ }\href
  {https://doi.org/10.1103/PhysRevD.100.064060} {\bibfield  {journal} {\bibinfo
   {journal} {Phys. Rev. D}\ }\textbf {\bibinfo {volume} {100}},\ \bibinfo
  {pages} {064060} (\bibinfo {year} {2019})},\ \Eprint
  {https://arxiv.org/abs/1906.04197} {arXiv:1906.04197 [gr-qc]} \BibitemShut
  {NoStop}%
\bibitem [{\citenamefont {Berti}\ \emph {et~al.}(2006)\citenamefont {Berti},
  \citenamefont {Cardoso},\ and\ \citenamefont {Will}}]{Berti:2005ys}%
  \BibitemOpen
  \bibfield  {author} {\bibinfo {author} {\bibfnamefont {E.}~\bibnamefont
  {Berti}}, \bibinfo {author} {\bibfnamefont {V.}~\bibnamefont {Cardoso}},\
  and\ \bibinfo {author} {\bibfnamefont {C.~M.}\ \bibnamefont {Will}},\
  }\bibfield  {title} {\bibinfo {title} {{On gravitational-wave spectroscopy of
  massive black holes with the space interferometer LISA}},\ }\href
  {https://doi.org/10.1103/PhysRevD.73.064030} {\bibfield  {journal} {\bibinfo
  {journal} {Phys. Rev. D}\ }\textbf {\bibinfo {volume} {73}},\ \bibinfo
  {pages} {064030} (\bibinfo {year} {2006})},\ \Eprint
  {https://arxiv.org/abs/gr-qc/0512160} {arXiv:gr-qc/0512160} \BibitemShut
  {NoStop}%
\bibitem [{\citenamefont {Barausse}\ \emph {et~al.}(2020)\citenamefont
  {Barausse} \emph {et~al.}}]{Barausse:2020rsu}%
  \BibitemOpen
  \bibfield  {author} {\bibinfo {author} {\bibfnamefont {E.}~\bibnamefont
  {Barausse}} \emph {et~al.},\ }\bibfield  {title} {\bibinfo {title}
  {{Prospects for Fundamental Physics with LISA}},\ }\href
  {https://doi.org/10.1007/s10714-020-02691-1} {\bibfield  {journal} {\bibinfo
  {journal} {Gen. Rel. Grav.}\ }\textbf {\bibinfo {volume} {52}},\ \bibinfo
  {pages} {81} (\bibinfo {year} {2020})},\ \Eprint
  {https://arxiv.org/abs/2001.09793} {arXiv:2001.09793 [gr-qc]} \BibitemShut
  {NoStop}%
\bibitem [{\citenamefont {Bellovary}\ \emph {et~al.}(2020)\citenamefont
  {Bellovary} \emph {et~al.}}]{Holley-Bockelmann:2020lzp}%
  \BibitemOpen
  \bibfield  {author} {\bibinfo {author} {\bibfnamefont {J.}~\bibnamefont
  {Bellovary}} \emph {et~al.} (\bibinfo {collaboration} {NASA LISA Study
  Team}),\ }\bibfield  {title} {\bibinfo {title} {{Getting Ready for LISA: The
  Data, Support and Preparation Needed to Maximize US Participation in
  Space-Based Gravitational Wave Science}},\ }\Eprint
  {https://arxiv.org/abs/2012.02650} {arXiv:2012.02650 [astro-ph.IM]}
  \BibitemShut {NoStop}%
\bibitem [{\citenamefont {Woodford}\ \emph {et~al.}(2019)\citenamefont
  {Woodford}, \citenamefont {Boyle},\ and\ \citenamefont
  {Pfeiffer}}]{Woodford:2019tlo}%
  \BibitemOpen
  \bibfield  {author} {\bibinfo {author} {\bibfnamefont {C.~J.}\ \bibnamefont
  {Woodford}}, \bibinfo {author} {\bibfnamefont {M.}~\bibnamefont {Boyle}},\
  and\ \bibinfo {author} {\bibfnamefont {H.~P.}\ \bibnamefont {Pfeiffer}},\
  }\bibfield  {title} {\bibinfo {title} {{Compact Binary Waveform
  Center-of-Mass Corrections}},\ }\href
  {https://doi.org/10.1103/PhysRevD.100.124010} {\bibfield  {journal} {\bibinfo
   {journal} {Phys. Rev. D}\ }\textbf {\bibinfo {volume} {100}},\ \bibinfo
  {pages} {124010} (\bibinfo {year} {2019})},\ \Eprint
  {https://arxiv.org/abs/1904.04842} {arXiv:1904.04842 [gr-qc]} \BibitemShut
  {NoStop}%
\bibitem [{\citenamefont {Mitman}\ \emph
  {et~al.}(2021{\natexlab{a}})\citenamefont {Mitman} \emph
  {et~al.}}]{Mitman:2021xkq}%
  \BibitemOpen
  \bibfield  {author} {\bibinfo {author} {\bibfnamefont {K.}~\bibnamefont
  {Mitman}} \emph {et~al.},\ }\bibfield  {title} {\bibinfo {title} {{Fixing the
  BMS frame of numerical relativity waveforms}},\ }\href
  {https://doi.org/10.1103/PhysRevD.104.024051} {\bibfield  {journal} {\bibinfo
   {journal} {Phys. Rev. D}\ }\textbf {\bibinfo {volume} {104}},\ \bibinfo
  {pages} {024051} (\bibinfo {year} {2021}{\natexlab{a}})},\ \Eprint
  {https://arxiv.org/abs/2105.02300} {arXiv:2105.02300 [gr-qc]} \BibitemShut
  {NoStop}%
\bibitem [{\citenamefont {Bondi}\ \emph {et~al.}(1962)\citenamefont {Bondi},
  \citenamefont {Van~der Burg},\ and\ \citenamefont {Metzner}}]{Bondi}%
  \BibitemOpen
  \bibfield  {author} {\bibinfo {author} {\bibfnamefont {H.}~\bibnamefont
  {Bondi}}, \bibinfo {author} {\bibfnamefont {M.~G.~J.}\ \bibnamefont {Van~der
  Burg}},\ and\ \bibinfo {author} {\bibfnamefont {A.~W.~K.}\ \bibnamefont
  {Metzner}},\ }\bibfield  {title} {\bibinfo {title} {Gravitational waves in
  general relativity, {VII}. {W}aves from axi-symmetric isolated system},\
  }\href {https://doi.org/10.1098/rspa.1962.0161} {\bibfield  {journal}
  {\bibinfo  {journal} {Proceedings of the Royal Society of London. Series A.
  Mathematical and Physical Sciences}\ }\textbf {\bibinfo {volume} {269}},\
  \bibinfo {pages} {21} (\bibinfo {year} {1962})}\BibitemShut {NoStop}%
\bibitem [{\citenamefont {Sachs}\ and\ \citenamefont {Bondi}(1962)}]{Sachs}%
  \BibitemOpen
  \bibfield  {author} {\bibinfo {author} {\bibfnamefont {R.~K.}\ \bibnamefont
  {Sachs}}\ and\ \bibinfo {author} {\bibfnamefont {H.}~\bibnamefont {Bondi}},\
  }\bibfield  {title} {\bibinfo {title} {Gravitational waves in general
  relativity {VIII}. {W}aves in asymptotically flat space-time},\ }\href
  {https://doi.org/10.1098/rspa.1962.0206} {\bibfield  {journal} {\bibinfo
  {journal} {Proceedings of the Royal Society of London. Series A. Mathematical
  and Physical Sciences}\ }\textbf {\bibinfo {volume} {270}},\ \bibinfo {pages}
  {103} (\bibinfo {year} {1962})}\BibitemShut {NoStop}%
\bibitem [{\citenamefont {Boyle}(2016{\natexlab{a}})}]{Boyle:2015nqa}%
  \BibitemOpen
  \bibfield  {author} {\bibinfo {author} {\bibfnamefont {M.}~\bibnamefont
  {Boyle}},\ }\bibfield  {title} {\bibinfo {title} {{Transformations of
  asymptotic gravitational-wave data}},\ }\href
  {https://doi.org/10.1103/PhysRevD.93.084031} {\bibfield  {journal} {\bibinfo
  {journal} {Phys. Rev. D}\ }\textbf {\bibinfo {volume} {93}},\ \bibinfo
  {pages} {084031} (\bibinfo {year} {2016}{\natexlab{a}})},\ \Eprint
  {https://arxiv.org/abs/1509.00862} {arXiv:1509.00862 [gr-qc]} \BibitemShut
  {NoStop}%
\bibitem [{\citenamefont {Moreschi}(1988)}]{Moreschi:1988pc}%
  \BibitemOpen
  \bibfield  {author} {\bibinfo {author} {\bibfnamefont {O.~M.}\ \bibnamefont
  {Moreschi}},\ }\bibfield  {title} {\bibinfo {title} {{Supercenter of Mass
  System at Future Null Infinity}},\ }\href
  {https://doi.org/10.1088/0264-9381/5/3/004} {\bibfield  {journal} {\bibinfo
  {journal} {Class. Quant. Grav.}\ }\textbf {\bibinfo {volume} {5}},\ \bibinfo
  {pages} {423} (\bibinfo {year} {1988})}\BibitemShut {NoStop}%
\bibitem [{\citenamefont {Moreschi}\ and\ \citenamefont
  {Dain}(1998)}]{Moreschi:1998mw}%
  \BibitemOpen
  \bibfield  {author} {\bibinfo {author} {\bibfnamefont {O.~M.}\ \bibnamefont
  {Moreschi}}\ and\ \bibinfo {author} {\bibfnamefont {S.}~\bibnamefont
  {Dain}},\ }\bibfield  {title} {\bibinfo {title} {{Rest frame system for
  asymptotically flat space-times}},\ }\href {https://doi.org/10.1063/1.532646}
  {\bibfield  {journal} {\bibinfo  {journal} {J. Math. Phys.}\ }\textbf
  {\bibinfo {volume} {39}},\ \bibinfo {pages} {6631} (\bibinfo {year}
  {1998})},\ \Eprint {https://arxiv.org/abs/gr-qc/0203075}
  {arXiv:gr-qc/0203075} \BibitemShut {NoStop}%
\bibitem [{\citenamefont {Dain}\ and\ \citenamefont
  {Moreschi}(2000)}]{Dain:2002mj}%
  \BibitemOpen
  \bibfield  {author} {\bibinfo {author} {\bibfnamefont {S.}~\bibnamefont
  {Dain}}\ and\ \bibinfo {author} {\bibfnamefont {O.~M.}\ \bibnamefont
  {Moreschi}},\ }\bibfield  {title} {\bibinfo {title} {{General existence proof
  for rest frame systems in asymptotically flat space-time}},\ }\href
  {https://doi.org/10.1088/0264-9381/17/18/305} {\bibfield  {journal} {\bibinfo
   {journal} {Class. Quant. Grav.}\ }\textbf {\bibinfo {volume} {17}},\
  \bibinfo {pages} {3663} (\bibinfo {year} {2000})},\ \Eprint
  {https://arxiv.org/abs/gr-qc/0203048} {arXiv:gr-qc/0203048} \BibitemShut
  {NoStop}%
\bibitem [{\citenamefont {Lim}\ \emph {et~al.}(2019)\citenamefont {Lim},
  \citenamefont {Khanna}, \citenamefont {Apte},\ and\ \citenamefont
  {Hughes}}]{Lim:2019xrb}%
  \BibitemOpen
  \bibfield  {author} {\bibinfo {author} {\bibfnamefont {H.}~\bibnamefont
  {Lim}}, \bibinfo {author} {\bibfnamefont {G.}~\bibnamefont {Khanna}},
  \bibinfo {author} {\bibfnamefont {A.}~\bibnamefont {Apte}},\ and\ \bibinfo
  {author} {\bibfnamefont {S.~A.}\ \bibnamefont {Hughes}},\ }\bibfield  {title}
  {\bibinfo {title} {{Exciting black hole modes via misaligned coalescences:
  II. The mode content of late-time coalescence waveforms}},\ }\href
  {https://doi.org/10.1103/PhysRevD.100.084032} {\bibfield  {journal} {\bibinfo
   {journal} {Phys. Rev. D}\ }\textbf {\bibinfo {volume} {100}},\ \bibinfo
  {pages} {084032} (\bibinfo {year} {2019})},\ \Eprint
  {https://arxiv.org/abs/1901.05902} {arXiv:1901.05902 [gr-qc]} \BibitemShut
  {NoStop}%
\bibitem [{\citenamefont {Stein}(2019)}]{Stein:2019mop}%
  \BibitemOpen
  \bibfield  {author} {\bibinfo {author} {\bibfnamefont {L.~C.}\ \bibnamefont
  {Stein}},\ }\bibfield  {title} {\bibinfo {title} {{qnm: A Python package for
  calculating Kerr quasinormal modes, separation constants, and
  spherical-spheroidal mixing coefficients}},\ }\href
  {https://doi.org/10.21105/joss.01683} {\bibfield  {journal} {\bibinfo
  {journal} {J. Open Source Softw.}\ }\textbf {\bibinfo {volume} {4}},\
  \bibinfo {pages} {1683} (\bibinfo {year} {2019})},\ \Eprint
  {https://arxiv.org/abs/1908.10377} {arXiv:1908.10377 [gr-qc]} \BibitemShut
  {NoStop}%
\bibitem [{\citenamefont {Goldberg}\ \emph {et~al.}(1967)\citenamefont
  {Goldberg}, \citenamefont {MacFarlane}, \citenamefont {Newman}, \citenamefont
  {Rohrlich},\ and\ \citenamefont {Sudarshan}}]{Goldberg:1966uu}%
  \BibitemOpen
  \bibfield  {author} {\bibinfo {author} {\bibfnamefont {J.~N.}\ \bibnamefont
  {Goldberg}}, \bibinfo {author} {\bibfnamefont {A.~J.}\ \bibnamefont
  {MacFarlane}}, \bibinfo {author} {\bibfnamefont {E.~T.}\ \bibnamefont
  {Newman}}, \bibinfo {author} {\bibfnamefont {F.}~\bibnamefont {Rohrlich}},\
  and\ \bibinfo {author} {\bibfnamefont {E.~C.~G.}\ \bibnamefont {Sudarshan}},\
  }\bibfield  {title} {\bibinfo {title} {{Spin s spherical harmonics and
  edth}},\ }\href {https://doi.org/10.1063/1.1705135} {\bibfield  {journal}
  {\bibinfo  {journal} {J. Math. Phys.}\ }\textbf {\bibinfo {volume} {8}},\
  \bibinfo {pages} {2155} (\bibinfo {year} {1967})}\BibitemShut {NoStop}%
\bibitem [{\citenamefont {Newman}\ and\ \citenamefont
  {Penrose}(1966)}]{Newman:1966ub}%
  \BibitemOpen
  \bibfield  {author} {\bibinfo {author} {\bibfnamefont {E.~T.}\ \bibnamefont
  {Newman}}\ and\ \bibinfo {author} {\bibfnamefont {R.}~\bibnamefont
  {Penrose}},\ }\bibfield  {title} {\bibinfo {title} {{Note on the
  Bondi-Metzner-Sachs group}},\ }\href {https://doi.org/10.1063/1.1931221}
  {\bibfield  {journal} {\bibinfo  {journal} {J. Math. Phys.}\ }\textbf
  {\bibinfo {volume} {7}},\ \bibinfo {pages} {863} (\bibinfo {year}
  {1966})}\BibitemShut {NoStop}%
\bibitem [{\citenamefont {London}\ and\ \citenamefont
  {Fauchon-Jones}(2019)}]{London:2018nxs}%
  \BibitemOpen
  \bibfield  {author} {\bibinfo {author} {\bibfnamefont {L.}~\bibnamefont
  {London}}\ and\ \bibinfo {author} {\bibfnamefont {E.}~\bibnamefont
  {Fauchon-Jones}},\ }\bibfield  {title} {\bibinfo {title} {{On modeling for
  Kerr black holes: Basis learning, QNM frequencies, and spherical-spheroidal
  mixing coefficients}},\ }\href {https://doi.org/10.1088/1361-6382/ab2f11}
  {\bibfield  {journal} {\bibinfo  {journal} {Class. Quant. Grav.}\ }\textbf
  {\bibinfo {volume} {36}},\ \bibinfo {pages} {235015} (\bibinfo {year}
  {2019})},\ \Eprint {https://arxiv.org/abs/1810.03550} {arXiv:1810.03550
  [gr-qc]} \BibitemShut {NoStop}%
\bibitem [{\citenamefont {Boyle}(2016{\natexlab{b}})}]{Boyle:2016tjj}%
  \BibitemOpen
  \bibfield  {author} {\bibinfo {author} {\bibfnamefont {M.}~\bibnamefont
  {Boyle}},\ }\bibfield  {title} {\bibinfo {title} {{How should spin-weighted
  spherical functions be defined?}},\ }\href
  {https://doi.org/10.1063/1.4962723} {\bibfield  {journal} {\bibinfo
  {journal} {J. Math. Phys.}\ }\textbf {\bibinfo {volume} {57}},\ \bibinfo
  {pages} {092504} (\bibinfo {year} {2016}{\natexlab{b}})},\ \Eprint
  {https://arxiv.org/abs/1604.08140} {arXiv:1604.08140 [gr-qc]} \BibitemShut
  {NoStop}%
\bibitem [{\citenamefont {{Zel'dovich}}\ and\ \citenamefont
  {{Polnarev}}(1974)}]{Zeldovich_1974}%
  \BibitemOpen
  \bibfield  {author} {\bibinfo {author} {\bibfnamefont {Y.~B.}\ \bibnamefont
  {{Zel'dovich}}}\ and\ \bibinfo {author} {\bibfnamefont {A.~G.}\ \bibnamefont
  {{Polnarev}}},\ }\bibfield  {title} {\bibinfo {title} {{Radiation of
  gravitational waves by a cluster of superdense stars}},\ }\href@noop {}
  {\bibfield  {journal} {\bibinfo  {journal} {Sov. Astron.}\ }\textbf {\bibinfo
  {volume} {18}},\ \bibinfo {pages} {17} (\bibinfo {year} {1974})}\BibitemShut
  {NoStop}%
\bibitem [{\citenamefont {Braginsky}\ and\ \citenamefont
  {Thorne}(1987)}]{Thorne_1987}%
  \BibitemOpen
  \bibfield  {author} {\bibinfo {author} {\bibfnamefont {V.~B.}\ \bibnamefont
  {Braginsky}}\ and\ \bibinfo {author} {\bibfnamefont {K.~S.}\ \bibnamefont
  {Thorne}},\ }\bibfield  {title} {\bibinfo {title} {Gravitational-wave bursts
  with memory and experimental prospects},\ }\href
  {https://doi.org/10.1038/327123a0} {\bibfield  {journal} {\bibinfo  {journal}
  {Nature}\ }\textbf {\bibinfo {volume} {327}},\ \bibinfo {pages} {123}
  (\bibinfo {year} {1987})}\BibitemShut {NoStop}%
\bibitem [{\citenamefont {Christodoulou}(1991)}]{Christodoulou_1991}%
  \BibitemOpen
  \bibfield  {author} {\bibinfo {author} {\bibfnamefont {D.}~\bibnamefont
  {Christodoulou}},\ }\bibfield  {title} {\bibinfo {title} {Nonlinear nature of
  gravitation and gravitational-wave experiments},\ }\href
  {https://doi.org/10.1103/PhysRevLett.67.1486} {\bibfield  {journal} {\bibinfo
   {journal} {Phys. Rev. Lett.}\ }\textbf {\bibinfo {volume} {67}},\ \bibinfo
  {pages} {1486} (\bibinfo {year} {1991})}\BibitemShut {NoStop}%
\bibitem [{\citenamefont {Thorne}(1992)}]{Thorne_1992}%
  \BibitemOpen
  \bibfield  {author} {\bibinfo {author} {\bibfnamefont {K.~S.}\ \bibnamefont
  {Thorne}},\ }\bibfield  {title} {\bibinfo {title} {Gravitational-wave bursts
  with memory: The {C}hristodoulou effect},\ }\href
  {https://doi.org/10.1103/PhysRevD.45.520} {\bibfield  {journal} {\bibinfo
  {journal} {Phys. Rev. D}\ }\textbf {\bibinfo {volume} {45}},\ \bibinfo
  {pages} {520} (\bibinfo {year} {1992})}\BibitemShut {NoStop}%
\bibitem [{\citenamefont {Mitman}\ \emph {et~al.}(2020)\citenamefont {Mitman},
  \citenamefont {Moxon}, \citenamefont {Scheel}, \citenamefont {Teukolsky},
  \citenamefont {Boyle}, \citenamefont {Deppe}, \citenamefont {Kidder},\ and\
  \citenamefont {Throwe}}]{Mitman:2020pbt}%
  \BibitemOpen
  \bibfield  {author} {\bibinfo {author} {\bibfnamefont {K.}~\bibnamefont
  {Mitman}}, \bibinfo {author} {\bibfnamefont {J.}~\bibnamefont {Moxon}},
  \bibinfo {author} {\bibfnamefont {M.~A.}\ \bibnamefont {Scheel}}, \bibinfo
  {author} {\bibfnamefont {S.~A.}\ \bibnamefont {Teukolsky}}, \bibinfo {author}
  {\bibfnamefont {M.}~\bibnamefont {Boyle}}, \bibinfo {author} {\bibfnamefont
  {N.}~\bibnamefont {Deppe}}, \bibinfo {author} {\bibfnamefont {L.~E.}\
  \bibnamefont {Kidder}},\ and\ \bibinfo {author} {\bibfnamefont
  {W.}~\bibnamefont {Throwe}},\ }\bibfield  {title} {\bibinfo {title}
  {{Computation of displacement and spin gravitational memory in numerical
  relativity}},\ }\href {https://doi.org/10.1103/PhysRevD.102.104007}
  {\bibfield  {journal} {\bibinfo  {journal} {Phys. Rev. D}\ }\textbf {\bibinfo
  {volume} {102}},\ \bibinfo {pages} {104007} (\bibinfo {year} {2020})},\
  \Eprint {https://arxiv.org/abs/2007.11562} {arXiv:2007.11562 [gr-qc]}
  \BibitemShut {NoStop}%
\bibitem [{\citenamefont {Grant}\ and\ \citenamefont
  {Nichols}(2021)}]{Grant:2021hga}%
  \BibitemOpen
  \bibfield  {author} {\bibinfo {author} {\bibfnamefont {A.~M.}\ \bibnamefont
  {Grant}}\ and\ \bibinfo {author} {\bibfnamefont {D.~A.}\ \bibnamefont
  {Nichols}},\ }\bibfield  {title} {\bibinfo {title} {{Persistent gravitational
  wave observables: Curve deviation in asymptotically flat spacetimes}},\
  }\Eprint {https://arxiv.org/abs/2109.03832} {arXiv:2109.03832 [gr-qc]}
  \BibitemShut {NoStop}%
\bibitem [{\citenamefont {Geroch}\ \emph {et~al.}(1973)\citenamefont {Geroch},
  \citenamefont {Held},\ and\ \citenamefont {Penrose}}]{GHP1973}%
  \BibitemOpen
  \bibfield  {author} {\bibinfo {author} {\bibfnamefont {R.}~\bibnamefont
  {Geroch}}, \bibinfo {author} {\bibfnamefont {A.}~\bibnamefont {Held}},\ and\
  \bibinfo {author} {\bibfnamefont {R.}~\bibnamefont {Penrose}},\ }\bibfield
  {title} {\bibinfo {title} {A space-time calculus based on pairs of null
  directions},\ }\href {https://doi.org/10.1063/1.1666410} {\bibfield
  {journal} {\bibinfo  {journal} {J. Math. Phys.}\ }\textbf {\bibinfo {volume}
  {14}},\ \bibinfo {pages} {874} (\bibinfo {year} {1973})}\BibitemShut
  {NoStop}%
\bibitem [{\citenamefont {Flanagan}\ and\ \citenamefont
  {Nichols}(2017)}]{Flanagan:2015pxa}%
  \BibitemOpen
  \bibfield  {author} {\bibinfo {author} {\bibfnamefont {E.~E.}\ \bibnamefont
  {Flanagan}}\ and\ \bibinfo {author} {\bibfnamefont {D.~A.}\ \bibnamefont
  {Nichols}},\ }\bibfield  {title} {\bibinfo {title} {{Conserved charges of the
  extended Bondi-Metzner-Sachs algebra}},\ }\href
  {https://doi.org/10.1103/PhysRevD.95.044002} {\bibfield  {journal} {\bibinfo
  {journal} {Phys. Rev. D}\ }\textbf {\bibinfo {volume} {95}},\ \bibinfo
  {pages} {044002} (\bibinfo {year} {2017})},\ \Eprint
  {https://arxiv.org/abs/1510.03386} {arXiv:1510.03386 [hep-th]} \BibitemShut
  {NoStop}%
\bibitem [{\citenamefont {Iozzo}\ \emph
  {et~al.}(2021{\natexlab{a}})\citenamefont {Iozzo}, \citenamefont {Boyle},
  \citenamefont {Deppe}, \citenamefont {Moxon}, \citenamefont {Scheel},
  \citenamefont {Kidder}, \citenamefont {Pfeiffer},\ and\ \citenamefont
  {Teukolsky}}]{iozzo2020improving}%
  \BibitemOpen
  \bibfield  {author} {\bibinfo {author} {\bibfnamefont {D.~A.~B.}\
  \bibnamefont {Iozzo}}, \bibinfo {author} {\bibfnamefont {M.}~\bibnamefont
  {Boyle}}, \bibinfo {author} {\bibfnamefont {N.}~\bibnamefont {Deppe}},
  \bibinfo {author} {\bibfnamefont {J.}~\bibnamefont {Moxon}}, \bibinfo
  {author} {\bibfnamefont {M.~A.}\ \bibnamefont {Scheel}}, \bibinfo {author}
  {\bibfnamefont {L.~E.}\ \bibnamefont {Kidder}}, \bibinfo {author}
  {\bibfnamefont {H.~P.}\ \bibnamefont {Pfeiffer}},\ and\ \bibinfo {author}
  {\bibfnamefont {S.~A.}\ \bibnamefont {Teukolsky}},\ }\bibfield  {title}
  {\bibinfo {title} {Extending gravitational wave extraction using weyl
  characteristic fields},\ }\href
  {http://dx.doi.org/10.1103/PhysRevD.103.024039} {\bibfield  {journal}
  {\bibinfo  {journal} {Phys. Rev. D}\ }\textbf {\bibinfo {volume} {103}}
  (\bibinfo {year} {2021}{\natexlab{a}})},\ \Eprint
  {https://arxiv.org/abs/2010.15200} {arXiv:2010.15200 [gr-qc]} \BibitemShut
  {NoStop}%
\bibitem [{\citenamefont {Moreschi}(1986)}]{Moreschi1986}%
  \BibitemOpen
  \bibfield  {author} {\bibinfo {author} {\bibfnamefont {O.~M.}\ \bibnamefont
  {Moreschi}},\ }\bibfield  {title} {\bibinfo {title} {{On angular momentum at
  future null infinity}},\ }\href {https://doi.org/10.1088/0264-9381/3/4/006}
  {\bibfield  {journal} {\bibinfo  {journal} {Class. Quantum Gravity}\ }\textbf
  {\bibinfo {volume} {3}},\ \bibinfo {pages} {503} (\bibinfo {year}
  {1986})}\BibitemShut {NoStop}%
\bibitem [{\citenamefont {Harris}\ \emph {et~al.}(2020)\citenamefont {Harris}
  \emph {et~al.}}]{Harris:2020xlr}%
  \BibitemOpen
  \bibfield  {author} {\bibinfo {author} {\bibfnamefont {C.~R.}\ \bibnamefont
  {Harris}} \emph {et~al.},\ }\bibfield  {title} {\bibinfo {title} {{Array
  programming with NumPy}},\ }\href {https://doi.org/10.1038/s41586-020-2649-2}
  {\bibfield  {journal} {\bibinfo  {journal} {Nature}\ }\textbf {\bibinfo
  {volume} {585}},\ \bibinfo {pages} {357} (\bibinfo {year} {2020})},\ \Eprint
  {https://arxiv.org/abs/2006.10256} {arXiv:2006.10256 [cs.MS]} \BibitemShut
  {NoStop}%
\bibitem [{\citenamefont {Virtanen}\ \emph {et~al.}(2020)\citenamefont
  {Virtanen} \emph {et~al.}}]{Virtanen:2019joe}%
  \BibitemOpen
  \bibfield  {author} {\bibinfo {author} {\bibfnamefont {P.}~\bibnamefont
  {Virtanen}} \emph {et~al.},\ }\bibfield  {title} {\bibinfo {title} {{SciPy
  1.0--Fundamental Algorithms for Scientific Computing in Python}},\ }\href
  {https://doi.org/10.1038/s41592-019-0686-2} {\bibfield  {journal} {\bibinfo
  {journal} {Nature Meth.}\ }\textbf {\bibinfo {volume} {17}},\ \bibinfo
  {pages} {261} (\bibinfo {year} {2020})},\ \Eprint
  {https://arxiv.org/abs/1907.10121} {arXiv:1907.10121 [cs.MS]} \BibitemShut
  {NoStop}%
\bibitem [{\citenamefont {Gao}\ and\ \citenamefont {Han}(2012)}]{NelderMead}%
  \BibitemOpen
  \bibfield  {author} {\bibinfo {author} {\bibfnamefont {F.}~\bibnamefont
  {Gao}}\ and\ \bibinfo {author} {\bibfnamefont {L.}~\bibnamefont {Han}},\
  }\bibfield  {title} {\bibinfo {title} {Implementing the nelder-mead simplex
  algorithm with adaptive parameters},\ }\href
  {https://doi.org/10.1007/s10589-010-9329-3} {\bibfield  {journal} {\bibinfo
  {journal} {Computational Optimization and Applications}\ }\textbf {\bibinfo
  {volume} {51}},\ \bibinfo {pages} {259} (\bibinfo {year} {2012})}\BibitemShut
  {NoStop}%
\bibitem [{\citenamefont {Iozzo}\ \emph
  {et~al.}(2021{\natexlab{b}})\citenamefont {Iozzo} \emph
  {et~al.}}]{Iozzo:2021vnq}%
  \BibitemOpen
  \bibfield  {author} {\bibinfo {author} {\bibfnamefont {D.~A.~B.}\
  \bibnamefont {Iozzo}} \emph {et~al.},\ }\bibfield  {title} {\bibinfo {title}
  {{Comparing Remnant Properties from Horizon Data and Asymptotic Data in
  Numerical Relativity}},\ }\href {https://doi.org/10.1103/PhysRevD.103.124029}
  {\bibfield  {journal} {\bibinfo  {journal} {Phys. Rev. D}\ }\textbf {\bibinfo
  {volume} {103}},\ \bibinfo {pages} {124029} (\bibinfo {year}
  {2021}{\natexlab{b}})},\ \Eprint {https://arxiv.org/abs/2104.07052}
  {arXiv:2104.07052 [gr-qc]} \BibitemShut {NoStop}%
\bibitem [{Ext()}]{ExtCCECatalog}%
  \BibitemOpen
  \href@noop {} {\bibinfo {title} {{SXS Ext-CCE Waveform Database}}},\ \bibinfo
  {howpublished}
  {\url{https://data.black-holes.org/waveforms/extcce_catalog.html}}\BibitemShut
  {NoStop}%
\bibitem [{SXS()}]{SXSCatalog}%
  \BibitemOpen
  \href@noop {} {\bibinfo {title} {{SXS Gravitational Waveform Database}}},\
  \bibinfo {howpublished}
  {\url{http://www.black-holes.org/waveforms}}\BibitemShut {NoStop}%
\bibitem [{SpE()}]{SpECCode}%
  \BibitemOpen
  \href@noop {} {}\bibinfo {howpublished}
  {\url{https://www.black-holes.org/code/SpEC.html}}\BibitemShut {NoStop}%
\bibitem [{\citenamefont {Iozzo}\ \emph
  {et~al.}(2021{\natexlab{c}})\citenamefont {Iozzo}, \citenamefont {Boyle},
  \citenamefont {Deppe}, \citenamefont {Moxon}, \citenamefont {Scheel},
  \citenamefont {Kidder}, \citenamefont {Pfeiffer},\ and\ \citenamefont
  {Teukolsky}}]{Iozzo:2020jcu}%
  \BibitemOpen
  \bibfield  {author} {\bibinfo {author} {\bibfnamefont {D.~A.~B.}\
  \bibnamefont {Iozzo}}, \bibinfo {author} {\bibfnamefont {M.}~\bibnamefont
  {Boyle}}, \bibinfo {author} {\bibfnamefont {N.}~\bibnamefont {Deppe}},
  \bibinfo {author} {\bibfnamefont {J.}~\bibnamefont {Moxon}}, \bibinfo
  {author} {\bibfnamefont {M.~A.}\ \bibnamefont {Scheel}}, \bibinfo {author}
  {\bibfnamefont {L.~E.}\ \bibnamefont {Kidder}}, \bibinfo {author}
  {\bibfnamefont {H.~P.}\ \bibnamefont {Pfeiffer}},\ and\ \bibinfo {author}
  {\bibfnamefont {S.~A.}\ \bibnamefont {Teukolsky}},\ }\bibfield  {title}
  {\bibinfo {title} {{Extending gravitational wave extraction using Weyl
  characteristic fields}},\ }\href
  {https://doi.org/10.1103/PhysRevD.103.024039} {\bibfield  {journal} {\bibinfo
   {journal} {Phys. Rev. D}\ }\textbf {\bibinfo {volume} {103}},\ \bibinfo
  {pages} {024039} (\bibinfo {year} {2021}{\natexlab{c}})},\ \Eprint
  {https://arxiv.org/abs/2010.15200} {arXiv:2010.15200 [gr-qc]} \BibitemShut
  {NoStop}%
\bibitem [{\citenamefont {Moxon}\ \emph {et~al.}(2020)\citenamefont {Moxon},
  \citenamefont {Scheel},\ and\ \citenamefont {Teukolsky}}]{Moxon:2020gha}%
  \BibitemOpen
  \bibfield  {author} {\bibinfo {author} {\bibfnamefont {J.}~\bibnamefont
  {Moxon}}, \bibinfo {author} {\bibfnamefont {M.~A.}\ \bibnamefont {Scheel}},\
  and\ \bibinfo {author} {\bibfnamefont {S.~A.}\ \bibnamefont {Teukolsky}},\
  }\bibfield  {title} {\bibinfo {title} {{Improved Cauchy-characteristic
  evolution system for high-precision numerical relativity waveforms}},\ }\href
  {https://doi.org/10.1103/PhysRevD.102.044052} {\bibfield  {journal} {\bibinfo
   {journal} {Phys. Rev. D}\ }\textbf {\bibinfo {volume} {102}},\ \bibinfo
  {pages} {044052} (\bibinfo {year} {2020})},\ \Eprint
  {https://arxiv.org/abs/2007.01339} {arXiv:2007.01339 [gr-qc]} \BibitemShut
  {NoStop}%
\bibitem [{\citenamefont {Moxon}\ \emph {et~al.}(2021)\citenamefont {Moxon},
  \citenamefont {Scheel}, \citenamefont {Teukolsky}, \citenamefont {Deppe},
  \citenamefont {Fischer}, \citenamefont {H\'ebert}, \citenamefont {Kidder},\
  and\ \citenamefont {Throwe}}]{Moxon:2021gbv}%
  \BibitemOpen
  \bibfield  {author} {\bibinfo {author} {\bibfnamefont {J.}~\bibnamefont
  {Moxon}}, \bibinfo {author} {\bibfnamefont {M.~A.}\ \bibnamefont {Scheel}},
  \bibinfo {author} {\bibfnamefont {S.~A.}\ \bibnamefont {Teukolsky}}, \bibinfo
  {author} {\bibfnamefont {N.}~\bibnamefont {Deppe}}, \bibinfo {author}
  {\bibfnamefont {N.}~\bibnamefont {Fischer}}, \bibinfo {author} {\bibfnamefont
  {F.}~\bibnamefont {H\'ebert}}, \bibinfo {author} {\bibfnamefont {L.~E.}\
  \bibnamefont {Kidder}},\ and\ \bibinfo {author} {\bibfnamefont
  {W.}~\bibnamefont {Throwe}},\ }\bibfield  {title} {\bibinfo {title} {{The
  SpECTRE Cauchy-characteristic evolution system for rapid, precise waveform
  extraction}},\ }\Eprint {https://arxiv.org/abs/2110.08635} {arXiv:2110.08635
  [gr-qc]} \BibitemShut {NoStop}%
\bibitem [{\citenamefont {Boyle}\ \emph {et~al.}(2020)\citenamefont {Boyle},
  \citenamefont {Iozzo},\ and\ \citenamefont {Stein}}]{scri_url}%
  \BibitemOpen
  \bibfield  {author} {\bibinfo {author} {\bibfnamefont {M.}~\bibnamefont
  {Boyle}}, \bibinfo {author} {\bibfnamefont {D.}~\bibnamefont {Iozzo}},\ and\
  \bibinfo {author} {\bibfnamefont {L.~C.}\ \bibnamefont {Stein}},\ }\href
  {https://doi.org/10.5281/zenodo.4041972} {\bibinfo {title} {moble/scri:
  v1.2}} (\bibinfo {year} {2020})\BibitemShut {NoStop}%
\bibitem [{\citenamefont {Boyle}(2013)}]{Boyle:2013nka}%
  \BibitemOpen
  \bibfield  {author} {\bibinfo {author} {\bibfnamefont {M.}~\bibnamefont
  {Boyle}},\ }\bibfield  {title} {\bibinfo {title} {{Angular velocity of
  gravitational radiation from precessing binaries and the corotating frame}},\
  }\href {https://doi.org/10.1103/PhysRevD.87.104006} {\bibfield  {journal}
  {\bibinfo  {journal} {Phys. Rev. D}\ }\textbf {\bibinfo {volume} {87}},\
  \bibinfo {pages} {104006} (\bibinfo {year} {2013})},\ \Eprint
  {https://arxiv.org/abs/1302.2919} {arXiv:1302.2919 [gr-qc]} \BibitemShut
  {NoStop}%
\bibitem [{\citenamefont {Boyle}\ \emph {et~al.}(2014)\citenamefont {Boyle},
  \citenamefont {Kidder}, \citenamefont {Ossokine},\ and\ \citenamefont
  {Pfeiffer}}]{Boyle:2014ioa}%
  \BibitemOpen
  \bibfield  {author} {\bibinfo {author} {\bibfnamefont {M.}~\bibnamefont
  {Boyle}}, \bibinfo {author} {\bibfnamefont {L.~E.}\ \bibnamefont {Kidder}},
  \bibinfo {author} {\bibfnamefont {S.}~\bibnamefont {Ossokine}},\ and\
  \bibinfo {author} {\bibfnamefont {H.~P.}\ \bibnamefont {Pfeiffer}},\
  }\bibfield  {title} {\bibinfo {title} {{Gravitational-wave modes from
  precessing black-hole binaries}},\ }\Eprint {https://arxiv.org/abs/1409.4431}
  {arXiv:1409.4431 [gr-qc]} \BibitemShut {NoStop}%
\bibitem [{\citenamefont {Deppe}\ \emph {et~al.}(2020)\citenamefont {Deppe},
  \citenamefont {Throwe}, \citenamefont {Kidder}, \citenamefont {Fischer},
  \citenamefont {Armaza}, \citenamefont {Bonilla}, \citenamefont {Hébert},
  \citenamefont {Kumar}, \citenamefont {Lovelace}, \citenamefont {Moxon},
  \citenamefont {O'Shea}, \citenamefont {Pfeiffer}, \citenamefont {Scheel},
  \citenamefont {Teukolsky}, \citenamefont {Anantpurkar}, \citenamefont
  {Boyle}, \citenamefont {Foucart}, \citenamefont {Giesler}, \citenamefont
  {Iozzo}, \citenamefont {Legred}, \citenamefont {Li}, \citenamefont {Macedo},
  \citenamefont {Melchor}, \citenamefont {Morales}, \citenamefont {Ramirez},
  \citenamefont {Rüter}, \citenamefont {Sanchez}, \citenamefont {Thomas},\
  and\ \citenamefont {Wlodarczyk}}]{CodeSpECTRE}%
  \BibitemOpen
  \bibfield  {author} {\bibinfo {author} {\bibfnamefont {N.}~\bibnamefont
  {Deppe}}, \bibinfo {author} {\bibfnamefont {W.}~\bibnamefont {Throwe}},
  \bibinfo {author} {\bibfnamefont {L.~E.}\ \bibnamefont {Kidder}}, \bibinfo
  {author} {\bibfnamefont {N.~L.}\ \bibnamefont {Fischer}}, \bibinfo {author}
  {\bibfnamefont {C.}~\bibnamefont {Armaza}}, \bibinfo {author} {\bibfnamefont
  {G.~S.}\ \bibnamefont {Bonilla}}, \bibinfo {author} {\bibfnamefont
  {F.}~\bibnamefont {Hébert}}, \bibinfo {author} {\bibfnamefont
  {P.}~\bibnamefont {Kumar}}, \bibinfo {author} {\bibfnamefont
  {G.}~\bibnamefont {Lovelace}}, \bibinfo {author} {\bibfnamefont
  {J.}~\bibnamefont {Moxon}}, \bibinfo {author} {\bibfnamefont
  {E.}~\bibnamefont {O'Shea}}, \bibinfo {author} {\bibfnamefont {H.~P.}\
  \bibnamefont {Pfeiffer}}, \bibinfo {author} {\bibfnamefont {M.~A.}\
  \bibnamefont {Scheel}}, \bibinfo {author} {\bibfnamefont {S.~A.}\
  \bibnamefont {Teukolsky}}, \bibinfo {author} {\bibfnamefont {I.}~\bibnamefont
  {Anantpurkar}}, \bibinfo {author} {\bibfnamefont {M.}~\bibnamefont {Boyle}},
  \bibinfo {author} {\bibfnamefont {F.}~\bibnamefont {Foucart}}, \bibinfo
  {author} {\bibfnamefont {M.}~\bibnamefont {Giesler}}, \bibinfo {author}
  {\bibfnamefont {D.~A.~B.}\ \bibnamefont {Iozzo}}, \bibinfo {author}
  {\bibfnamefont {I.}~\bibnamefont {Legred}}, \bibinfo {author} {\bibfnamefont
  {D.}~\bibnamefont {Li}}, \bibinfo {author} {\bibfnamefont {A.}~\bibnamefont
  {Macedo}}, \bibinfo {author} {\bibfnamefont {D.}~\bibnamefont {Melchor}},
  \bibinfo {author} {\bibfnamefont {M.}~\bibnamefont {Morales}}, \bibinfo
  {author} {\bibfnamefont {T.}~\bibnamefont {Ramirez}}, \bibinfo {author}
  {\bibfnamefont {H.~R.}\ \bibnamefont {Rüter}}, \bibinfo {author}
  {\bibfnamefont {J.}~\bibnamefont {Sanchez}}, \bibinfo {author} {\bibfnamefont
  {S.}~\bibnamefont {Thomas}},\ and\ \bibinfo {author} {\bibfnamefont
  {T.}~\bibnamefont {Wlodarczyk}},\ }\href
  {https://doi.org/10.5281/zenodo.4290405} {\bibinfo {title} {{SpECTRE}}}
  (\bibinfo {year} {2020})\BibitemShut {NoStop}%
\bibitem [{\citenamefont {Mitman}\ \emph
  {et~al.}(2021{\natexlab{b}})\citenamefont {Mitman} \emph
  {et~al.}}]{Mitman:2020bjf}%
  \BibitemOpen
  \bibfield  {author} {\bibinfo {author} {\bibfnamefont {K.}~\bibnamefont
  {Mitman}} \emph {et~al.},\ }\bibfield  {title} {\bibinfo {title} {{Adding
  gravitational memory to waveform catalogs using BMS balance laws}},\ }\href
  {https://doi.org/10.1103/PhysRevD.103.024031} {\bibfield  {journal} {\bibinfo
   {journal} {Phys. Rev. D}\ }\textbf {\bibinfo {volume} {103}},\ \bibinfo
  {pages} {024031} (\bibinfo {year} {2021}{\natexlab{b}})},\ \Eprint
  {https://arxiv.org/abs/2011.01309} {arXiv:2011.01309 [gr-qc]} \BibitemShut
  {NoStop}%
\bibitem [{\citenamefont {Sarbach}\ and\ \citenamefont
  {Tiglio}(2001)}]{Sarbach:2001qq}%
  \BibitemOpen
  \bibfield  {author} {\bibinfo {author} {\bibfnamefont {O.}~\bibnamefont
  {Sarbach}}\ and\ \bibinfo {author} {\bibfnamefont {M.}~\bibnamefont
  {Tiglio}},\ }\bibfield  {title} {\bibinfo {title} {{Gauge invariant
  perturbations of Schwarzschild black holes in horizon penetrating
  coordinates}},\ }\href {https://doi.org/10.1103/PhysRevD.64.084016}
  {\bibfield  {journal} {\bibinfo  {journal} {Phys. Rev. D}\ }\textbf {\bibinfo
  {volume} {64}},\ \bibinfo {pages} {084016} (\bibinfo {year} {2001})},\
  \Eprint {https://arxiv.org/abs/gr-qc/0104061} {arXiv:gr-qc/0104061}
  \BibitemShut {NoStop}%
\bibitem [{\citenamefont {Regge}\ and\ \citenamefont
  {Wheeler}(1957)}]{Regge:1957td}%
  \BibitemOpen
  \bibfield  {author} {\bibinfo {author} {\bibfnamefont {T.}~\bibnamefont
  {Regge}}\ and\ \bibinfo {author} {\bibfnamefont {J.~A.}\ \bibnamefont
  {Wheeler}},\ }\bibfield  {title} {\bibinfo {title} {{Stability of a
  Schwarzschild singularity}},\ }\href
  {https://doi.org/10.1103/PhysRev.108.1063} {\bibfield  {journal} {\bibinfo
  {journal} {Phys. Rev.}\ }\textbf {\bibinfo {volume} {108}},\ \bibinfo {pages}
  {1063} (\bibinfo {year} {1957})}\BibitemShut {NoStop}%
\bibitem [{\citenamefont {Zerilli}(1970)}]{Zerilli:1970se}%
  \BibitemOpen
  \bibfield  {author} {\bibinfo {author} {\bibfnamefont {F.~J.}\ \bibnamefont
  {Zerilli}},\ }\bibfield  {title} {\bibinfo {title} {{Effective potential for
  even parity Regge-Wheeler gravitational perturbation equations}},\ }\href
  {https://doi.org/10.1103/PhysRevLett.24.737} {\bibfield  {journal} {\bibinfo
  {journal} {Phys. Rev. Lett.}\ }\textbf {\bibinfo {volume} {24}},\ \bibinfo
  {pages} {737} (\bibinfo {year} {1970})}\BibitemShut {NoStop}%
\bibitem [{\citenamefont {Boyle}\ and\ \citenamefont
  {Mroue}(2009)}]{Boyle:2009vi}%
  \BibitemOpen
  \bibfield  {author} {\bibinfo {author} {\bibfnamefont {M.}~\bibnamefont
  {Boyle}}\ and\ \bibinfo {author} {\bibfnamefont {A.~H.}\ \bibnamefont
  {Mroue}},\ }\bibfield  {title} {\bibinfo {title} {{Extrapolating
  gravitational-wave data from numerical simulations}},\ }\href
  {https://doi.org/10.1103/PhysRevD.80.124045} {\bibfield  {journal} {\bibinfo
  {journal} {Phys. Rev. D}\ }\textbf {\bibinfo {volume} {80}},\ \bibinfo
  {pages} {124045} (\bibinfo {year} {2009})},\ \Eprint
  {https://arxiv.org/abs/0905.3177} {arXiv:0905.3177 [gr-qc]} \BibitemShut
  {NoStop}%
\bibitem [{\citenamefont {Forteza}\ and\ \citenamefont
  {Mourier}(2021)}]{Forteza:2021wfq}%
  \BibitemOpen
  \bibfield  {author} {\bibinfo {author} {\bibfnamefont {X.~J.}\ \bibnamefont
  {Forteza}}\ and\ \bibinfo {author} {\bibfnamefont {P.}~\bibnamefont
  {Mourier}},\ }\bibfield  {title} {\bibinfo {title} {{High-overtone fits to
  numerical relativity ringdowns: beyond the dismissed n=8 special tone}},\
  }\Eprint {https://arxiv.org/abs/2107.11829} {arXiv:2107.11829 [gr-qc]}
  \BibitemShut {NoStop}%
\bibitem [{\citenamefont {Varma}\ \emph {et~al.}(2019)\citenamefont {Varma},
  \citenamefont {Field}, \citenamefont {Scheel}, \citenamefont {Blackman},
  \citenamefont {Kidder},\ and\ \citenamefont {Pfeiffer}}]{Varma:2018mmi}%
  \BibitemOpen
  \bibfield  {author} {\bibinfo {author} {\bibfnamefont {V.}~\bibnamefont
  {Varma}}, \bibinfo {author} {\bibfnamefont {S.~E.}\ \bibnamefont {Field}},
  \bibinfo {author} {\bibfnamefont {M.~A.}\ \bibnamefont {Scheel}}, \bibinfo
  {author} {\bibfnamefont {J.}~\bibnamefont {Blackman}}, \bibinfo {author}
  {\bibfnamefont {L.~E.}\ \bibnamefont {Kidder}},\ and\ \bibinfo {author}
  {\bibfnamefont {H.~P.}\ \bibnamefont {Pfeiffer}},\ }\bibfield  {title}
  {\bibinfo {title} {{Surrogate model of hybridized numerical relativity binary
  black hole waveforms}},\ }\href {https://doi.org/10.1103/PhysRevD.99.064045}
  {\bibfield  {journal} {\bibinfo  {journal} {Phys. Rev. D}\ }\textbf {\bibinfo
  {volume} {99}},\ \bibinfo {pages} {064045} (\bibinfo {year} {2019})},\
  \Eprint {https://arxiv.org/abs/1812.07865} {arXiv:1812.07865 [gr-qc]}
  \BibitemShut {NoStop}%
\bibitem [{\citenamefont {Campbell}\ and\ \citenamefont
  {Morgan}(1971)}]{Campbell:1970ww}%
  \BibitemOpen
  \bibfield  {author} {\bibinfo {author} {\bibfnamefont {W.~B.}\ \bibnamefont
  {Campbell}}\ and\ \bibinfo {author} {\bibfnamefont {T.}~\bibnamefont
  {Morgan}},\ }\bibfield  {title} {\bibinfo {title} {{Debye Potentials for the
  Gravitational Field}},\ }\href {https://doi.org/10.1016/0031-8914(71)90074-7}
  {\bibfield  {journal} {\bibinfo  {journal} {Physica}\ }\textbf {\bibinfo
  {volume} {53}},\ \bibinfo {pages} {264} (\bibinfo {year} {1971})}\BibitemShut
  {NoStop}%
\bibitem [{\citenamefont {Guyon}\ and\ \citenamefont
  {Elisseeff}(2003)}]{Guyon2003AnIT}%
  \BibitemOpen
  \bibfield  {author} {\bibinfo {author} {\bibfnamefont {I.}~\bibnamefont
  {Guyon}}\ and\ \bibinfo {author} {\bibfnamefont {A.}~\bibnamefont
  {Elisseeff}},\ }\bibfield  {title} {\bibinfo {title} {An introduction to
  variable and feature selection},\ }\href
  {https://www.jmlr.org/papers/v3/guyon03a.html} {\bibfield  {journal}
  {\bibinfo  {journal} {J. Mach. Learn. Res.}\ }\textbf {\bibinfo {volume}
  {3}},\ \bibinfo {pages} {1157} (\bibinfo {year} {2003})}\BibitemShut
  {NoStop}%
\bibitem [{\citenamefont {Isi}\ \emph {et~al.}(2019)\citenamefont {Isi},
  \citenamefont {Giesler}, \citenamefont {Farr}, \citenamefont {Scheel},\ and\
  \citenamefont {Teukolsky}}]{Isi:2019aib}%
  \BibitemOpen
  \bibfield  {author} {\bibinfo {author} {\bibfnamefont {M.}~\bibnamefont
  {Isi}}, \bibinfo {author} {\bibfnamefont {M.}~\bibnamefont {Giesler}},
  \bibinfo {author} {\bibfnamefont {W.~M.}\ \bibnamefont {Farr}}, \bibinfo
  {author} {\bibfnamefont {M.~A.}\ \bibnamefont {Scheel}},\ and\ \bibinfo
  {author} {\bibfnamefont {S.~A.}\ \bibnamefont {Teukolsky}},\ }\bibfield
  {title} {\bibinfo {title} {{Testing the no-hair theorem with GW150914}},\
  }\href {https://doi.org/10.1103/PhysRevLett.123.111102} {\bibfield  {journal}
  {\bibinfo  {journal} {Phys. Rev. Lett.}\ }\textbf {\bibinfo {volume} {123}},\
  \bibinfo {pages} {111102} (\bibinfo {year} {2019})},\ \Eprint
  {https://arxiv.org/abs/1905.00869} {arXiv:1905.00869 [gr-qc]} \BibitemShut
  {NoStop}%
\bibitem [{\citenamefont {Hunter}(2007)}]{Hunter:2007ouj}%
  \BibitemOpen
  \bibfield  {author} {\bibinfo {author} {\bibfnamefont {J.~D.}\ \bibnamefont
  {Hunter}},\ }\bibfield  {title} {\bibinfo {title} {{Matplotlib: A 2D Graphics
  Environment}},\ }\href {https://doi.org/10.1109/MCSE.2007.55} {\bibfield
  {journal} {\bibinfo  {journal} {Comput. Sci. Eng.}\ }\textbf {\bibinfo
  {volume} {9}},\ \bibinfo {pages} {90} (\bibinfo {year} {2007})}\BibitemShut
  {NoStop}%
\end{thebibliography}%

\end{document}